\documentclass[%
 aip,
 amsmath,amssymb,
 reprint,
]{revtex4-1}

\usepackage{graphicx}
\usepackage{dcolumn}
\usepackage{bm}
\usepackage{floatrow}
\usepackage{mathptmx,lipsum}
\usepackage{physics}
\usepackage{floatrow}
\usepackage{amsmath}
\usepackage{relsize}
\usepackage{bbold}
\usepackage{xcolor}
\setcitestyle{square}
\usepackage{verbatim}
\usepackage{mathtools,txfonts,dsfont}
\usepackage[most]{tcolorbox}
\usepackage{environ}
\usepackage{hyperref}
\usepackage[caption=false,position=top]{subfig}
\usepackage{makecell}
\usepackage{svg}
\usepackage{soul}
\soulregister\citep7
\soulregister\ref7
\soulregister\eqref7
\makeatletter
\def\blfootnote{\gdef\@thefnmark{}\@footnotetext}
\makeatother

\definecolor{palerobineggblue}{rgb}{0.59, 0.87, 0.82}
\definecolor{palepink}{rgb}{0.98, 0.85, 0.87}
\definecolor{palecornflowerblue}{rgb}{0.67, 0.8, 0.94}

\begin{document}

\preprint{AIP/123-QED}

\title[]{Introduction to Quantum Optimal Control for Quantum Sensing with Nitrogen-Vacancy Centers in Diamond}
\author{\textbf{Phila Rembold}}
 \thanks{These authors contributed equally to this work.}
 \affiliation{Institute for Theoretical Physics, University of Cologne, D-50937 Cologne, Germany}
 \affiliation{Forschungszentrum J{\"u}lich GmbH, Peter Gr{\"u}nberg Institute -  Quantum Control (PGI-8), D-52425 J{\"u}lich, Germany}
 \affiliation{Dipartimento di Fisica e Astronomia ”G. Galilei”, Universit{\`a} degli Studi di Padova, I-35131 Padua, Italy}
\author{\textbf{Nimba Oshnik}}
 \thanks{These authors contributed equally to this work.}
 \affiliation{University of Kaiserslautern, Department of physics, Erwin Schr{\"o}dinger Strasse, 67663 Kaiserslautern, Germany}
 \affiliation{Universit{\"a}t des Saarlandes, Faculty of Natural Sciences and Technology, Physics, Campus E2.6, D-66123 Saarbr{\"u}cken, Germany}
\author{Matthias M. M{\"u}ller}
 \affiliation{Forschungszentrum J{\"u}lich GmbH, Peter Gr{\"u}nberg Institute - Quantum Control (PGI-8), D-52425 J{\"u}lich, Germany}
\author{Simone Montangero}
 \affiliation{Dipartimento di Fisica e Astronomia ”G. Galilei”, Universit{\`a} degli Studi di Padova, I-35131 Padua, Italy}
\affiliation{Istituto Nazionale di Fisica Nucleare (INFN), Sezione di Padova, I-35131 Padua, Italy}
\author{Tommaso Calarco}
 \affiliation{Institute for Theoretical Physics, University of Cologne, D-50937 Cologne, Germany}
 \affiliation{Forschungszentrum J{\"u}lich GmbH, Peter Gr{\"u}nberg Institute -  Quantum Control (PGI-8), D-52425 J{\"u}lich, Germany}
\author{Elke Neu}
\email{elkeneu@physik.uni-saarland.de; nruffing@rhrk.uni-kl.de}
\affiliation{University of Kaiserslautern, Department of physics, Erwin Schr{\"o}dinger Strasse, 67663 Kaiserslautern, Germany}
\affiliation{Universit{\"a}t des Saarlandes, Faculty of Natural Sciences and Technology, Physics, Campus E2.6, D-66123 Saarbr{\"u}cken, Germany}

\date{\today}

\begin{abstract}
Diamond based quantum technology is a fast emerging field with both scientific and technological importance. With the growing knowledge and experience concerning diamond based quantum systems, comes an increased demand for performance. Quantum optimal control (QOC) provides a direct solution to a number of existing challenges as well as a basis for proposed future applications. Together with a swift review of QOC strategies, quantum sensing and other relevant quantum technology applications of nitrogen-vacancy (NV) centers in diamond, we give the necessary background to summarize recent advancements in the field of QOC assisted quantum applications with NV centers in diamond.
\end{abstract}

\maketitle

\tableofcontents

\begin{quotation}
The nitrogen-vacancy (NV) center\citep{Wrachtrup1993,Doherty2013} is one of the major platforms in the evolving field of quantum technologies. Its remarkable stability, long spin coherence time, and optical properties make it especially attractive for quantum applications. Like any other quantum system, however, it is subject to experimental imperfections and limitations. Quantum optimal control~\citep{Glaser2015} (QOC) aims to improve the efficiency of system manipulation under such constraints.
In this introductory review article, we aim to provide an overview of the NV center applications that can be improved with QOC and outline the modus operandi for their implementation.\\
This review has been organised as follows: Section \ref{sec:Nitrogen-vacancyCenter} introduces NV centers in diamond and their most useful properties. Section \ref{sec:Applications} reviews the basic quantum sensing techniques, as well as a brief account of possible quantum information and computation applications of NV centers. Section \ref{sec:Theory} breaks down the principles and methods of QOC theory. This includes an introduction on the structure of QOC problems and an overview of some of the most common numerical QOC algorithms. Section \ref{sec:QOC_forNVs} reviews the control techniques discussed in section \ref{sec:Theory} applied to the system and methods discussed in section \ref{sec:Nitrogen-vacancyCenter} and \ref{sec:Applications}, respectively. This review has been written such that the reader can refer to section \ref{sec:Nitrogen-vacancyCenter} and section \ref{sec:Theory} more or less independently. Finally, the Appendices give a description of the relevant Hamiltonians for the NV center spin system.
\end{quotation}

 \begin{tcolorbox}[breakable, enhanced,colback=palecornflowerblue,boxrule=0pt,title=Examples]
    Throughout this review, illustrating examples are given in colored text boxes to distinguish them from the general text.
\end{tcolorbox}

\section{\label{sec:Nitrogen-vacancyCenter}The Nitrogen-Vacancy Center}

\begin{figure}[h!]
    \includegraphics[width = \textwidth]{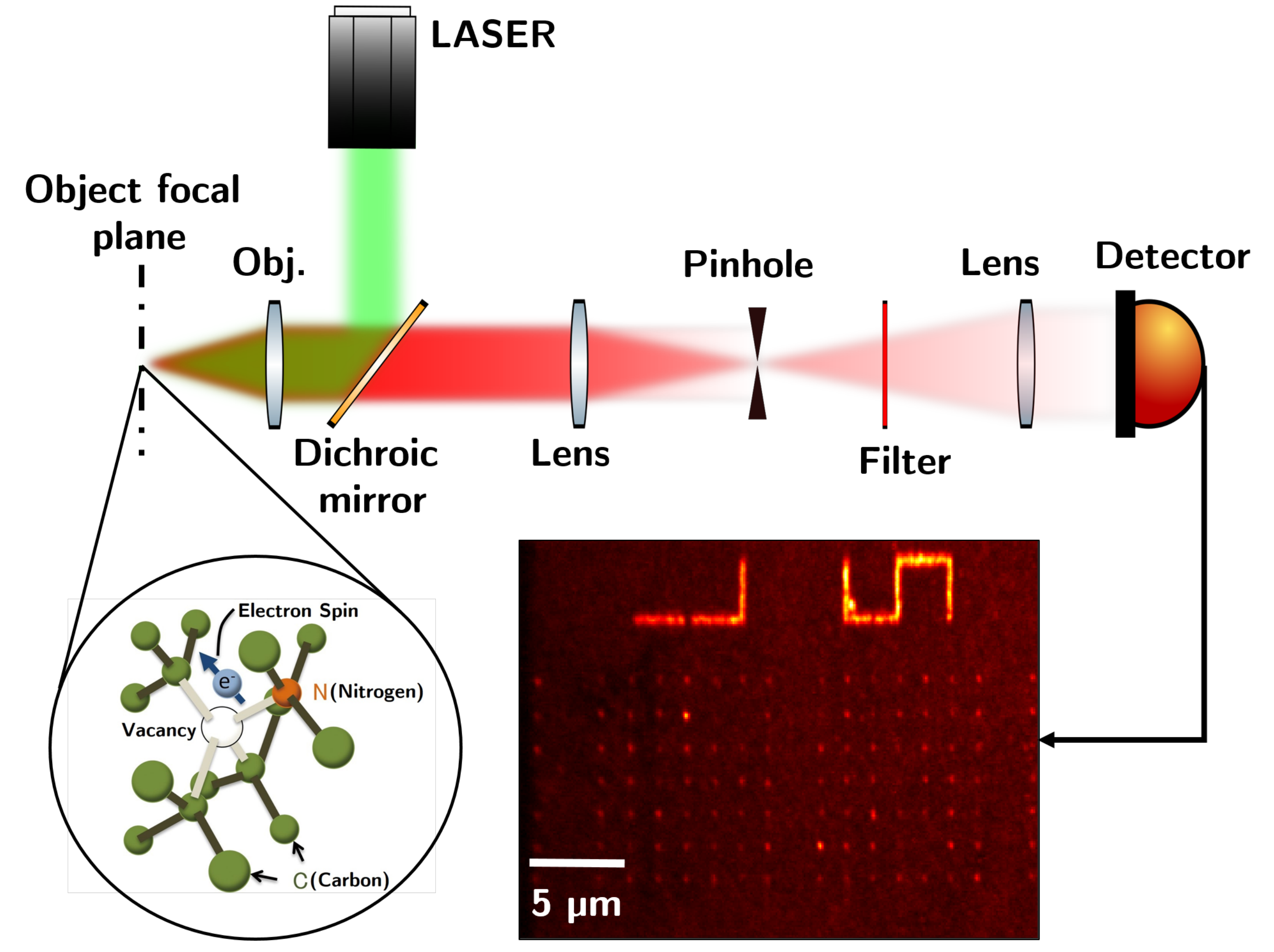}
    \caption{Sketch of a typical confocal setup used for experiments with NV centers in diamond. A laser (green) optically addresses the NV centers in the objective focal plane. The photoluminescence signal (red) is filtered and collected for analysis. The image in the bottom right panel shows a confocal scan of a diamond sample with a regular square pattern of diamond nanopillars and markers containing NV centers. The image on the lower left side shows the NV$^-$ center lattice structure. Inside the typical diamond lattice structure, two adjacent carbon atoms are replaced by a nitrogen atom and a vacancy. The NV axis, joining the nitrogen atom and the vacancy, can have four possible orientations in the unit cell. (bottom left panel [in circle] reproduced with permission from AAPPS Bulletin, 25 (1), 12 (2015)\citep{Itoh2015})}
    \label{Fig.1}
\end{figure}

\begin{figure}[h!]
    \includegraphics[width = \textwidth]{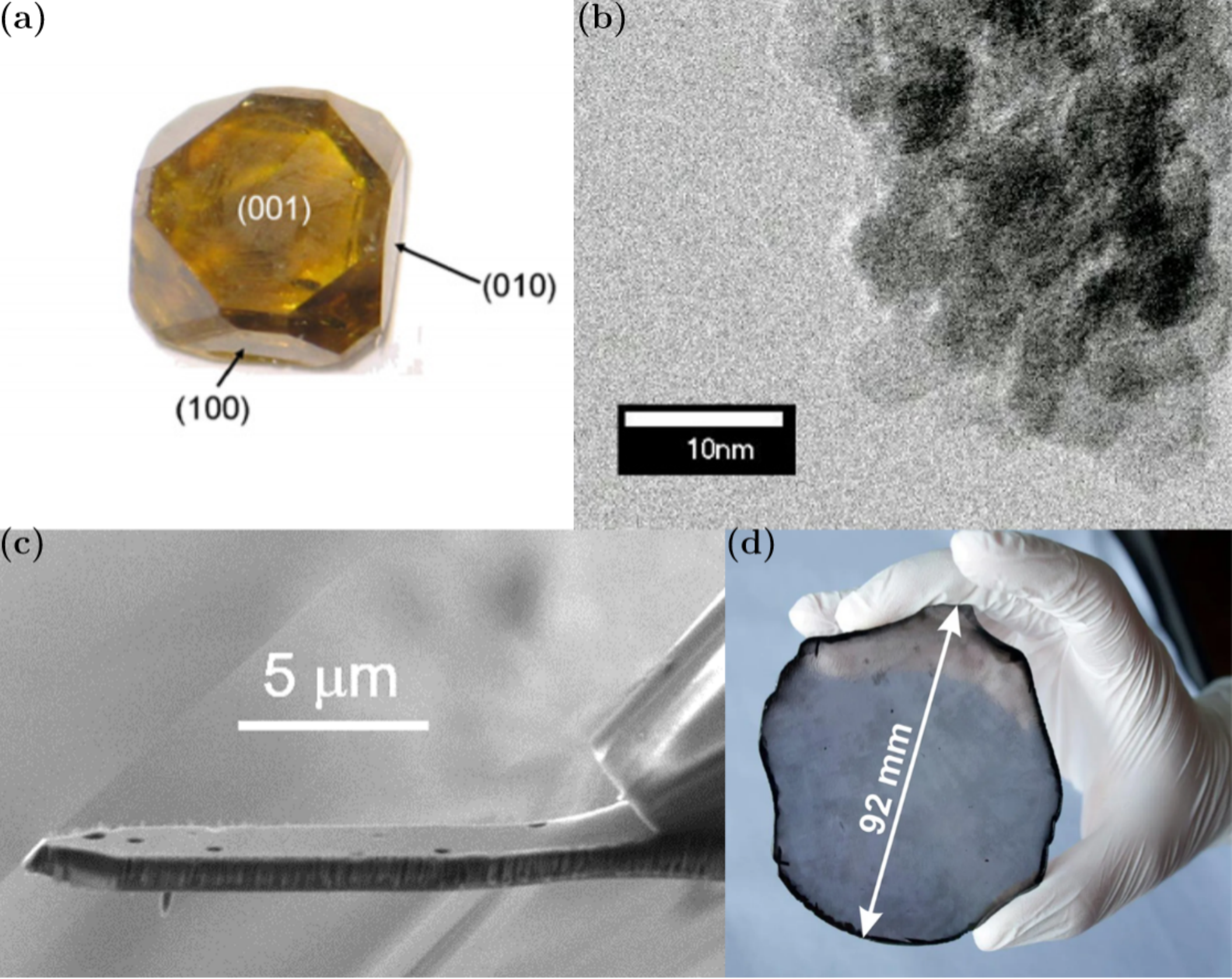}
    \caption{Diamonds for NV center-based applications: (a) HPHT grown single crystal diamond with labelled crystal facets\citep{Balmer_2009} (mm sized crystal).  The yellow color arises due to nitrogen impurities (Ib diamond). Reprinted with permission from Journal of Physics: Condensed Matter 21, 364221 (2009). Copyright 2009 by IOP Publishing. (b) Detonation nanodiamonds. Adapted and modified under a CC BY-SA 3.0 license from Reference \cite{Wikimedia} c) Diamond based scanning probe with single NV center at the tip for nanoscale sensing. (d) CVD grown diamond wafer (heteroepitaxy). Adapted and Modified under a CC BY 4.0 license from Scientific Reports 7, 1–8 (2017). }
    \label{Fig.2}
\end{figure}

Diamonds host a variety of point defects~\citep{Aharonovich2011,Jelezko2006} i.e.\ lattice sites where one of the carbon atoms is replaced with a different atom or vacancy. The appearance of color in pure diamond is due to the presence of particular optically active point defects. Ib diamonds, for example, are yellow (see Fig.~\ref{Fig.2}a) due to single substitutional nitrogen impurities. Such fluorescent point defects have a unique spectral signature and are called color-centers. Among them are the silicon-vacancy center \citep{Clark1995}, the germanium-vacancy center \citep{Iwasaki2015}, the tin-vacancy center \citep{Iwasaki2017}, the NV center, and several others which have been extensively studied over the past two decades.\citep{Aharonovich2011,Iakoubovskii2001} It is noteworthy that the color centers in diamond emit photoluminescence (PL), which is bright enough to enable observation of single defect sites with a confocal microscope, e.g. in regular patterns of nanostructures (see Fig.~\ref{Fig.1}, and caption for details). To add the capability for spin manipulation, microwaves are applied to the color centers under investigation.

The subject of this review, the NV center, is formed when one of the four carbon atoms in the unit cell of the diamond lattice is replaced by a nitrogen atom, accompanied by the formation of a neighboring vacancy site, shown in Fig.~\ref{Fig.1}. NV centers have three known energetically stable states; NV$^-$, NV$^0$, and NV$^+$. In this review, we limit our focus to NV$^-$ (from here on simply called NV). It is the most promising of the three for applications in quantum metrology and quantum information because of its magneto-optical properties.
In the following sections, we briefly describe these useful characteristics of the NV center for the matter of clarity and self-sufficiency with a main focus on optimal control applications. For more details, we refer the reader to more in-depth reviews on NV centers and related applications~\citep{Doherty2013,Barry2019,Radtke2019}.

\subsection{General Properties of NV Centers in Diamond}
\label{sec:GeneralProperties}

\begin{figure}[h!]
    \centering
    \includegraphics[width = \textwidth]{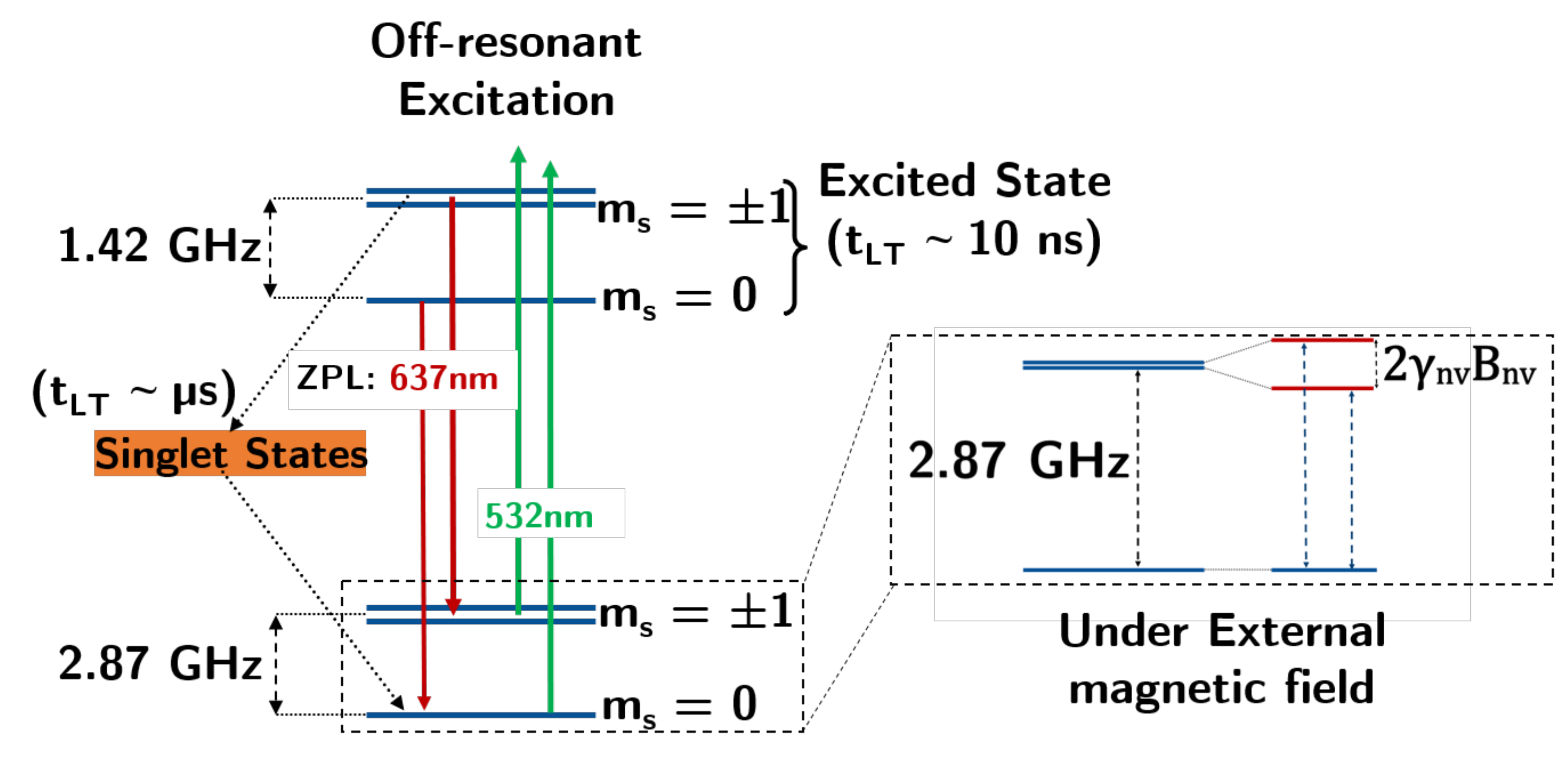}
    \caption{NV center electronic structure. The zero-phonon line (ZPL) is represented by red arrows ($\lambda = 637$~nm), green arrows signify off-resonant excitation, and the dashed arrows indicate the possible non-radiative decay channels. The spin state lifetimes are given as $t_{LT}$, and the in-box level scheme shows the Zeeman splitting of the ground-levels under the influence of a static external magnetic field $B_{nv}$.}
    \label{Fig .3}
\end{figure}

As a host material, diamond itself has a number of useful properties, including hardness, high refractive index, and non-toxicity. Various techniques can be used to manufacture NV center containing diamond (Fig.~\ref{Fig.2}). The main fabrication methods include high pressure-high temperature (HPHT) synthesis, detonation synthesis and chemical vapor deposition\citep{Maertz2010,Dwyer2013} (CVD). While detonation synthesis only leads to nanodiamonds (see Fig.~\ref{Fig.2}b). HPHT and CVD produce diamonds of various morphologies including bulk, single crystals. Very recently, up-scaling single crystal CVD diamond to wafer scales (10 cm) has been accomplished via heteroepitaxy\citep{Schreck2017}(see Fig.~\ref{Fig.2}d).
Ion implantation\citep{Meijer2008} is commonly used to create NV centers in diamond, other methods include doping the crystal during the growth process to incorporate the desired impurity. The availability of a broad range of materials fosters a spectrum of applications including scanning probe diamond tips (Fig. ~\ref{Fig.2}c) with single NV centers for nanoscale sensing,\citep{Degen2008,Balasubramanian2008,Maletinsky2012,Taylor2008,Hingant2014,Hall2009} engineered NV ensemble based sensors,\citep{Osterkamp2019} wide field imaging with NV center ensembles in bulk diamond,\citep{Steinert2013} diamonds coupled to cavities\citep{Fujiwara2015,Wolters2012,AliMomenzadeh2016,Jensen2014} as well as biological applications\citep{Schirhagl2014,Mohan2010,Fu2007,Kucsko2013} in different environments, temperature ranges and pressures.\\
Despite the many advances in material synthesis techniques, diamonds have inherent noise sources that limit the performance of NV center based applications. In diamonds with exotic geometries like the scanning probes (Fig.~\ref{Fig.2}) as well as systems with shallow NV centers, even the external environment plays a big role. Accordingly the protection from both inherent and external noise has a high priority for the NV center community to improve performance. Exploring techniques of QOC (section~\ref{sec:Theory}) promises to make  the aforementioned spin systems more robust against unwanted noise.\\
Also important for its applicability is the NV center's electronic structure\citep{Lenef1996,Rogers2015,Doherty2011} (Fig.~\ref{Fig .3}): The vacancy site contains six electrons (three dangling carbon electrons, two nitrogen electrons, and one electron trapped from a nearby donor in the crystal lattice as shown in Fig.~\ref{Fig.1}) which form a spin triplet system with $m_s = \pm 1, 0$. The electrons are tightly bound to the defect making the NV center a truly atomic sized systems.

 \begin{figure*}
    \centering

\subfloat[\label{Fig .4a}]{%
         \includegraphics[width=0.32\textwidth]{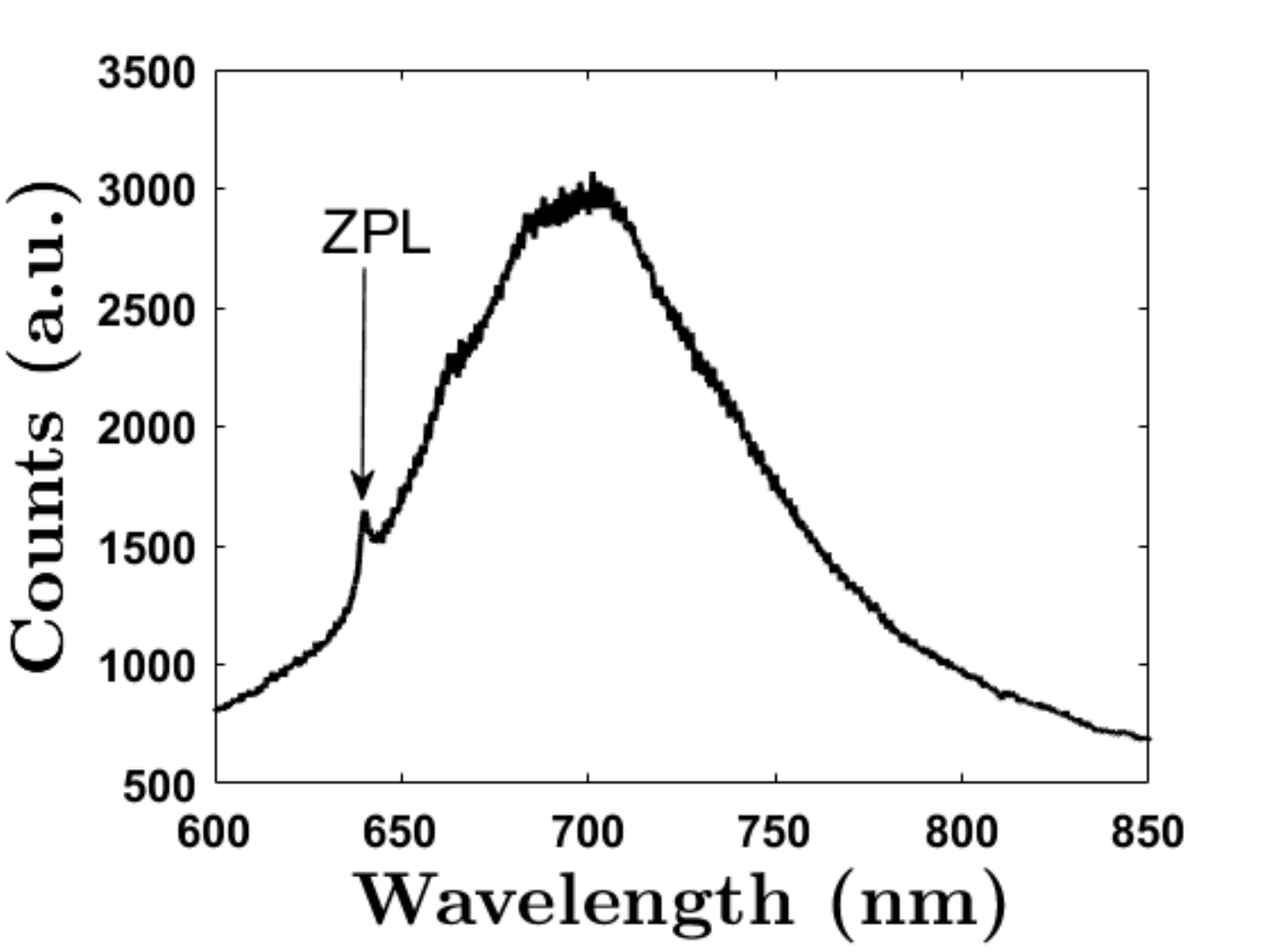}
}
\subfloat[\label{Fig .4b}]{%
        \includegraphics[width=0.32\textwidth]{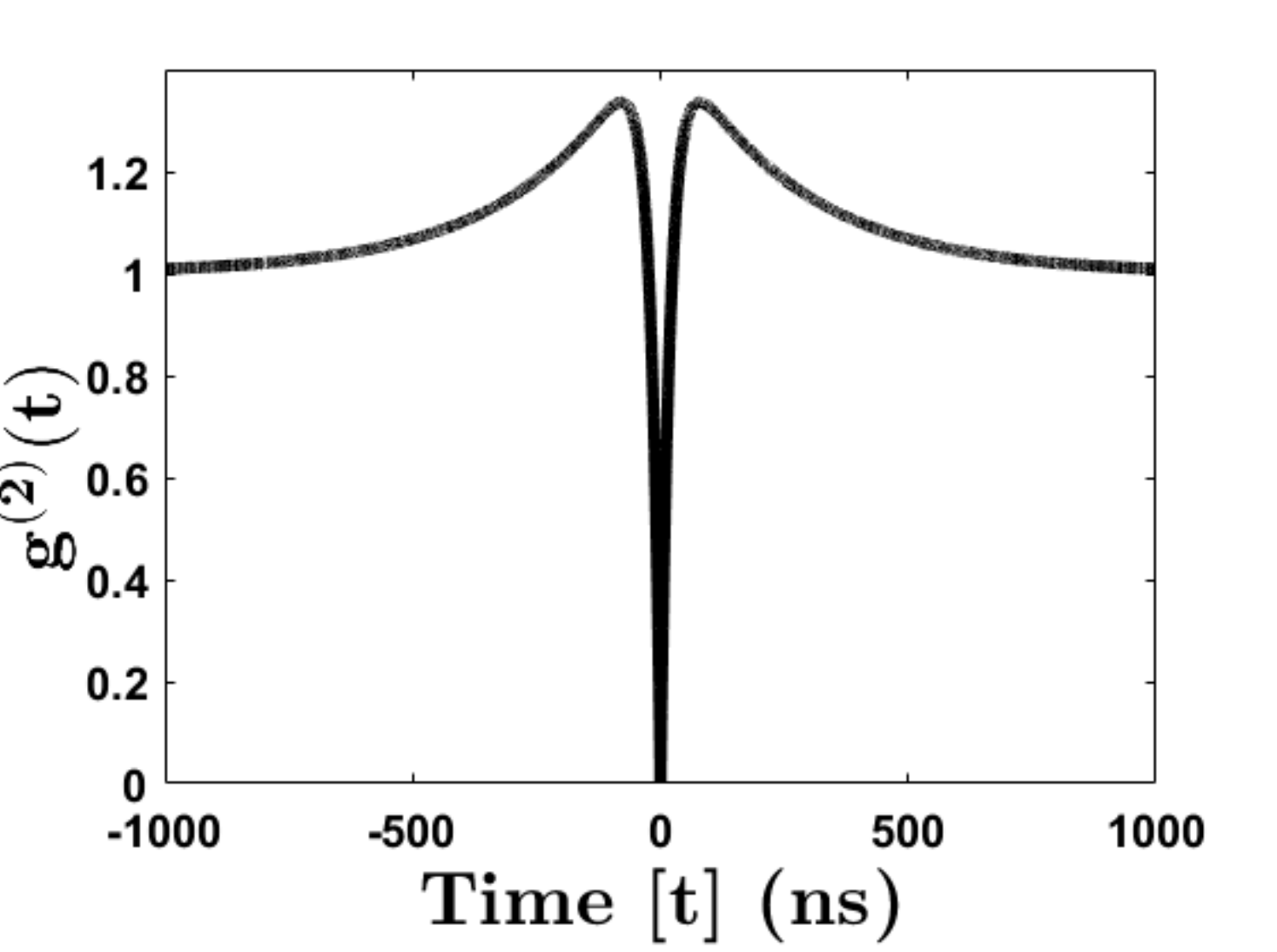}
}

\subfloat[\label{Fig .4c}]{%
         \includegraphics[width=0.32\textwidth]{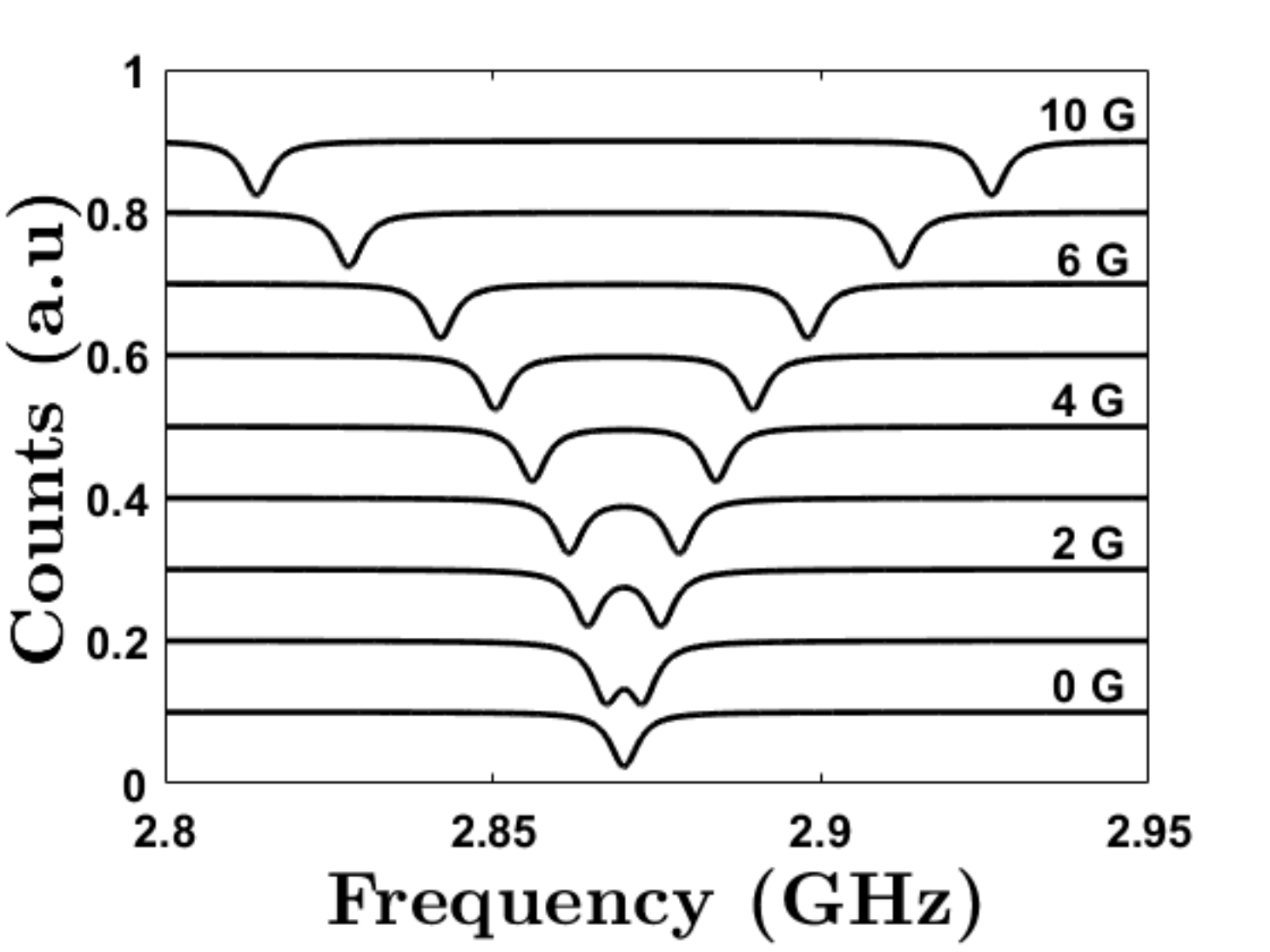}
}
\subfloat[\label{Fig .4d}]{%
         \includegraphics[width=0.32\textwidth]{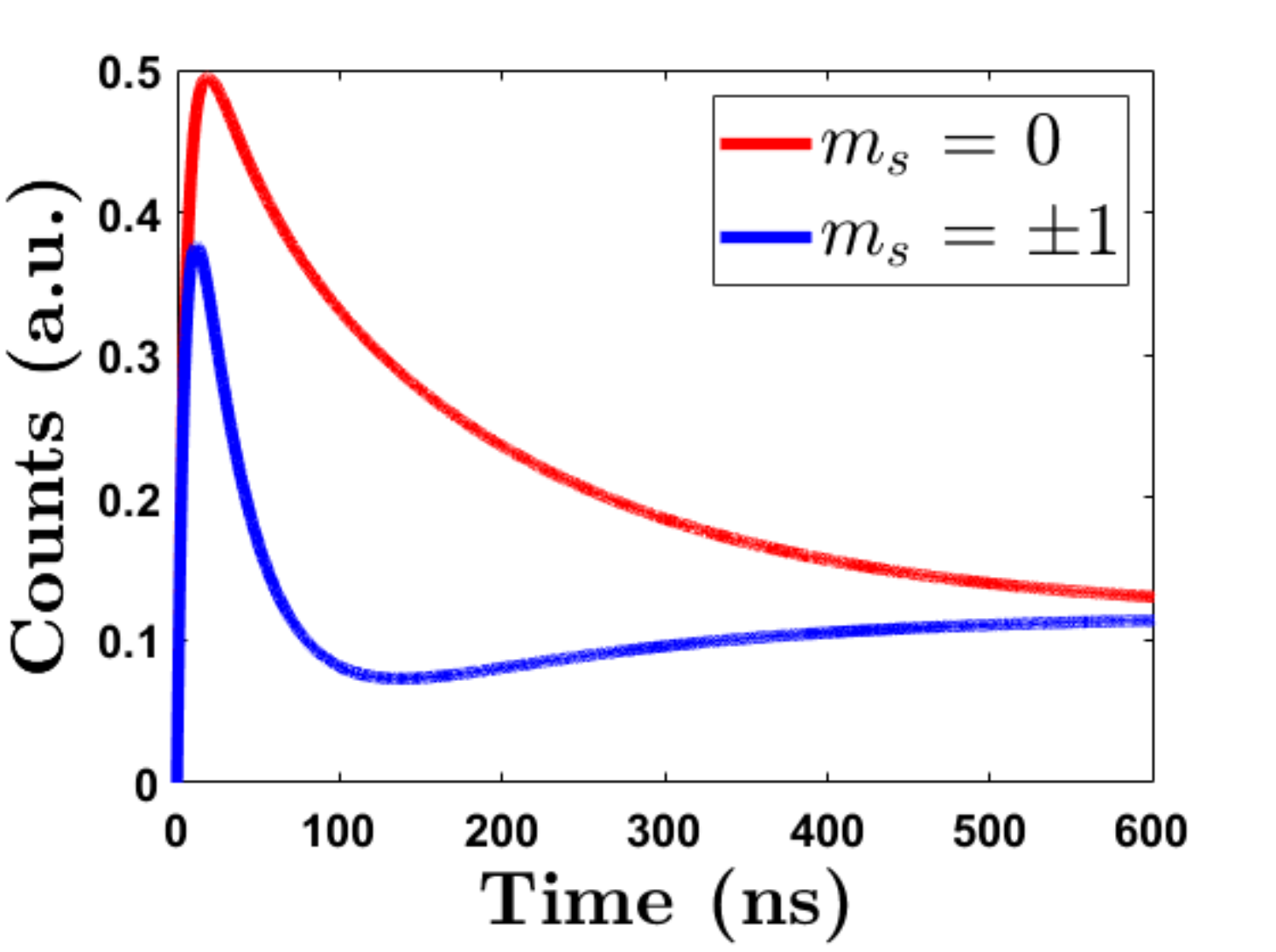}
}

\caption{Distinguishing properties of the NV center (at room temperature). The y-axis (counts) represents the PL signal from the NV centers; (a) Emission spectrum of the NV center, the zero-phonon line (ZPL) corresponds to the purely radiative 637 nm transition from excited to ground-level; (b) simulated second-order correlation function of photons emitted from a single NV center show antibunching effect; (c) simulated Zeeman splitting under an external magnetic field; the right-hand vertical axis gives the approximate values of the magnetic field along the NV axis; (d) simulated spin dependent fluorescence; time trace of the state-dependent photoluminescence from the NV center; the $m_s=\pm1$ state (blue curve) is darker compared to the $m_S=0$ state (red curve) because of population shelving in the intermediate metastable state. (Courtesy: Richard Nelz [(a),(b),(c)] and Simone Magaletti [(d)])}\label{Fig .4}

\end{figure*}

As depicted in Fig.~\ref{Fig .3}, the excited and ground state spin triplets are separated by 1.945~eV, corresponding to a purely radiative transition (zero-phonon line [ZPL] in Fig.~\ref{Fig .4a}) at 637~nm. Additionally, intermediate singlet states occur energetically in-between the triplet states. At thermal equilibrium, the spin population in the ground states is governed by the Maxwell-Boltzmann distribution law.\citep{MBD} Upon irradiation with a green laser ($\lambda = 532$~nm), the population in the ground states is transferred to the excited levels through a combination of radiative absorption and non-radiative relaxation processes, involving the excited levels of the NV center and the conduction band of the diamond crystal. The excitation process is followed by direct radiative decay as well as non-radiative decay through the intermediate states. The radiative transitions between the ground and excited states are spin preserving, the decay via the intermediate states however, is not (although it is spin dependent, nonetheless). The decay rate from the excited level $m_s=\pm1$ spin states to the intermediate singlet state is comparable to the rate of the direct radiative decay to the ground state. For the excited level $m_s=0$ spin state, the decay via the singlet state is negligible. This results in a majority of the population being pumped into the $m_s = 0$ ground-level spin state under laser illumination.\\
The transition frequency in absence of an external field between the degenerate $m_s= \pm1$ and the $m_s = 0$ spin states in the ground-level is 2.871~GHz. An external magnetic field lifts the degeneracy of the $m_s=\pm1$ states as depicted in Fig.~\ref{Fig .3} and experimentally observed in Fig.~\ref{Fig .4c} (Zeeman splitting). If the magnetic field is sufficiently strong the $m_s=\pm1$ states are well separated in frequency allowing for the spin system to be approximated as a qubit, with one of the $m_s = \pm1$ states and the $m_s = 0$ state forming the two qubit levels.\\
\begin{tcolorbox}[breakable, enhanced,colback=palecornflowerblue,boxrule=0pt,title= Rabi Oscillations]
    Let us consider an example (also see the example in section~\ref{sec:DefiningAControlPorblem}), approximating the NV center as a two level system with the ground state $\ket{0}$ and excited state $\ket{1}$. This assumption holds when the $m_s=\pm1$ spin states are non-degenerate under the influence of a static magnetic field along the NV axis (the axis connecting the nitrogen atom and the vacancy in the lattice). It is common practice to consider the NV axis as the quantization axis (Fig.~\ref{Fig.1}). The system can be initialised in the $m_s=0$ ground state with a green laser pulse. Subsequently, a MW field of frequency $\omega_{nv}$ is applied to resonantly drive the transition $\ket{0} \leftrightarrow\ket{1}$, leading to a continuous population transfer. The observed coherent oscillation (Rabi flopping) of the population between the spin states is shown in the figure below on the left hand side.\\
    The figure on the right shows the Bloch sphere representation  of the NV qubit spin state. The north pole of the Bloch sphere corresponds to the pure $\ket{1}$ state, whereas the south pole indicates the pure $\ket{0}$ state. The Bloch vector (red arrow) represents the spin-state of the system at a given instance, $\hat{x}$ and $\hat{y}$ indicate the polarisation of the driving field in the rotating frame (see Appendix~\ref{app:RWA}). Pulses rotate the state vector with a speed dependent on the driving field strength around an axis dependent on its phase (blue curve). A rotation by 180$^\circ$, in this case fully transferring the population from $m_s=0$ to $m_s=1$ is referred to as a $\pi$-pulse. $\pi/2$-pulses rotate the state-vector by 90$^\circ$. In this case, they would create a superposition state between $m_s=0$ and $m_s=1$. The Hamiltonian for this system in the rotating wave approximation (see Appendix~\ref{app:RWA}) is given by Eq.~\eqref{eqn:Hamiltonian}. If the drive is constantly applied, the system undergoes Rabi oscillations with a frequency $\Omega$ that is proportional to the strength of the applied microwave field.\\
    The decay of the amplitude of oscillation is owed to the inherent decoherence and decay processes in the spin system.
    \begin{center}
    \includegraphics[width = \textwidth]{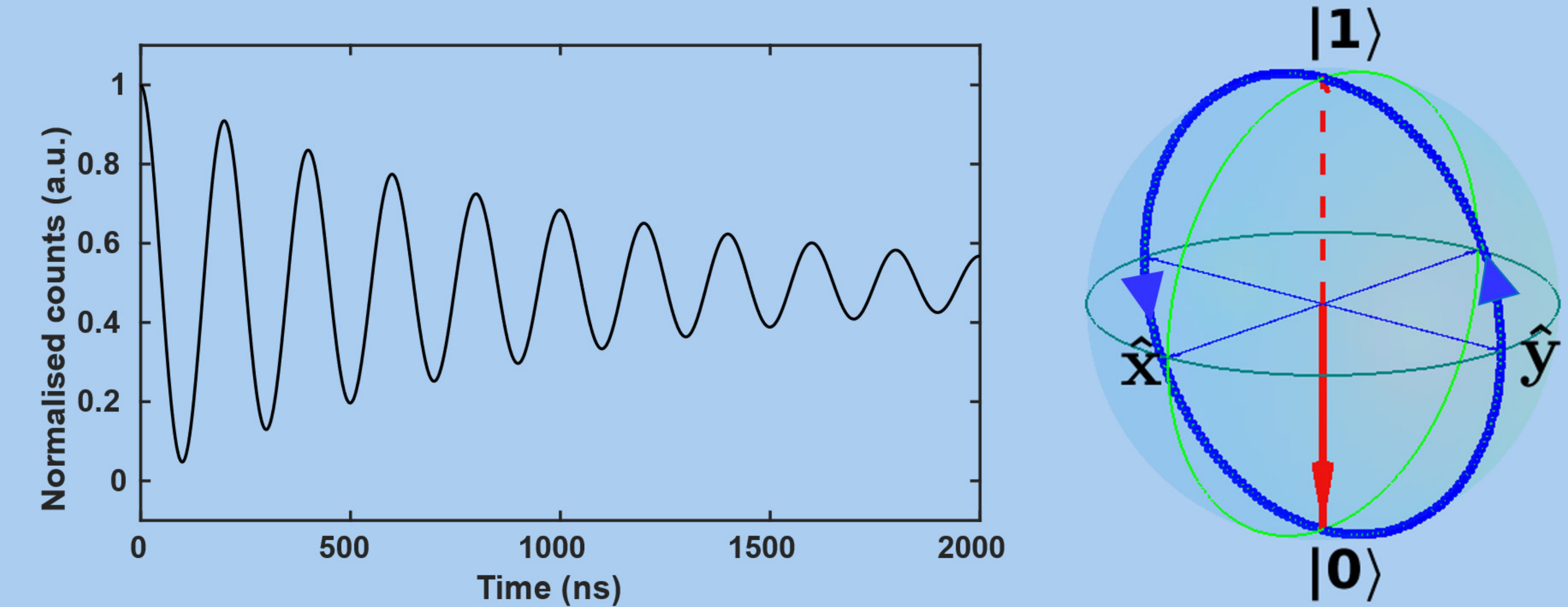}
    \end{center}
\end{tcolorbox}
Following points summarise the properties that render NV centers well suited for quantum technology applications:
 
\begin{itemize}
    \item The NV center has a magnetically active ground-level spin triplet state with the $m_s = \pm1$ states exhibiting Zeeman splitting (Fig.~\ref{Fig .3}) in the presence of external fields. This sensitivity towards external fields is the key to the majority of sensing applications.
    \item In addition, the transition between the ground-level spin states can be coherently manipulated with microwaves.
    \item The ground-level spin states also exhibit spin dependent fluorescence due to selection rules and the transition rates mentioned above. It ensures a clear readout contrast between the two qubit levels, which is crucial for quantum applications (see Fig.~\ref{Fig .4d}). Alternatively, other readout schemes based on NV charge state detection are also possible. 
    \item The $m_s=0$ spin state of the ground-level triplet can be efficiently initialised via an off-resonant laser excitation, this qubit initialisation is a pivotal property for all quantum sensing and quantum computation applications.
    \item The nuclear spins in the NV center cluster exhibit dynamic nuclear polarisation,~\citep{Jacques2009} opening up the possibility of nuclear spin based quantum computation applications.
    \item All these properties are exploitable at room as well as cryogenic temperature.
    \item Additionally, diamond is a bio-compatible host, which is crucial for life science applications. 
\end{itemize}

Various other properties (Fig.~\ref{Fig .4}), e.g.\ stable emission of single photons, also make these spin systems highly suitable for quantum sensing and quantum information applications.

\subsection{\label{sec:SpinHamiltonian}Spin Hamiltonian}

Before we explore the vast field of applications of the NV center, it is useful to study the governing Hamiltonian for the spin system in order to understand the theoretical basis of the applications. The NV center spin in the diamond lattice undergoes a variety of interactions because of external magnetic and electric fields, as well as the crystal strain field, and other spins in the crystal lattice. The spin Hamiltonian for the NV center's ground-level spin triplet can be obtained in the basis of the $m_s=\pm1$ and $m_s=0$ spin operators. It can be derived by perturbative expansion of the full Hamiltonian of the system in terms of the spin operators (see Appendix~\ref{app:SpinHam} for more details), such that the thermal degrees of freedom average out.\citep{Levitt,Stoneham2001} Assuming the NV axis to be the quantization axis, the spin Hamiltonian can be written as
\begin{widetext}
\begin{align}
   \hat{H} =& \overbrace{\hbar D\left[\hat{S}_Z^2-\frac{2}{3}\right] + \hbar E(\hat{S}_X^2 - \hat{S}_Y^2)}^\text{zero-field term} + \overbrace{\hbar\gamma_{nv}\Vec{B}\cdot\hat{\Vec{S}}}^\text{magnetic interaction} +
   \overbrace{\hbar\delta_{\parallel}\mathcal{E}_Z\left[\hat{S}_Z^2-\frac{2}{3}\right] - \hbar\delta_{\perp}\left[\mathcal{E}_X(\hat{S}_X\hat{S}_Y+\hat{S}_Y\hat{S}_X) + \mathcal{E}_Y(\hat{S}_X^2 - \hat{S}_Y^2)  \right]}^\text{electric interaction}\nonumber \\
   &+\hbar\mathlarger{\mathlarger{\sum}}_{i=1}^{n}\left(\underbrace{\hat{\Vec{S}}\mathcal{N}_i\hat{\Vec{I}}_i}_\text{hyperfine interactions}+
   \underbrace{\gamma_i\Vec{B}\cdot\hat{\Vec{I}}_i}_\text{nuclear Zeeman interactions}+\underbrace{\mathcal{Q}_i\hat{I}_{Z,i}^2}_\text{nuclear quadrupole interactions}\right). \label{eqn:fullHamiltonian}
\end{align}
\end{widetext}
The first term in the Hamiltonian above is the zero-field term, i.e.\,  the system Hamiltonian in the absence of external fields, where $\hbar$ is the reduced Planck constant, $\hat{\Vec{S}} = (\hat{S}_X,\; \hat{S}_Y,\; \hat{S}_Z)^\intercal$ are the spin operators (see Appendix~\ref{app:SpinHam}), $D$ is the axial and $E$ the non-axial zero-field parameter. At room temperature and ambient pressure $D\approx$ 2.87~GHz varies by around $-80~kHz/K$ and $1.5kHz/bar$ with temperature and pressure changes respectively (see sections~\ref{sec:Thermometery} and \ref{sec:PSSO}).\\
The second term in the Hamiltonian represents the magnetic field interaction, where $\gamma_{nv} = 2\pi \times28$~MHz/mT is the gyromagnetic ratio for the NV center spin and $\Vec{B}$ is the external magnetic field.\\
The electric field interaction with the NV center spin is represented by the third term in the Hamiltonian. $\Vec{\mathcal{E}}=\left(\mathcal{E}_X,\: \mathcal{E}_Y,\: \mathcal{E}_Z\right)^\intercal$ is the effective electric field vector, representing contributions from external electric fields as well as the crystal field, arising from crystal strain and pressure (see section \ref{sec:PSSO}). $\delta_{\parallel}$ and $\delta_{\perp}$ are the axial and transverse coupling constants respectively arising from the symmetry of the crystal structure at the site of the NV center. In comparison to the magnetic field, the electric field has a much weaker interaction with the NV spin, with $\delta_{\parallel}$ at around $0.17Hz/(V m^{-1})$ and $\delta_{\perp}$ of the order of $10^{-3}Hz/(V m^{-1})$.\\
The NV center spin interacts with other spin systems in the host crystal and its surroundings. The most dominant interactions are with the $^{15}$N or $^{14}$N nucleus in the NV center and the $^{13}$C nuclear spins in the surrounding diamond lattice. These spin-spin interactions are represented by the hyperfine interaction terms in the Hamiltonian. $\mathcal{N}_i$ is the hyperfine interaction tensor between the electron and the $i{^\text{th}}$ nuclear spin and $\hat{\Vec{I}}_i$ is the nuclear spin operator. There are two main contributions to the hyperfine interaction tensor: First, the isotropic Fermi contact interaction that arises from the interaction of the electron cloud with the nearby nucleus (see Appendix~\ref{app:SpinHam}). Second, the anisotropic magnetic dipole interactions of the NV spin with distant nuclear spins. The magnitude of the latter interaction decays as $1/r^3$ with the distance $r$ from the nuclear spin and is comparatively weaker than the Fermi contact interaction terms.\\
On top of the hyperfine coupling, the nuclear spins also interact with the external magnetic field. This is represented as the nuclear Zeeman interaction term, where $\gamma_i$ is the gyromagnetic ratio of the corresponding nuclear spin $i$.\\
Finally, nuclei like $^{14}$N can exhibit quadrupole splitting; this is represented by the last term in the Hamiltonian. Here $\mathcal{Q}_i$ represents the quadrupole splitting. As an example, the quadrupole splitting for the $^{14}$N nuclei inside the diamond lattice has been estimated to be around 5~MHz\citep{Fischer2013}.\\
The reader is advised to refer to Appendix~\ref{app:SpinHam} for a detailed description of these Hamiltonian terms and their origins.

\section{\label{sec:Applications}Quantum Technology Applications with NV Centers}

We now proceed with the exploration of the versatility of NV centers in terms of their potential applications in quantum technologies. As quantum sensors, they have been utilised for magnetic and electric field sensing (section \ref{sec:Magneticfieldsensing} and \ref{sec:Electricfieldsensing}), thermometery (section \ref{sec:Thermometery}), strain/pressure measurements (section \ref{sec:PSSO}), orientation tracking and more. NV centers have further found application in quantum information related experiments (section \ref{sec:QuantumInformation}). In addition, the NV center is also a reliable non-classical source of single photons, and has been utilised in various single photon experiments.\citep{Kurtsiefer2000,Brouri2000}

\subsection{\label{sec:QauntumSensingApplications} Quantum Sensing Applications}

\begin{figure}[h!]
    \centering
    \includegraphics[width = \textwidth]{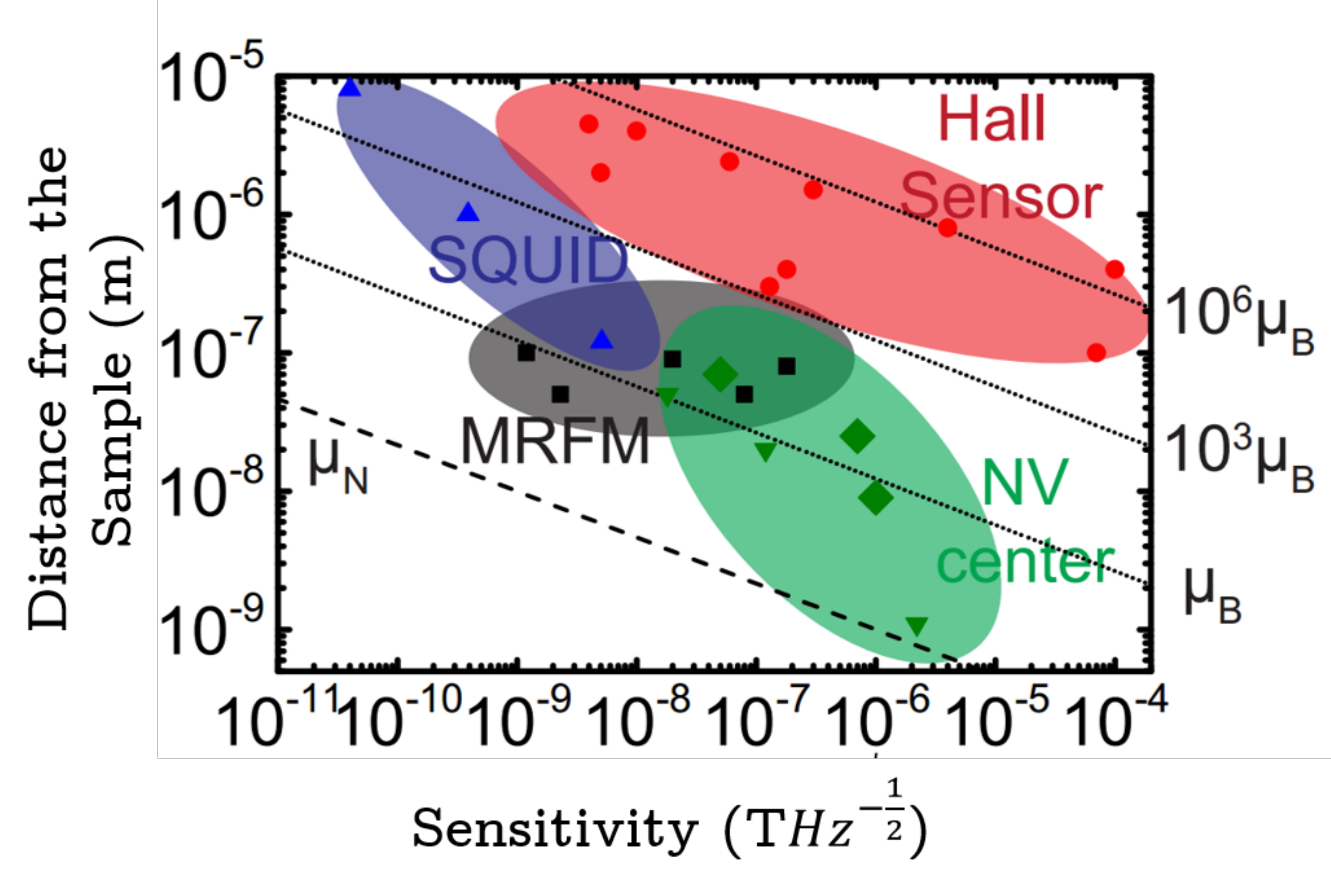}
    \caption{Comparison of scanning magnetometers for nanoscale sensing. The plot shows experimentally demonstrated magnetic field sensitivities for Hall sensors,\citep{Boero2003} scanning SQUID sensors,\citep{Vasyukov2013} magnetic resonance force microscopes\citep{Rugar2004} and NV center based sensors\citep{Grinolds2013} as a function of the sensor-sample distance (markers represent experimentally obtained values, for details refer to Appel et al.\citep{Appel2017}). Diagonal lines indicate the theoretical threshold for the detection of 1,  10$^3$ and 10$^6$ electron spins ($\mu_B$) within one second and a single nuclear spin (dashed, $\mu_N$) in the same time. Adapted and modified from P. Appel, Scanning nanomagnetometry: Probing magnetism with single spins in diamond, Ph.D. thesis, University of Basel (2017), under the terms of a CC BY-NC-ND 4.0 license.}
    \label{Fig .5}
\end{figure}

In principle, NV centers possess the necessary properties to be practical quantum sensors.\citep{Degen2017} The efficient spin state initialisation/readout and the sensitivity to various physical parameters have resulted in a variety of diamond based sensing applications. As diamonds are simple to handle in terms of logistics, maintenance, and manipulation in comparison to other (atomic) quantum systems, their application has received a lot of attention. The main focus has been on using the quantum nature of the defect center's spin to detect classical physical signals (called classical detection, in contrast to quantum detection which is based on the use of entanglement~\citep{Degen2017}). In several sensing applications NV centers have proven to outperform their counterparts. Fig.~\ref{Fig .5} shows one such example, juxtaposing different scanning magnetometers that are commonly implemented for sensing. A summary of the different measurable parameters with NV center-based sensors is provided in Table~\ref{Tab .1}. The sensing methods rely on efficient and coherent spin manipulation, whereby experimental imperfections limit their success. In this section, we briefly discuss different sensing protocols, experimental limitations and other challenges. A review of the current techniques based on quantum optimal control (QOC) (see section~\ref{sec:Theory}) to overcome/circumvent the described obstacles and efficiently use the NV centers as quantum sensors is given in section \ref{sec:Optimalcontrolforquantumsensing}. For completeness, we also outline related applications in quantum information and computation. 

\subsubsection{\label{sec:Magneticfieldsensing} Magnetic Field Sensing}

Magnetic field sensing is the most common application of NV centers as quantum sensors. As illustrated in the full Hamiltonian Eq.~\eqref{eqn:fullHamiltonian}, the external magnetic field interacts directly with the NV spin. In addition to continuous wave spectroscopy experiments (Fig.~\ref{Fig .4c}), several NMR/EPR based spin manipulation techniques\citep{RevModPhys.76.1037} with microwave pulses can be applied to quantify the magnitude of the interaction as well as the directional dependence of the external field to be studied. These techniques pave the way for vector magnetometery,\citep{Balasubramanian2008} magnetic field imaging at nanoscale,\citep{Maletinsky2012} detection of single spins,\citep{Zhao2012} and a number of other applications. All these methods rely on efficient initialisation (laser), manipulation (microwaves) and readout (typically laser, although several other possibilities exist\citep{Shields2015,Bourgeois2015}, see Barry et al.\citep{Barry2019} for details) of the spin system.\\
The most basic sensing protocol is the direct measurement of the Zeeman splitting (see section \ref{sec:SpinHamiltonian}) under the influence of an external magnetic field (Fig.~\ref{Fig .4c}). This technique requires the application of a continuous, off-resonant laser to the spin system. The MW frequency is swept, driving the $m_s=0 \rightarrow m_s=\pm1$ transition when the resonance condition is met. This in turn results in a measurable decrease in the fluorescence from the NV center (see Fig.~\ref{Fig .4c}). Gruber et al.\citep{Gruber1997} demonstrated the first of these optically detected magnetic resonance (\textbf{ODMR}) experiments with a single NV center, showing that the splitting of resonant peaks (dips) gives information about the external magnetic field.\\
Non-polarized (thus randomly flipping) spins proximal to the NV center create fluctuating magnetic fields at the position of the NV center. The interaction strength of the NV with the fields fluctuating at the transition frequency between $m_s=0 \rightarrow m_s=\pm1$ can be directly inferred from the lifetime ($T_1$) of the spin states. Such \textbf{relaxometry} techniques have in particular been used for life science applications, \citep{Steinert2013} magnetic noise sensing,\citep{Tetienne2013} and surface and material studies~\citep{Pelliccione2014}.

\begin{figure}[h!]
         \centering
         \includegraphics[width=\textwidth]{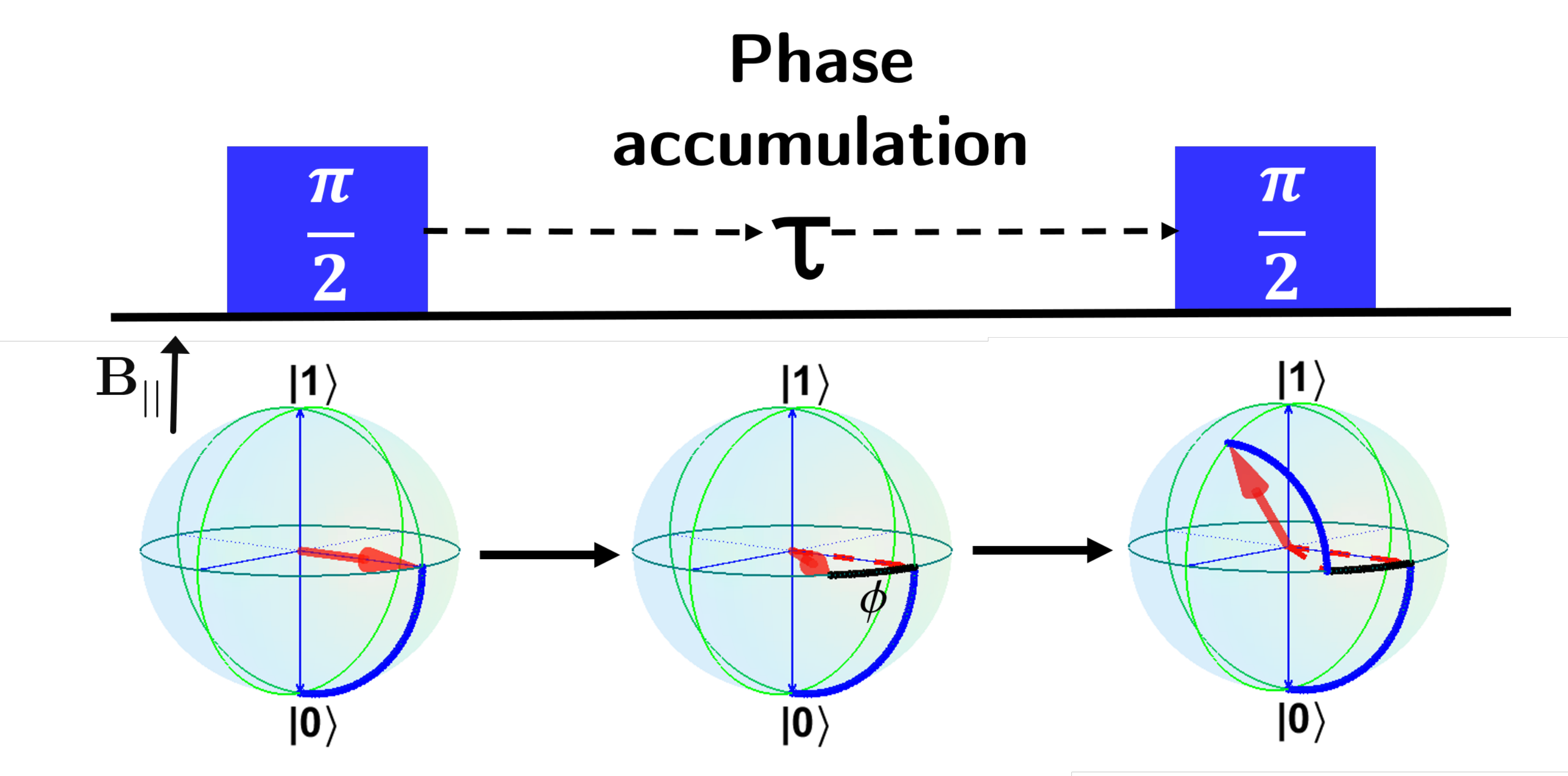}
         \caption{Ramsey spin manipulation sequence for DC magnetic field sensing. The NV axis is considered to be the quantization axis. The parallel component $B_\parallel$, of the external DC magnetic field is the signal under consideration. Resonant MW control pulses are used to apply $\frac{\pi}{2}$-rotations. The first pulse rotates the spin from the ground state into a superposition in the equatorial plane (blue curve). During the free precession time $\tau$ the magnetic field component $B_\parallel$ induces a phase $\phi$ (black curve). When the spin is rotated again with a second pulse, the phase can be inferred by projecting the spin along the quantisation axis. The required measurement is done by repeating the process a number of times and reading out whether the NV center is in the ground or excited state after each repetition. The distribution of measurements then gives the projection along the quantisation axis.}
         \label{Fig .6}
\end{figure}{}

\begin{figure}[!ht]
\subfloat[Spin-echo pulse sequence\label{Fig .7a}]{%
    \centering{\includegraphics[width=0.9\textwidth]{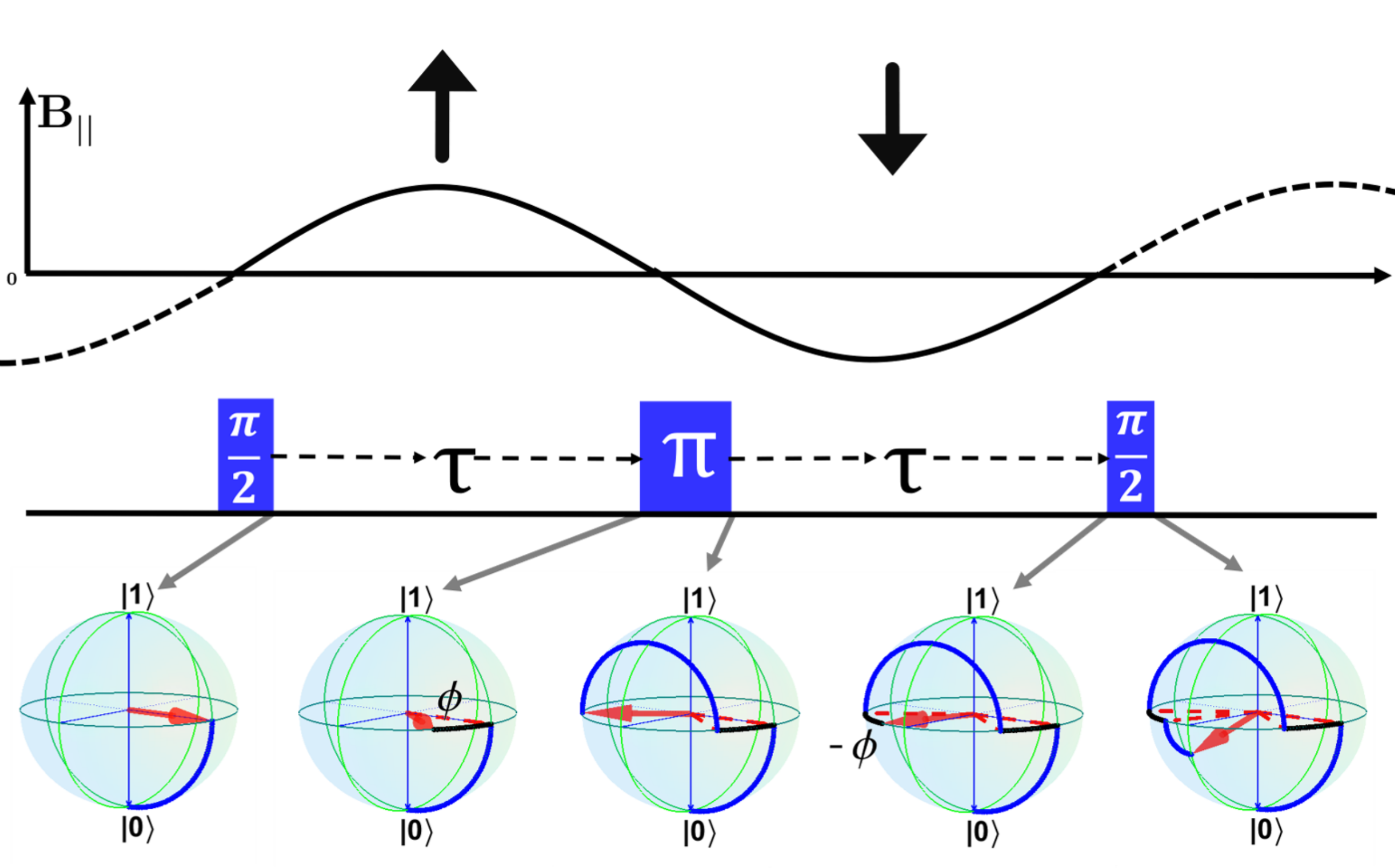}}
}\hfill
\subfloat[State population vs the precession time\label{Fig .7b}]{%
    \centering{\includegraphics[width=0.9\textwidth]{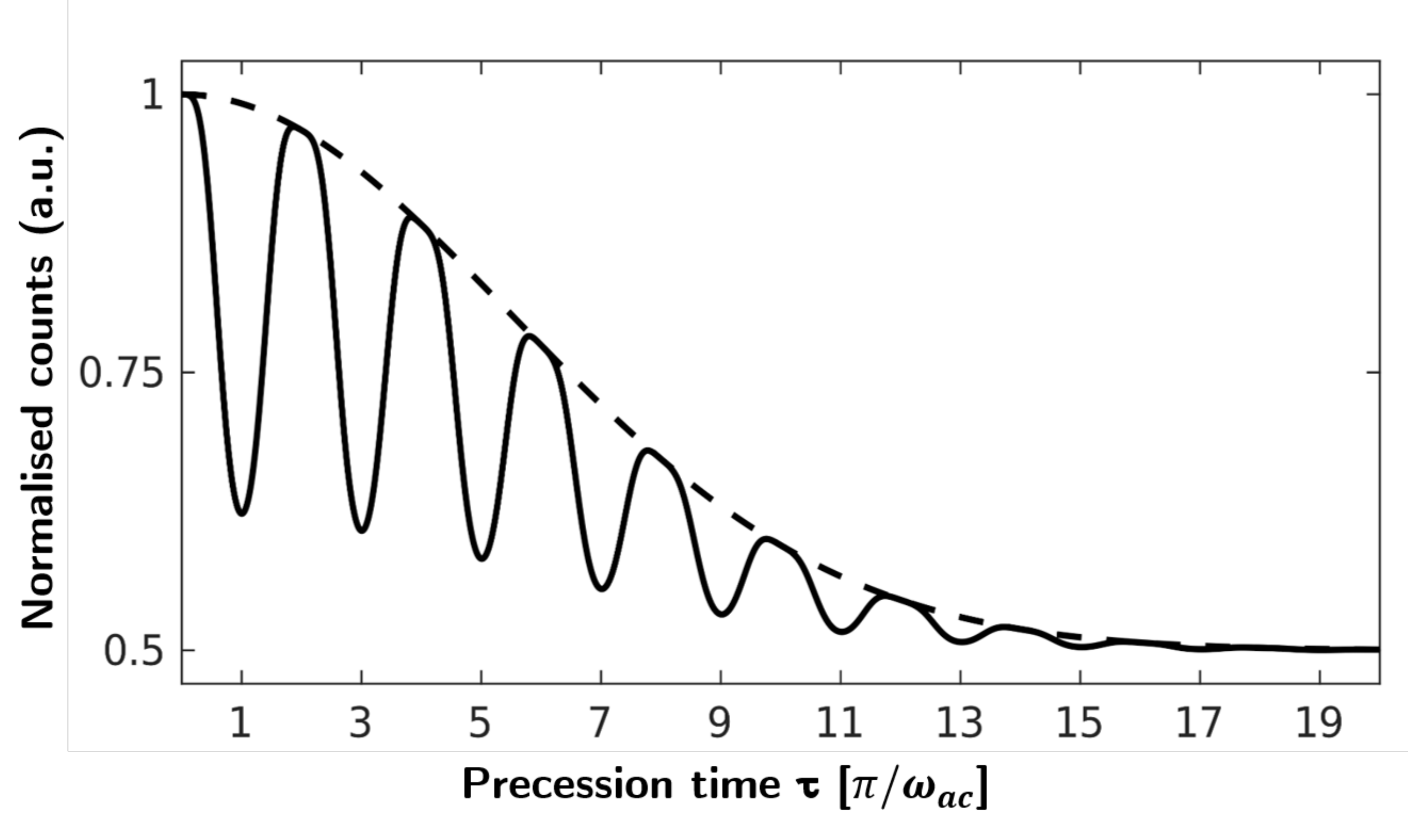}}
}

  \caption{AC field sensing: (a) Spin-echo pulse sequence for AC field sensing. $B_\parallel$ is the magnetic field component along the NV axis which is under consideration, the arrows on the top indicate the direction of the magnetic field vector during the given period. As in Fig.~\ref{Fig .6}, at the bottom, the red vector represents the state of the system. The first pulse rotates the spin by 90$^\circ$ from the ground state into a superposition in the equatorial plane (blue curve). During the free precession time $\tau$ the parallel magnetic field component $B_\parallel$ induces a phase $\phi$ (black curve). The $\pi$-pulse then flips the direction of precession, proceeded by another phase accumulation over a time period of equal length. Finally, a $\frac{\pi}{2}$-pulse rotates the spin again and the magnetic field strength can be inferred by projecting the spin along the quantisation axis. In case of a DC field the phase contributions are of opposite sign for the two free precession periods and hence are subtracted. However, for fields oscillating with frequency $\omega_\text{ac}=\pi/\tau$, the final projection gives a measurement of its amplitude similar to the Ramsey sequence in Fig.~\ref{Fig .6}. (b) Population in the state $\ket{0}$ after applying spin-echo sequences with different precession times $\tau$. Whenever the frequency of the external field satisfies the condition $\omega_\text{ac}=k\pi/\tau$ with odd $k$, the fidelity is reduced. The overall envelope decay of the population depends on the decoherence time $T_2$.\label{Fig .7}}
\end{figure}

Another common method for DC magnetometery is the \textbf{Ramsey interference} pulse technique\citep{PhysRev.78.695} depicted in Fig.~\ref{Fig .6}. Balasubramanian et al.\citep{Balasubramanian2009} first experimentally demonstrated this method with NV centers using ultrapure CVD grown single crystal diamond as a host material. In theory, the Ramsey sequence is the spin equivalent of an optical Mach-Zehnder interferometer. It is based on measuring the phase induced by external perturbations on a superposition state of the spin system. The ``mirrors'' and ``beam splitters'' for this interference experiments are formed by the $\frac{\pi}{2}$-pulses applied at MW frequency (Fig.~\ref{Fig .6}). The induced phase $\phi$ is directly proportional to the external magnetic field component $B_\parallel(t)$:
\begin{equation}
    \phi =\gamma_{nv}\int_0^{\tau}B_\parallel(t)dt,
\end{equation}
where $\tau$ is the free spin precession duration.
The phase accumulation arises from the Larmor precession of the spin about the magnetic field vector.\\
Additional pulses can be used to cancel out the phase induced by the DC field, so that the phase information can be used for measuring AC magnetic fields (Fig.~\ref{Fig .7}).

\begin{figure}[h!]
         \centering
         \includegraphics[width=\textwidth]{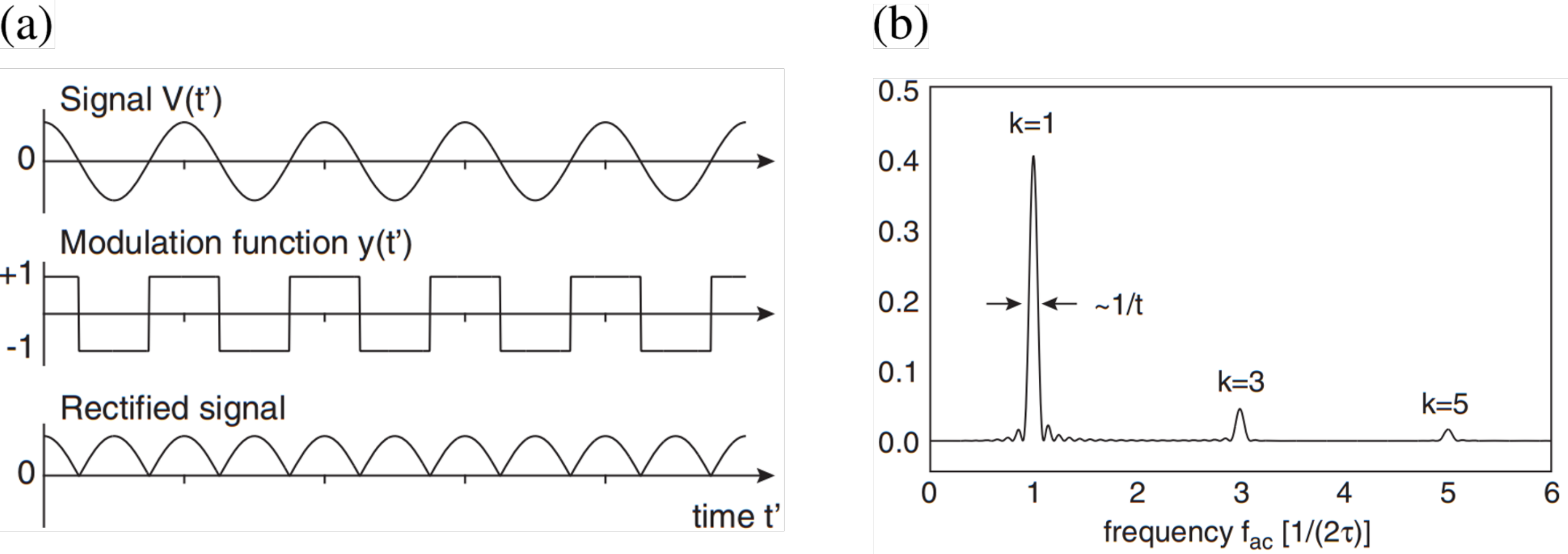}
         \caption{Dynamical decoupling sequence: (a) The protocol consists of a train of $\pi$-pulses. The frequency that is measured depends on the spacing between the pulses. The modulation function reflects the state of the NV center at every point of the sequence. By multiplying the signal with the modulation function, one gets the rectified signal, whose integral is proportional to the phase accumulation. (b) The Fourier transform of the modulation function quantifies the weighing function for the detectable frequencies using the given sequence. Please note that the $\pi$-pulses are assumed to be instantaneous such that the phase variation during the pulse is negligible (hard pulses). The peaks ($k$) in the plot correspond to the different harmonic orders of the filter function. Adapted and modified with permission from C. L. Degen, F. Reinhard, and P. Cappellaro, Reviews of Modern Physics 89, 1–39 (2017). Copyright 2017 by the American Physical Society.}
         \label{Fig .8}
\end{figure}

This idea is exploited in the \textbf{spin-echo} pulse sequence\citep{PhysRev.80.580} (Fig.~\ref{Fig .7}). The additional $\pi$-pulse refocuses the signal such that the total phase accumulation in this case is
\begin{equation}
    \phi=\gamma_{nv}\left[\int_0^{\tau}B_\parallel(t)dt - \int_{\tau}^{2\tau}B_\parallel(t)dt\right].
\end{equation}
When the time period of the AC field is an even multiple of the free precession time $\tau$, the two terms for phase accumulation above are equal and cancel each other (see Fig.~\ref{Fig .7b}). This way the phase induced by slowly varying fields over the entire sequence is zero. Such spin manipulation experiments with NV centers have been independently reported by several authors in the last decade.\citep{Jelezko_2004,Childress_2006,Balasubramanian2009,Maze2008}\\
The spin-echo protocol with a single refocusing $\pi$-pulse is the basic building block of the \textbf{dynamical decoupling (DD)} sequences\citep{Macquarrie2015} (Fig.~\ref{Fig .8}) that make use of multiple refocusing $\pi$-pulses. Decoupling the system from DC or slowly varying fields is accompanied by coupling to a given frequency signal (Fig.~\ref{Fig .7}). The general phase accumulation for a series of $\pi$ pulses over a duration of time $t$ can be estimated as
\begin{equation}
    \phi_se =\gamma_{nv}\int_0^{t}B_\parallel(t')\mathcal{M}(t')dt', 
\end{equation}{}
\noindent where $\mathcal{M}$ is known as the modulation function and reflects whether the phase accumulated over the specific period $\tau$ in the sequence is positive or negative (Fig.~\ref{Fig .8}). 

{}

In the case of a broadband external field (periodically or randomly oscillating), the contribution to the total phase induced by different frequency components depends on the so-called weighing function (Fig.~\ref{Fig .8}b) for the given pulse sequence. These weighing functions are similar to those of narrow-band frequency filters, i.e.\ they enhance signals from resonant frequencies and suppress non-resonant external field contributions.\cite{Viola1999} Decoupling the spin system from the noisy spin bath in the diamond crystal enhances the coherence time of the spin states.\citep{Bar-Gill2013} It also improves the sensitivity towards the targeted signal which is limited by the spin dephasing ($T_2^*$) and spin decoherence ($T_2$) times. For more information on the nature, usefulness and application of these DD-sequences and filter functions for quantum sensing, the reader may refer to the extensive review on the topic by Degen et al.\citep{Degen2017}. QOC (section~\ref{sec:Theory}) offers an alternative approach to enhance the coherence time for systems with high frequency noise such as shallow NV centers. In addition, the hard pulse approach (i.e. using rectangular pulses) is subjected to cumulative errors with respect to the phase and duration of the pulses, that become significant for longer DD sequences. QOC provides an elegant solution to avoid such errors, which is useful for sensing applications as well as quantum information and computation applications (section~\ref{sec:QuantumInformation}) that require high fidelity gate operations. \\
As an indirect application of magnetic field sensing, gyroscopic measurements with NV centers have been performed.~\citep{McGuinness2011,Ledbetter2012}

\subsubsection{\label{sec:Electricfieldsensing}Electric Field Sensing}

The coupling of the magnetic field to the NV spin is much stronger than that of the electric field. Hence, electric field sensing techniques\citep{VanOort1990} are experimentally more challenging. However, the external electric field directly influences the spin, as discernible from the Hamiltonian in Eq.~\eqref{eqn:fullHamiltonian}. 

\begin{figure}[h!]
        \centering
         \includegraphics[width=\textwidth]{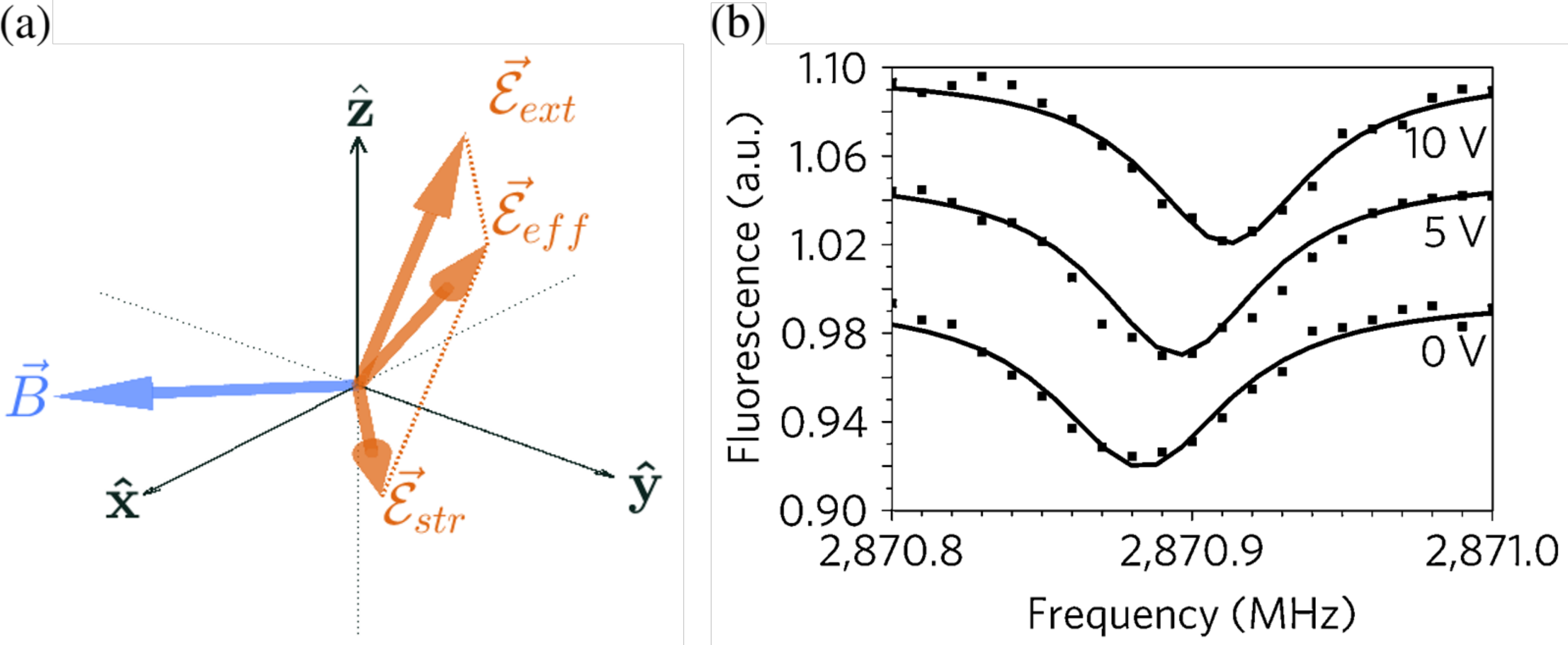}
         \caption{Electric field sensing with NV centers: (a) Careful alignment of the external magnetic field $\Vec{B}$ (orthogonal to the $\hat{z}$- and hence to the NV-axis) is required to measure the effect of the electric field.  $\Vec{\mathcal{E}}_{ext}$ represents the external electric field and $\Vec{\mathcal{E}}_{str}$ the non-axial crystal strain field. $\Vec{\mathcal{E}}_{eff}$ indicates the resulting effective electric field experienced by the spin. (b) The plots show the Stark effect shifting the resonant frequencies between the NV center spin states for different electric field strengths. For details on the experiment see Dolde et al.\citep{Dolde2011}. Reprinted with permission from F. Dolde, H.  Fedder, M. W. Doherty, T.  N ̈obauer, F. Rempp, G. Balasubramanian, T. Wolf, F. Reinhard, L. C. L. Hollenberg, F. Jelezko, and J. Wrachtrup, Nature Physics 7, 459–463 (2011). Copyright 2011.}
         \label{Fig .9}
\end{figure}{}

Comparing their coupling, the magnetic field's effect (represented by $\gamma_{nv}$) is of the order of a few GHz/T\citep{Balasubramanian2008}, whereas the analogous constant of the electric field $\mathcal{E}_I$ is of sub-Hz/Vm$^{-1}$-order\citep{VanOort1990}, making the coupling very weak, as indicated in section \ref{sec:SpinHamiltonian}. Nevertheless, electric fields are also detected quantitatively by the phase induced during the pulse sequence. Hence, similar spin manipulation protocols employed for magnetic field sensing are used for detecting electric fields. As the signal is much weaker than for magnetic fields, the challenge lies in separating it from the effects of other strong undesirable interactions.\\
Dolde et al.\citep{Dolde2011} exploited the interplay between magnetic Zeeman effect, Stark effect and crystal strain field in their carefully devised experiment to suppress the magnetic field effects on the NV spin, making the shifts in the electric field transitions more prominently detectable via spin manipulation protocols (Fig.~\ref{Fig .9}).

\subsubsection{\label{sec:Thermometery}Thermometery}

\begin{figure}[h!]
         \centering
         \includegraphics[width=\textwidth]{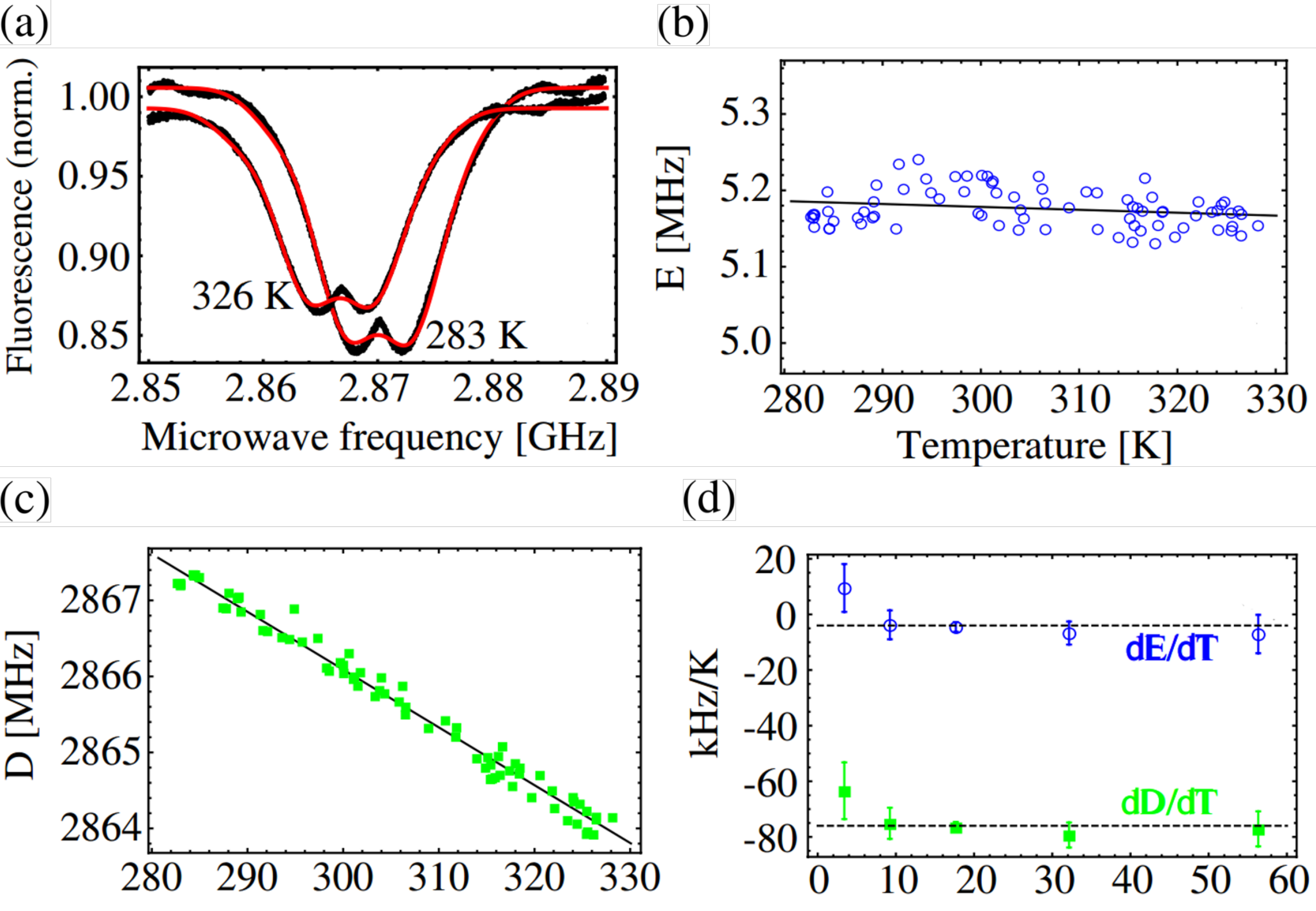}
         \caption{Thermometery with NV centers: (a) Zero-field magnetic resonance spectra for different temperatures (fit in solid red); (b) dependence of the non-axial zero-field parameter $E$ on temperature; (c) dependence of the axial zero-field parameter $D$ on temperature; (d) temperature dependence of $D$ and $E$ vs the laser intensity used for initialisation and readout of the spin state. Adapted and modified with permission from V. M. Acosta, E. Bauch, M. P. Ledbetter, A. Waxman, L.-S. Bouchard, and D. Budker, Physical Review Letters 104, 070801 (2010). Copyright 2010 by the American Physical Society.}
         \label{Fig .10}
\end{figure}{}

\begin{table*}
\begin{minipage}{\textwidth}
    \caption{\label{Tab .1}Summary of quantum sensing with NV centers. $\eta_{B}$ and $\eta_{\mathcal{E}}$ are the sensitivities with respect to the magnetic field $B$ and the electric field $\mathcal{E}$, respectively. $\sigma$ is the standard deviation in the measurement (related to the detected noise), $S$ is the detected signal, and $\mathit{t_m}$ is the measurement time (limited by the coherence time of the system). $D$ and $E$ are the axial and non-axial zero-field parameters, respectively.}
    \begin{ruledtabular}
    \begin{tabular}{lccc}
    Sensed parameter&Parameter/sensitivity dependence &Reported sensitivities&Reference\\
    \hline
    &&&\\
    Magnetic field (B)&$ \eta_{B} = \frac{\sigma}{\left[\pdv*{S}{B}\right]_{max}}\sqrt{\mathit{t_m}}$&pT-$\mu$T/$\sqrt{\text{Hz}}$&Balasubramanian et al., \citep{Balasubramanian2009} Webb et al., \citep{Webb2019}\\
    &&&Neumamm et al.\citep{PhysRevX.5.041001}\\
    Electric field ($\mathcal{E}$)&$ \eta_{\mathcal{E}} = \frac{\sigma}{\left[\pdv*{S}{\mathcal{E}}\right]_{max}}\sqrt{\mathit{t_m}}$&$\approx 100$ V/cm/$\sqrt{\text{Hz}}$&Dolde et al. \citep{Dolde2011}\\
    &&&\\
    Temperature ($\mathcal{T}$)&$\fdv{D}{\mathcal{T}}$, $\fdv{E}{\mathcal{T}}$ & 10-100 KHz/K, $-1.4 \times10^{-4}$Hz/K& Dolde et al., \citep{Dolde2014} Acosta et al.\citep{Acosta2010}\\
    &&&\\
    Pressure (P)& $\fdv{D}{P}$ &$10^5 - 10^6$ Pa/$\sqrt{\text{Hz}}$&Doherty et al.\citep{Doherty2014}\\
    \hline
    \end{tabular}
    \end{ruledtabular}
    \end{minipage}
\end{table*}

Unlike the electric and magnetic fields, temperature is not directly coupled to the spin system (strictly speaking spin-phonon coupling is still temperature dependent, however temperature as a parameter does not directly couple to the spin). Instead, temperature fluctuations have a direct effect on the crystal field as they affect the lattice constant of the crystal. This in turn influences the zero-field splitting parameters discussed in section \ref{sec:SpinHamiltonian}. Acosta et al.\citep{Acosta2010} reported the dependence of the parameters $D$ and $E$ on temperature, demonstrating that temperatures changes can be directly quantified from ODMR spectra (Fig.~\ref{Fig .10}). They found that the axial zero-field splitting parameter is more sensitive to temperature changes than the non-axial parameter.
Neumann et al.\citep{Neumann2013} utilised the D-Ramsey spin manipulation protocol to suppress the effect of external magnetic fields, paving the way to an enhanced measurement of temperature changes.

\subsubsection{\label{sec:PSSO} Pressure and Strain Sensing}
Similar to the temperature, other physical quantities like pressure and strain affect the crystal field. Changes in these quantities can be studied by their coupling to the zero-field splitting parameter. 

\begin{figure}[h!]
         \centering
         \includegraphics[width=0.9\textwidth]{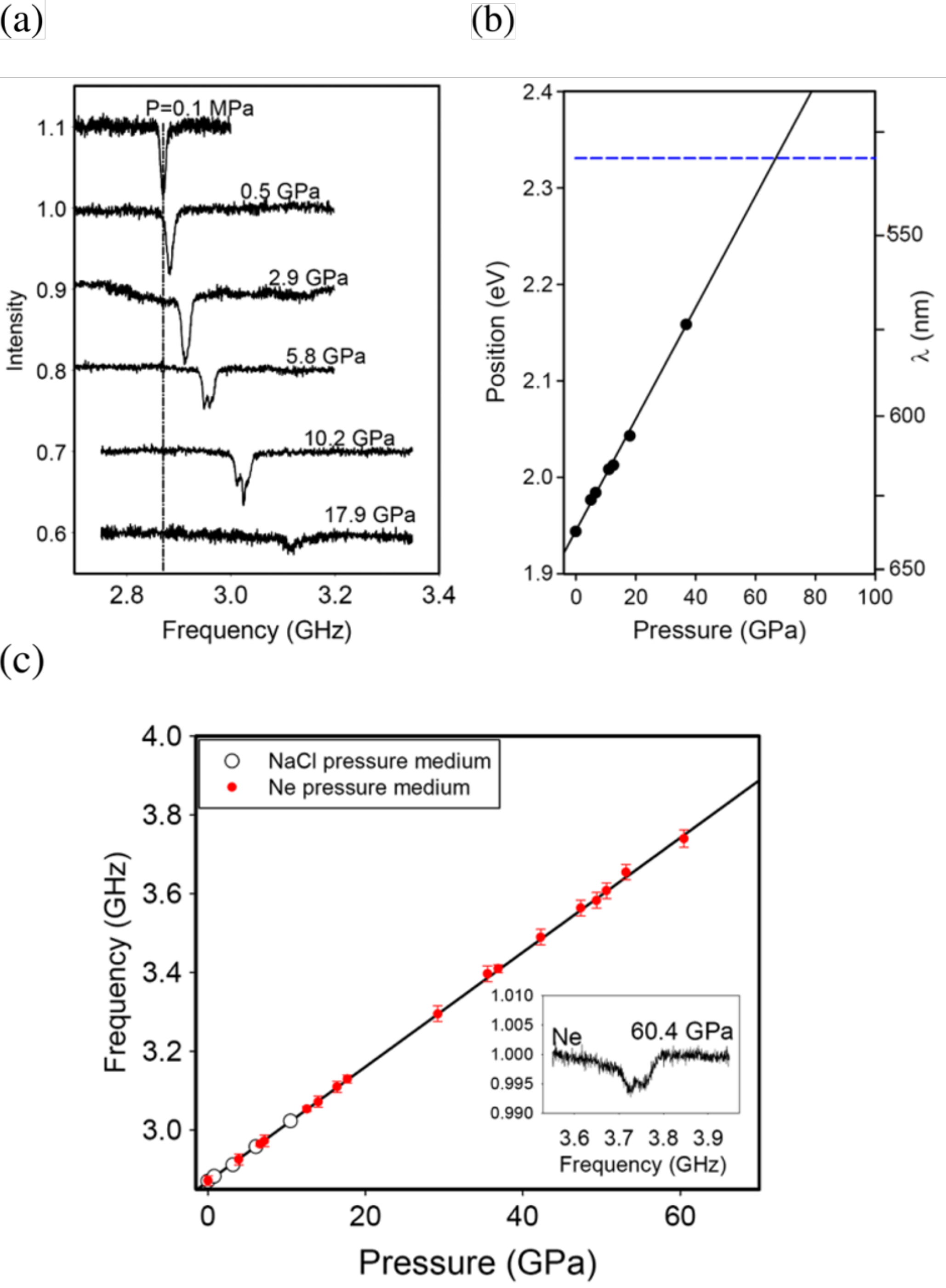}
         \caption{(a) Effect of external pressure on the ODMR spectrum of NV centers (dashed line: resonance frequency at normal atmospheric pressure); (b) shift of the ZPL vs external pressure; (c) shift of the zero-field parameter $D$ vs external pressure. Adapted and modified with permission from M. W.  Doherty, V. V. Struzhkin, D. A. Simpson, L. P. McGuinness, Y. Meng, A. Stacey, T. J. Karle, R. J. Hemley, N. B. Manson, L. C. Hollenberg, and S. Prawer, Physical Review Letters 112, 047601 (2014). Copyright 2014 by the American Physical Society.}
         \label{Fig .11}
\end{figure}

Doherty et al.\citep{Doherty2014} reported the effect of hydrostatic pressure on the  ZPL and the ground state ODMR of NV centers at room temperature in a hydrostatic pressure medium (Fig.~\ref{Fig .11}). Crystal strain variations lead to changes in the effective electric field as described in section \ref{sec:SpinHamiltonian}. The reader is referred to Maze et al.\cite{Balasubramanian2008} for a theoretical description of the effect of strain on the NV center's electronic structure. Teissier et al.\citep{Teissier2014} reported NV spin-strain coupling in diamond cantilever devices.

\subsection{Quantum Information and Computation Applications}
\label{sec:QuantumInformation}

The same characteristics that make NV centers a promising sensor, also qualify them in the fields of quantum information and computation.\citep{Childress2013}
As described in section \ref{sec:GeneralProperties}, the NV electron spin interacts with the internal N nuclear spin of the NV center and $^{13}$C nuclear spins in its close proximity (NV center cluster).
Abobeih et al.~\citep{Abobeih2019} showed that even a much more precise characterisation is possible: They imaged a 27-nuclear spin cluster using a single NV center. Considering the surrounding nuclei, the resulting hyperfine structure~\citep{Dreau2014} (see section \ref{sec:SpinHamiltonian} and Appendix~\ref{app:SpinHam}) forms a more complex quantum mechanical systems than a single electronic qubit. In fact, the majority of quantum computation applications can in principle, be performed with a nuclear spin qubit, using the NV center electron as a quantum bus qubit for initialisation and readout.\citep{Fischer2013,Liu2017}\\
This technique has lead to various NV-based applications such as quantum error correction,\citep{Waldherr2014,Taminiau2014,Cramer2016} quantum algorithms,\citep{Shi2010,Xu2017} quantum simulation,\citep{Wang2015,Kong2016}
fault-tolerant quantum repeaters for long-distance communication,\citep{childress2006}
and quantum memory.\citep{Fuchs2011,Maurer2012,Hensen2015}\\
Entanglement between optical photons and the NV spin has also been demonstrated experimentally\citep{Togan2010}. In rare cases two NV centers might be close enough to each other for dipole-dipole interaction\citep{Dolde2014}. This provides another possible source of entanglement and is further described in Appendix~\ref{app:SpinHam}. In the proceeding paragraphs, we describe applications mainly regarding the use of NV centers as quantum registers, entangling gates with NV centers, and use of NV center qubits to perform quantum error correction.\\

\begin{figure}[h!]
    \centering
    \includegraphics[width = 0.8\textwidth]{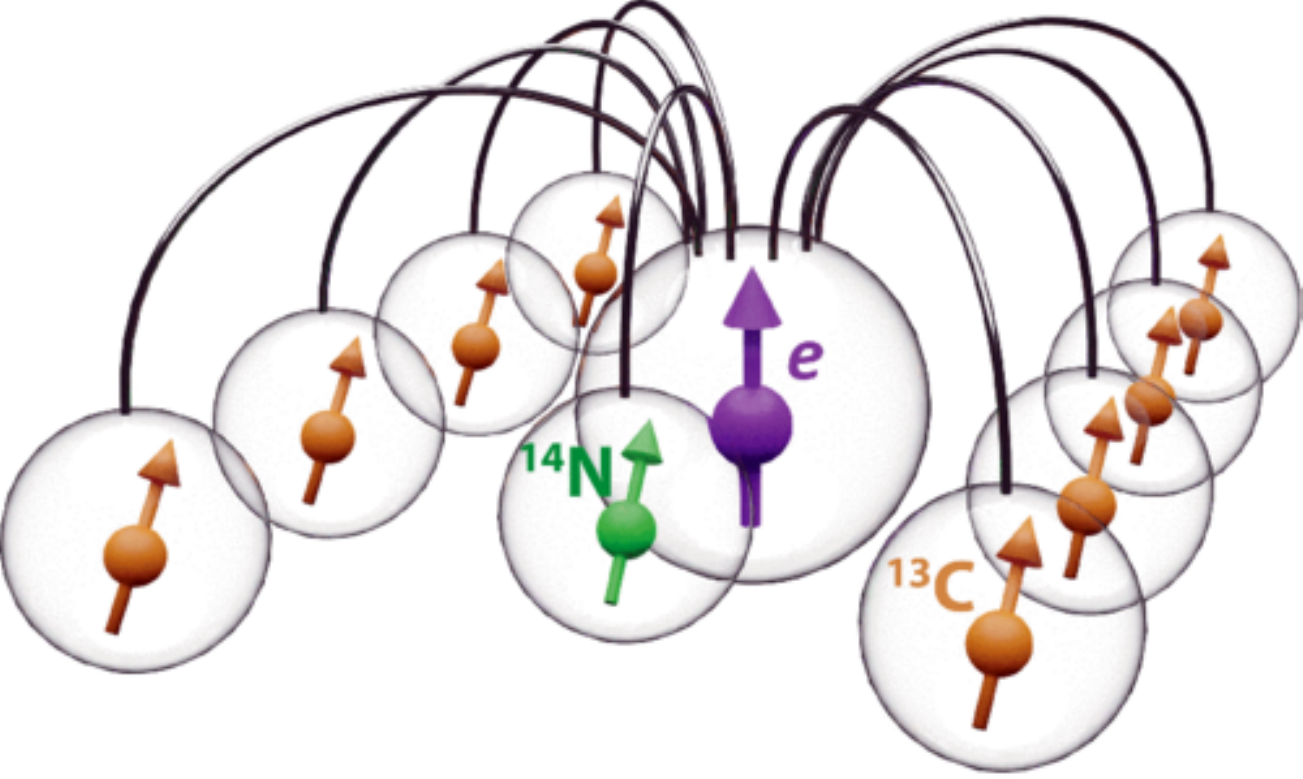}
    \caption{Illustration of a ten-qubit spin register based on an NV center as demonstrated by Brandley et al. \citep{Bradley2019} The electron spin of a single NV center in diamond acts as a central qubit along with one $^{14}$N and eight $^{13}$C nuclear spin qubits. Adapted and modified from C. E. Bradley, J. Randall, M. H. Abobeih, R. C. Berrevoets, M. J. Degen, M. A. Bakker, M.  Markham, D. J. Twitchen, and T.  H. Taminiau, Physical Review X9, 031045 (2019), under the terms of the Creative Commons Attribution 4.0 International License.}
    \label{Fig .12}
\end{figure}

Quantum registers are at the heart of quantum computation techniques (Fig.~\ref{Fig .12}). They contain the coupled set of qubits which are used to perform quantum computations. The nuclear spins in the NV center cluster exhibit long coherence times at room temperature, which is an essential feature for spin system based quantum registers. Recently, Bradely et al.\citep{Bradley2019} presented a quantum register based on the electron spin of an NV center together with nine nuclear spins depicted in Fig.~\ref{Fig .12}. To efficiently implement gates in such a large system, they simultaneously used dynamical decoupling based gates on the NV center electron spin and selective phase-controlled driving of the nuclear spins. As a result they saw coherence times from ten seconds up to a minute at room temperature. In a different approach, Neumann et al. \citep{Neumann2010} demonstrated the working concept of a diamond-based spin register containing two optically addressable NV center spin qubits coupled to each other in an ultrapure and isotopically engineered CVD grown diamond. NV center based spin registers have also been used for quantum simulation of molecules.\citep{Wang2015}\\

\begin{figure*}[ht]
    \centering
    \includegraphics[width=0.8\textwidth]{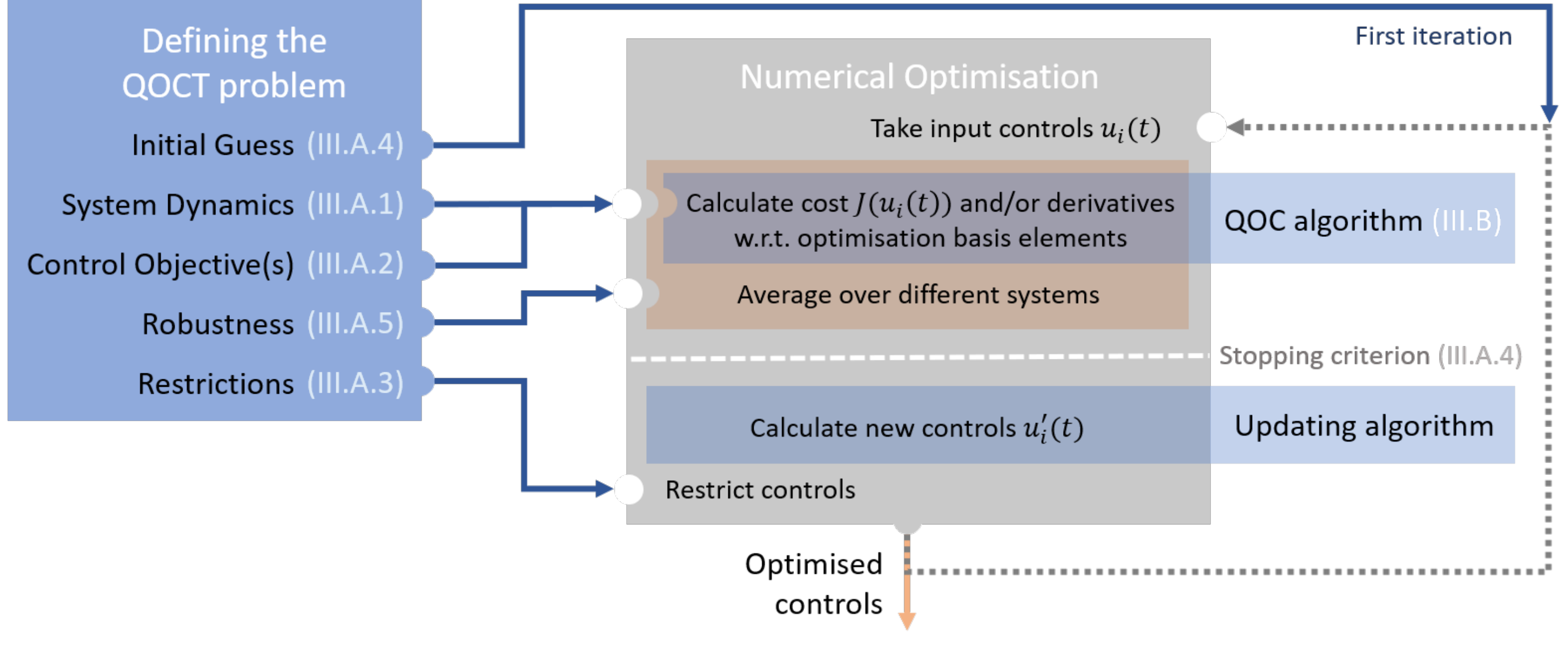}
    \caption{Schematic drawing of a generic QOC optimisation. The box on the left contains the elements that define a basic QOC problem with blue solid arrows connecting them to the algorithm. The grey box at the center illustrates the optimisation algorithm itself, with the dotted grey arrow indicating its iterative nature. The cost function $J$ is calculated from the controls $u_i(t)$ and used to update the controls. In parenthesis the relevant sections in the review paper are indicated where applicable.}
    \label{Fig .13}
\end{figure*}

Similar to the way quantum registers form the underlying structure, entanglement generation enables the connections required for any quantum information technology (entanglement also plays a vital role in several quantum sensing applications, for more information the reader is advised to refer to Degen et al.~\citep{Degen2017}). For example, Dolde et al.\citep{Dolde2013} demonstrated NV-NV spin entanglement at room temperature, while Neumann et al. \citep{Neumann2008} demonstrated entangled two- and three-particle quantum states with $^{13}$C nuclear spins in an NV center based quantum register.\\
Another essential concept for fault-tolerant quantum computation is quantum error correction (QEC). As mentioned above, NV-based spin registers may provide systems with long coherence times and desired entanglement properties. These are prerequisites to implement standard QEC protocols. Taminiau et al.\citep{Taminiau2014} demonstrated error correction with a spin register of nuclear spins from the NV center cluster, using the NV center electron spin as the control qubit. In another work, Unden et al.\citep{Unden2016} demonstrated enhanced quantum sensitivity with metrology techniques based on iterative quantum error correction. Casanova et al.\citep{Casanova2016} performed a complete set of universal quantum gates with an NV center-based spin register, providing a proof of principle for NV center based quantum computation.\\
To obtain better control over the qubit system, several QOC based methods have been implemented so far. They helped to overcome decoherence effects caused by the environment, and to circumvent experimental limitations. We will review some of these works in section \ref{sec:QOC_QI}.\\
Furthermore, NV centers have been long known as a non classical photon source\citep{Kurtsiefer2000,Beveratos2001}, which was applied in single and two photon interference experiments.\citep{Sipahigil2012, Bernien2012} The NV center has also been used for Bell's inequality test related experiments.\citep{Hensen2015}

\section{Optimal Control Theory}
\label{sec:Theory}

Without the ability to precisely manipulate quantum systems, researching their properties and applying them for quantum technologies is almost impossible. Quantum Optimal Control (QOC) theory \citep{Brif2010, Glaser2015} improves the shape of dynamical controls (typically electromagnetic field pulses) to achieve a certain goal to maximum precision. The section starts with the details of defining a QOC problem in section \ref{sec:DefiningAControlPorblem} followed by a description of different numerical optimisation tools in section \ref{sec:NumericalQOCAlgorithms} and concluded by a brief discussion of the limits of QOC (section \ref{sec:LimitsOfControl}). The first part is structured according to the schematic in Fig.~\ref{Fig .13}.\\
In the field of Nuclear Magnetic Resonance (NMR), pulse shaping is used since the 1980s\citep{McDonald1991} and many of the arguments for pulse shape optimisation\citep{Khaneja2005} equally apply to NV centers, which we will focus on. 
In many cases, the time scales defining the decay of NV centers are large compared to the control time. In that case, it is sufficient to study closed system dynamics. Indeed, specific open system techniques such as population suppression and the exploitation of useful dissipation processes are often not applicable to the problems considered in this review.~\citep{Koch2016} Hence, we limit ourselves to a closed system description.

\subsection{Defining a Control Problem}
\label{sec:DefiningAControlPorblem}

The principles of QOC theory derive from early extremisation problems such as Johann Bernoulli's brachistochrone curve problem.\citep{Costabel1988} Similarly, QOC problems are formulated through a system of equations, which broadly defines three things: First, the \textbf{system dynamics}, i.e.\ the theoretically obtained description reflecting the system's behaviour, for example given by the Hamiltonian. Alternatively, this first equation might be replaced by a description through the experiment itself. Second and third, the \textbf{control objectives} and \textbf{control space restrictions}. The objectives on the one hand set the goal of the optimisation, like e.g.\ high fidelity for the transfer to a target state. The control space restrictions on the other hand limit the resources that may be used to reach the desired goal. Together, the three aspects are combined into a so-called control landscape. Each set of controls will result in a different value of the ``cost function'' $J$, a measure for how close the system is to reaching the objective. In case of a minimisation (and throughout this review we will always assume minimisations, unless stated otherwise), each valley corresponds to a locally optimal combination of controls. The goal of the optimisation can now be easily defined as reaching the lowest point in the landscape.\\
We will now discuss in more detail these three ingredients of QOC problems as well as the initial guess, stopping criteria and robustness.

\subsubsection{System Dynamics}
\label{sec:SystemDynamics}

One way to characterise the evolution of a closed quantum system with time dependent controls is through Schr{\"o}dinger's equation. The system Hamiltonian is usually split into two parts; the drift Hamiltonian $\hat{H}^d$, which is constant and cannot be manipulated, and the control Hamiltonians $\hat{H}^c_i$ which are multiplied with time-dependent coefficients $u_i(t)$ called ``control pulses''. The full Hamiltonian then reads
\begin{equation}
    \hat{H}=\hat{H}^d+\sum_i u_i(t) \hat{H}^c_i.
\end{equation}
Please note that system dynamics for control problems may also be defined through Lindblad-operators and even for non-Markovian dynamics (for a review on open systems QOC see Koch\citep{Koch2016}).\\

\begin{tcolorbox}[breakable, enhanced,colback=palecornflowerblue,boxrule=0pt,title=Drift and Control Hamiltonian]
    As an example, let us consider a NV center, approximated as a qubit with the ground state $\ket{0}$ and excited state $\ket{1}$. In this simple consideration the goal will be to create a high-fidelity $\frac{\pi}{2}_x$-rotation 
    similar to the system in Frank et al.\citep{Frank2017}\\
    A static magnetic field $B_{\parallel}$ is applied in $z$-direction (the quantisation-/NV-axis) and a circularly polarised microwave field $\Vec{B}_\perp$ with an amplitude $B_\perp(t)$, frequency $\omega_{mw}$ and phase $\varphi$ is applied orthogonal to $B_{\parallel}$. Let us define the gyromagnetic ratio of the NV center as $\gamma_{nv}$. The rotating frame of $\Vec{B}_\perp$ then gives the Hamiltonian
    \begin{equation}
    \begin{split}
        \hat{H}_\text{RWA}/\hbar&=\Delta\hat{\boldsymbol{s}}_Z + \Omega(t)\left(\hat{\boldsymbol{s}}_X \cos\varphi(t) + \hat{\boldsymbol{s}}_Y \sin\varphi(t)\right)\\
        &=\Delta\hat{\boldsymbol{s}}_Z + u_1(t)\hat{\boldsymbol{s}}_X + u_2(t)\hat{\boldsymbol{s}}_Y,
    \end{split}
    \label{eqn:Hamiltonian}
    \end{equation}
    where $\Delta=\omega_{nv}-\omega_{mw}$ is the detuning, $\omega_{nv}=B_{\parallel}\gamma_{nv}$ the NV's resonant frequency, $\Omega(t)=B_\perp \gamma_{nv}$ is the Rabi frequency and $\hat{\Vec{\boldsymbol{s}}}$ are the spin operators in the $\ket{0},\ket{1}$ basis. A derivation of this Hamiltonian can be found in Appendix~\ref{app:RWA}.\\
    We can easily identify the drift Hamiltonian $\hat{H}^d=\Delta\hat{\boldsymbol{s}}_Z$. Let us assume, that both the Rabi frequency $\Omega(t)$ and the phase of the magnetic field $\varphi(t)$ can be manipulated dynamically. The control Hamiltonians may then be identified as $\hat{H}^c_1=\hat{\boldsymbol{s}}_X$ and $\hat{H}^c_2=\hat{\boldsymbol{s}}_Y$ and the control pulses as $u_1(t)=\Omega(t)\cos\varphi(t)$ and $u_2(t)=\Omega(t)\sin\varphi(t)$.
\end{tcolorbox}

Once the system has evolved, it is time to test whether the goals have been reached by checking the control objectives.\\
It should be noted at this point that the rotating wave approximation, as presented in Appendix~\ref{app:RWA}, is widely used to simplify the NV center's Hamiltonian. While it is useful, when the Rabi frequency is much lower than the NV center's resonant frequency, it can have a detrimental effect on a simulation's accuracy, if the Rabi frequency is of a similar scale as the qubit transition. In fact, Scheuer et al.~\citep{Scheuer2014} have shown how the inaccurate use of the RWA can affect the outcomes of optimal control procedures designed for NV centers.

\begin{figure}[h!]
    \centering
    \includegraphics[width=\textwidth]{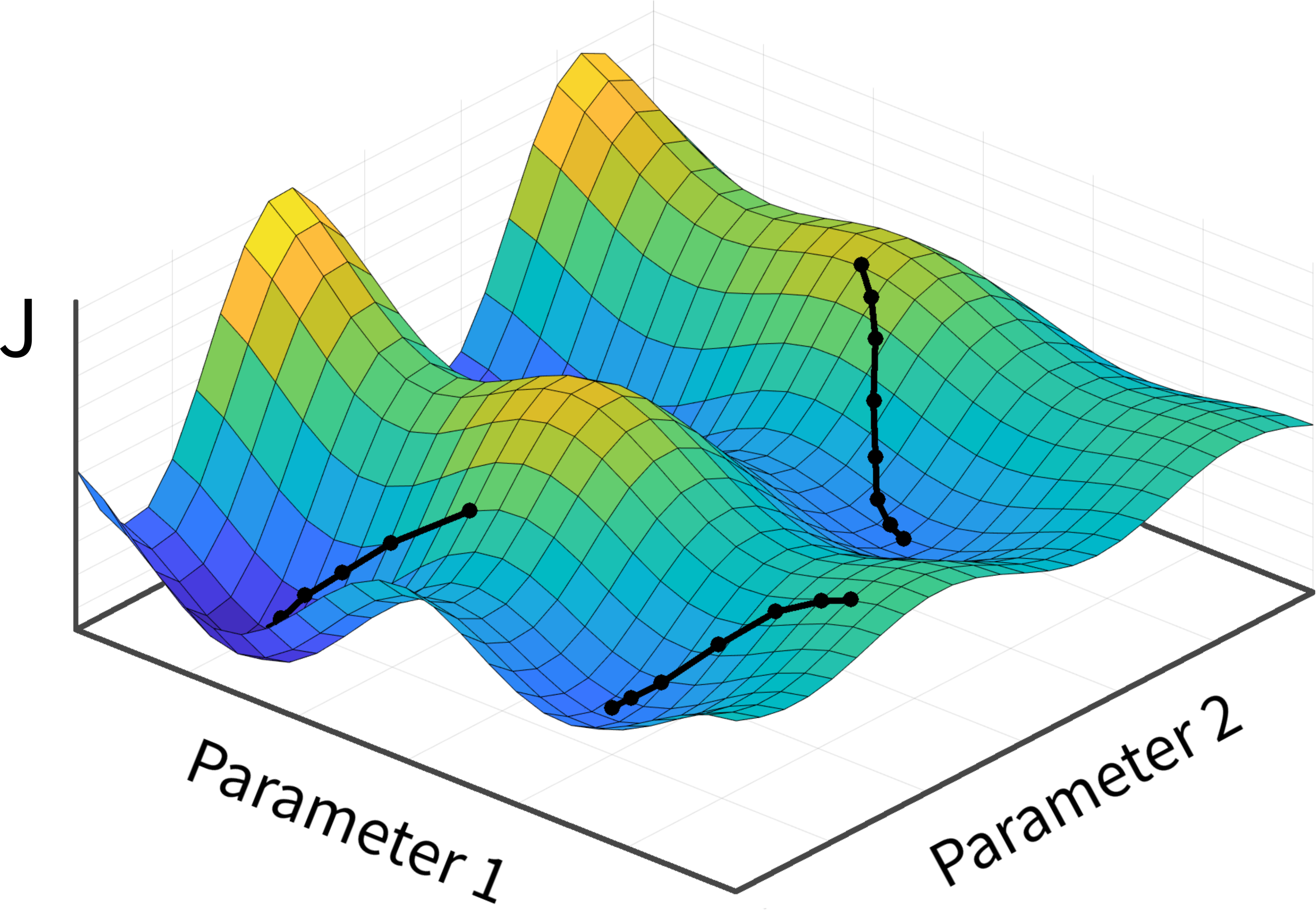}
    \caption{Example of a QOC landscape. Considering two control parameters, we may represent the cost function $J$ as a surface dependent on the set of controls. The minima correspond to locally optimal control coordinates. Each black path represents a local optimisation starting from a different initial guess.}
    \label{Fig .14}
\end{figure}

\subsubsection{Control Objective(s)}
\label{sec:ControlObjectives}

The cost function (or figure of merit) $J$ defines what is minimised in any QOC problem. This way it describes the goal of the optimisation (terminal cost) and optionally the control limits through penalty-terms (running costs). 
The terminal costs are determined at the final time of the system's evolution. They quantify for instance the distance between the final state and the desired goal state in the relevant Hilbert space. The running costs are usually related to the restrictions on the control pulses, for example the limited power of a microwave source. In this review, the cost function is defined to be zero, when all objectives are met and to be greater than zero, when they are not met. Note that the running costs have a similar role as the control space restrictions that will be discussed in section~\ref{sec:ControlSpaceRestrictions}.\\
In the following, we will briefly describe some of the most relevant cost functions found in relation to NV centers in the literature. For more examples of NV center applications, please refer to section \ref{sec:Applications} and to section \ref{sec:QOC_forNVs} for examples specifically combining them with QOC.\\

\begin{itemize}
    \item \textbf{State to state transfer} (terminal cost)\\
        State to state transfer is the most common optimisation objective and has been used in many papers \citep{Konzelmann2018,Haberle2013,Frank2017,Tsurumoto2019}. The infidelity is a measure for the distance between two states $\ket{\phi_t}$ and $\ket{\phi(T)}$: If they are equal, it gives zero, if they are orthogonal, it has a value of one.
        The infidelity can be used directly to define the cost function $J_\text{state}$
        \begin{equation}
            J_\text{state}=1-|\braket{\phi_t}{\phi(T)}|^2,
        \end{equation}
        describing the distance between $\ket{\phi(T)}$, the final state of the system at time $T$, and $\ket{\phi_t}$, the target state.
        An alternative way to formulate the transfer is fixing the global phase using $J_\text{state}=1-Re\{\braket{\phi(T)}{\phi_t}\}$.
    \item \textbf{Unitary gate optimisation} (terminal cost)\\
        To measure the distance between the unitary $U(T)$ produced by the controls and the target gate $U_t$, we define the cost function
        \begin{equation}
        \begin{split}
            J_\text{gate}&=1-\frac{1}{N_0^2}\left|\Tr(U_t^\dagger U(T))\right|^2\\
            &=1-\frac{1}{N_0^2}\left|\sum_{i=1}^{N_0}\bra{\zeta_i}U_t^\dagger\ket{\phi_i(T)}\right|^2.
        \end{split}
        \end{equation}
        In the second line, the gate fidelity is defined through the $N_0$ basis states $\ket{\zeta_i}$ of the initial system and their propagated version $\ket{\phi_i(T)}=U(T)\ket{\zeta_i}$. Similarly to $J_\text{state}$, we can also define a global phase dependent version of this cost function, $J_\text{gate}=1-\frac{1}{N_0}\Re\{\text{Tr}(U_t^\dagger U(T))\}$. Examples for its application can be found in references \citep{Said2009,Montangero2007,Goerz2015,Palao2003}.
    \item \textbf{Sensitivity} (terminal cost)\\
        In contrast to the previous examples, the sensitivity does not directly contain information about the system. Instead, it quantifies the amount of information about a parameter $\theta$ (e.g.\ the magnetic field) that may be derived from a set of measurements.\\
        The sensitivity may be defined as the variance $(\Delta \theta)^2$ of the parameter estimate $\theta_0$ obtained from $N_M$ measurements. Each measurement produces the expectation value of some positive-operator-values measure (POVM)\footnote{A POVM is defined by a set of Hermitian operators which produce a positively-valued expectation values with a normalised probability distribution.\citep{nielsen00}} $\Theta$.
        The probability to measure $\theta(\ket{\phi})=x$ probing a wavefunction $\ket{\phi}$ is then given by the expectation value $p(x|\theta)=\bra{\phi}\Theta \ket{\phi}=\Tr(\rho \Theta)$, with $\rho=\ketbra{\phi}$.\\
        The cost function should, however, not contain information about the outcome of the measurement, but rather about its precision.
        The lower bound of $\Delta\theta$ is given by the Cram{\'e}r-Rao bound
        \begin{equation}
            (\Delta\theta)^2\ge\frac{1}{N_M F(\theta_0)},
            \label{eqn:Sensitivity}
        \end{equation}
        a value which is inversely proportional to the Fisher information $F(\theta)$, calculated by
        \begin{equation}
            \begin{split}
                F(\theta)&=\int dx \frac{1}{p(x|\theta)} \left(\frac{\partial p(x|\theta)}{\partial \theta}\right)^2\\
                &=\sum^{N_x}_i \frac{1}{p(x_i|\theta)} \left(\frac{\partial p(x_i|\theta)}{\partial \theta}\right)^2,
            \end{split}
            \label{eqn:FisherInfo}
        \end{equation}
        where the second line is specifically related to a discrete number of possible measurement outcomes $N_x$.\\
        One may interpret the Fisher information as the curvature of the logarithmic probability distribution: If it is completely flat, hence giving no information, $F(\theta_0)=0$, if it is strongly peaked, indicating a clear parameter estimate, $F(\theta_0)\gg0$.\\
        The corresponding cost function may be defined as 
        \begin{equation}
            J_\text{Fisher}=\frac{1}{N_M F(\theta)}.
        \end{equation}
        Reviews introducing Fisher information in the context of quantum sensing and metrology were written by Degen et al.\citep{Degen2017} and Pezze et al.\citep{Pezze2018} The original paper relating Fisher information and quantum mechanics was published in 1994 by Braunstein and Caves.\citep{Braunstein1994} Applications of Fisher information as a part of optimal control can be found in references.\citep{Muller2018,Poggiali2018}
    \item \textbf{Limited power} (running cost)\\
        The power of a control pulse is typically calculated as $P_i=\int_0^T |u_i(t)|^2dt$. To limit $P_i$ to a reasonable range $P_i\in[0,P_\text{lim}]$ a penalty term can be introduced which adds a high cost to $J$, if a certain limit is crossed. 
        \begin{equation}
            J_\text{power}=\kappa(P_i)=\kappa\left( \int_0^T |u_i(t)|^2 dt\right).
        \end{equation}
        The function $\kappa(P_i)$ should give very little to no penalty, if the power is within the acceptable range $\kappa(P_i\ll P_\text{lim})\rightarrow 0$ and a high penalty, if it is out of range $\kappa(P_i\gg P_\text{lim})\rightarrow \infty$. These criteria can be satisfied by a wide variety of functions and it depends on the chosen system. Examples can be found in a number of references. \citep{Caneva2011, Khaneja2005, Kobzar2008,Nobauer2015}
    \item \textbf{Limited bandwidth} (running cost)\\
        There is a number of ways to limit the bandwidth of the controls. One solution is to gently punish any quickly oscillating solutions through
        \begin{equation}
            J_\text{bandw}=\epsilon \int_0^T\left(\frac{\partial u(t)}{\partial t}\right)^2dt,
        \end{equation}
        where $\epsilon$ is some small factor~\citep{Goodwin2018}.
        It should be noted that this expression, does not give strict bounds in terms of bandwidth.
        An alternative, stricter approach is to punish fast oscillations, only if they lie outside a pre-defined filter function as described by Sch{\"a}fer et al.\citep{Schaefer2019} and Kosloff's group.\citep{Pezeshki2008,Werschnik2007}
        A completely different approach is to restrict the basis of the control pulses. This is possible with certain algorithms of the (d)CRAB family including GROUP, and GOAT and will be further discussed in section \ref{sec:NumericalQOCAlgorithms}.
\end{itemize}

There are many more possible terminal costs, each describing a different control problem including partial state transfer, taking into account the full density function, maximising entanglement,\citep{Omran2019} or adjusting a certain observable.\citep{Riviello2017} Similarly, equally many different running costs exist e.g.\ to avoid populating fast decaying states.\citep{Palao2008}

\begin{tcolorbox}[breakable, enhanced,colback=palecornflowerblue,boxrule=0pt,title=Gate Optimisation]
    In the experiment by Frank et al.\citep{Frank2017} the control objective was to optimise a unitary defined as the Hadamard gate
    \begin{gather*}
            U_t=\frac{1}{\sqrt{2}}
        \begin{bmatrix}
          1 & -i \\
          -i & 1 \\
        \end{bmatrix}.
    \end{gather*}
    Hence the cost function may be defined as
    \begin{equation}
        J=J_\text{gate}=1-\frac{1}{4}\left|\Tr(U_t^\dagger U(T))\right|^2.
    \end{equation}
    We can see that this is a good cost function as it is minimal when $U(T)=U_t$ at the final time $T$ (up to a global phase). If we were to also include a bandwidth limitation on the two control pulses $u_1(t)$ and $u_2(t)$, we may simply sum up different cost terms. The resulting cost function, where $\epsilon_i$ are some small factors, would be
    \begin{equation}
        \begin{split}
            J=&J_\text{gate}+J_\text{bandw}=1-\frac{1}{4}\left|\Tr(U_t^\dagger U(T))\right|^2\\
            &+\epsilon_1\int_0^T\left(\frac{\partial u_1(t)}{\partial t}\right)^2 dt+
            \epsilon_2\int_0^T\left(\frac{\partial u_2(t)}{\partial t}\right)^2dt.
        \end{split}
    \end{equation}
\end{tcolorbox}
Running costs favour acceptable types of controls, as opposed to physically impossible ones, but if stricter limits are required, control space restrictions might be the more suitable mean of limitation.

\subsubsection{Control Space Restrictions}
\label{sec:ControlSpaceRestrictions}

While the running costs (see section \ref{sec:ControlObjectives}) can only passively punish controls which lie outside the achievable frame, control space restrictions actively change the controls to only allow what is experimentally achievable. One might imagine them as a horizontal squeezing and stretching of the control landscape or as the introduction of hard walls (see Fig.~\ref{Fig .14}), opposed to a vertical distortion induced by running costs.\\
\begin{tcolorbox}[breakable, enhanced,colback=palecornflowerblue,boxrule=0pt,title=Restricting the Control Amplitude]
    As an example, let us consider the amplitude of a control pulse $u_i(t)$ that should be restricted to $u_i^\text{max}$. The pulse could be cut off at the beginning of each iteration according to
    \[
        \tilde{u}_i(t) =
        \begin{cases}
            u_i(t) ,& \text{if } -u_i^\text{max}<u_i(t)<u_i^\text{max}\\
            u_i^\text{max},& \qquad \text{if } u_i(t)\geq u_i^\text{max}\\
           - u_i^\text{max},& \qquad \text{if } u_i(t)\leq -u_i^\text{max}.
        \end{cases}
    \]
    This form ensures maximum exploitation of the amplitude space but is not differentiable, hence it requires that the control pulse is cut off during an extra step (Fig.~\ref{Fig .13}) before the cost function and/or gradient is evaluated.\citep{SKINNER2004}\\
    An alternative approach is mapping the control pulse to a restricted subspace using a continuous function. For example by replacing it with $\tilde{u}_i(t) = u_i^\text{max} \sin(u_i(t))$.\citep{Machnes_2018}\\
    Another common example for the application of mapping are shape functions. They restrict the overall shape of the pulse, which is useful, if e.g. the experiment requires a smoothly rising and falling control pulse with $\Gamma(0)=\Gamma(T)=0$, such that $u_i'(t) = \Gamma(t) u_i(t)$.
\end{tcolorbox}

\subsubsection{Initial Guess and Stopping Criterion}
\label{sec:InitAndStop}

Numerical optimal control techniques are based on iterative algorithms, which require a starting point (called ``initial guess'') and a clearly defined situation to stop at i.e.\ the stopping criterion. The optimisation will in most cases find the the closest local minimum to the initial guess (see examples in Fig.~\ref{Fig .14}). Accordingly, it is often helpful to try a number of initial guesses to find which one is closest to the global minimum.\\
The stopping criterion is simpler to define: It might be based on the maximum number of iterations (limited computation time or experimental run time), a measure for convergence or the clear definition of a goal.

\subsubsection{Robustness}

Usually, there is some discrepancy between the theoretical model and the experiment. In an optimisation, this can be taken into account to ensure that the optimised pulses will work in the presence of such a discrepancy by averaging over cost functions for slightly different systems. Let us consider each system being described by the Hamiltonian $\hat{H}_i$, then by taking into account $N_\text{rob}$ different versions, the cost function becomes
\begin{equation}
    J_\text{robust}=\frac{1}{N_\text{rob}}\sum_{i=1}^{N_\text{rob}} J(H_i).
\end{equation}
\begin{tcolorbox}[breakable, enhanced,colback=palecornflowerblue,boxrule=0pt,title=Robustness Against Detuning]
    The resonance line of the NV center has a finite width and can be described by the normalised distribution $f(\omega)$. Off-center NV centers can however still be described with the Hamiltonian $\hat{H}$ in Eq.~\eqref{eqn:Hamiltonian} by adjusting the static magnetic field $B_{||}$ and hence the detuning $\Delta$. One may average the cost over $N_\text{det}$ different detunings $\Delta_i$ to get a robust cost function
    \begin{equation}
        J=\frac{1}{N_\text{det}}\sum_{i=1}^{N_\text{det}} J(\hat{H}(\Delta_i))f(B_{||}\gamma_{nv}-\Delta_i).
    \end{equation}
    We can see that this cost function will only reach zero, if all $J(\hat{H}(\Delta_i))$ are zero, ensuring robustness against the detuning. By including the probability distribution $f(B_{||}\gamma_{nv})$, we ensure that the optimisation favours solutions centered on the average detuning.\\
    Due to field inhomogeneities in $B_\perp$, the Rabi frequency can also have a finite distribution when considering an ensemble of NV centers. In many optimisations, both detuning and Rabi errors are accounted for simultaneously.~\citep{Kobzar2012,Kobzar2008}
\end{tcolorbox}

\subsection{Numerical QOC Algorithms}
\label{sec:NumericalQOCAlgorithms}

Once the problem has been defined, an algorithm is required to systematically test possible solutions minimising the cost.~\citep{Glaser2015} In this review, we will only describe \textbf{numerical} optimisation algorithms as they have produced promising results and a variety of packages exist to implement them (see section \ref{sec:OCPackages} for more details).\\
There are, however, alternative strategies, such as geometrical optimal control \citep{Bonnard2003,Boscain2004} (GOC) and shortcuts to adiabaticity~\citep{Guery-Odelin2019,Torrontegui2013} (STA). They usually rely on a deeper analytical analysis of the control problem and hence access a smaller solution space than QOC, but can nevertheless be effective. One example is the direct application of Pontryagin’s minimum principle (PMP) which falls under the category of GOC. It has been shown to provide time optimal evolution for NV centers.~\citep{Avinadav2014,Ansel2018} Similarly, STA has been used to implement specific gates on NV centers~\citep{Kleissler2018} and protect them from decoherence.~\citep{Kolbl2019}\\
Section \ref{sec:DefiningAControlPorblem} started by mentioning the brachistochrone problem, whose solution is usually obtained via an analytical variational approach. In the case of quantum mechanical problems however, one would produce a set of nonlinear equations which, in most cases, cannot be solved analytically. Instead, a variational approach combined with numerical solving was introduced by Konnov and Krotov,~\cite{Konnov1999_global_methods} and Sklarz and Tannor~\cite{Sklarz2002} and further adapted by Ohtsuki et al.~\citep{Ohtsuki2004} Since then, different attempts have been made to numerically solve this class of problems.\\
In general, two families of QOC algorithms can be identified: Gradient-free \citep{Doria2011} and gradient-based \citep{Khaneja2005}.
Gradient-based algorithms determine the derivatives of the cost function with respect to the control pulses to find an improved solution. These methods are usually efficient as they make use of all the available information. Gradient-free methods on the other hand can be applied directly to experiments or to complicated problems, where the gradients are not straightforwardly calculated. In this review, the direct experimental implementation is referred to as closed-loop, while a purely simulation based optimisation is called open-loop.\\
We will start by looking at the working principle of gradient-based optimisation algorithms, before exploring their gradient-free counterparts.

\subsubsection{Gradient-based Optimisation}
\label{sec:GradientbasedOptimisation}

To understand how gradient-based algorithms work, let us first consider the effect of a small change $\Delta u$ in some control $u(t)$ on the cost function $J(u(t))$ (see section \ref{sec:ControlObjectives}). If the change $\Delta u$ is small enough, we can approximate
\begin{equation}
\label{eqn:update}
    J(u(t)+\Delta u) \approx J(u(t)) + \Delta u \frac{\partial J}{\partial u(t)}.
\end{equation}
We can now deduce the properties $\Delta u$ should have to decrease $J$. Indeed it enables us to make small changes and update $u(t)$ iteratively. Consider $\Delta u = -\epsilon \frac{\partial J}{\partial u(t)}$, where $\epsilon$ is a small positive factor. In this case, the new cost function becomes
\begin{equation}
\begin{split}
    J(u(t)+\Delta u) &\approx J(u(t)) - \epsilon \left( \frac{\partial J}{\partial u(t)}\right)^2\\
    &< J(u(t)) 
\end{split}
\end{equation}
which is smaller than the previous value, implying an optimisation.\\
In order to avoid functional derivatives and iteratively improve the cost, the control function $u(t)$ needs to be split up into time independent control parameters $u^{(k)}$. According to the simple updating algorithm above, the new control parameters ${u^{(k)}}'$ become
\begin{equation}
    {u^{(k)}}'=u^{(k)}-\epsilon\frac{\partial J}{\partial u^{(k)}}.
\end{equation}
More advanced updating algorithms promise faster convergence. Eq.~\eqref{eqn:update} could for example be extended to second-order.~\citep{Konnov1999_global_methods,Reich2012} A popular method approximating the second-order term from the first-order term is the L-BFGS (limited-memory Broyden-Fletcher-Goldfarb-Shanno) quasi-Newton method algorithm.~\citep{NocedalWright2006_book, DeFouquieres2011}\\
In the following, we describe the working principles of the GRadient Ascent Pulse Engineering (GRAPE) algorithm as an example for the whole class of algorithms. It was originally designed for Nuclear Magnetic Resonance (NMR) \citep{Khaneja2005} but has since found various applications with NV centers. 
For another comprehensive explanation, we refer to Saywell et al.\citep{Saywell2019}
\\\\

\begin{figure}[h!]
    \centering
    \includegraphics[width=\textwidth]{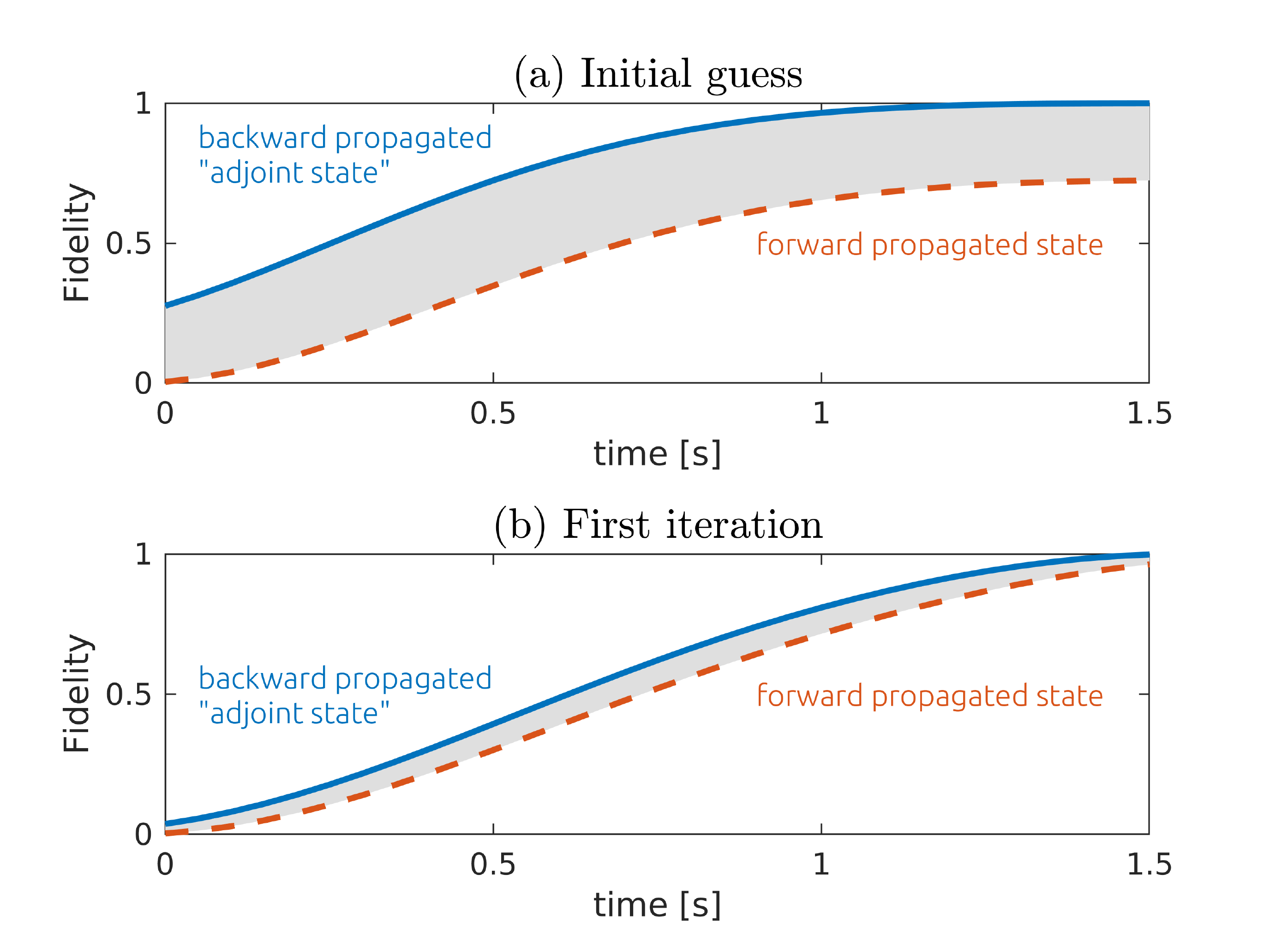}
    \caption{The principle of GRAPE optimisation. A state $\ket{\phi(t)}$ was propagated for a time $T=1.5$s according to the Hamiltonian in Eq.~\eqref{eqn:Hamiltonian}. The fidelity $|\braket{\phi(t)}{\phi_t}|^2$ is plotted as a function of time. The upper panel (a) shows the fidelities resulting from the initial guess for the control pulse of a quantum process. The large grey area indicates that the forward propagated state (dashed orange) and the adjoint state (solid blue) do not match, i.e.\ the target state is not reached. By calculating the derivatives w.r.t.\ the different time slices, an updated control pulse was found using GRAPE producing the lower panel (b) with a clearly improved fidelity.}
    \label{Fig. 15}
\end{figure}

\begin{tcolorbox}[breakable, enhanced,colback=palecornflowerblue,boxrule=0pt,title=GRAPE Optimisation of a State Transfer]
    We start by defining an exemplary cost function $J$, where $J=0$ implies the target state $\phi_t$ is reached at a time $T$ (also see section \ref{sec:ControlObjectives}):
    \begin{equation}
    \begin{split}
        J &=J_\text{state} =1-|\braket{\phi_t}{\phi(T)}|^2\\
          &=1-|\bra{\phi_t}U(T)\ket{\phi_0}|^2,\\
        \text{where } & U(T) = \hat{T} \exp({\int_0^T -\frac{i}{\hbar}\hat{H}(t) dt}),\\
    \end{split}
    \end{equation}
    The initial state is defined as $\ket{\phi(0)}=\ket{\phi_0}$ and $\hat{T}$ is the time ordering operator.\\
    In order to take the derivatives of $J$, we choose the piece-wise constant control basis, chopping up the control pulses into $N$ small slices $u_i^{(k)}$ of width $\Delta t$. This gives a new way to formulate the propagator $U(T)$:
    \begin{equation}
    \begin{split}
        U(T) = \hat{T}\prod_k \exp({-\frac{i\Delta t}{\hbar}(\hat{H}^d+\sum_i u_i^{(k)} \hat{H}^c_i)}) = \hat{T}\prod_k U^{(k)}
    \end{split}
    \end{equation}
    It should be noted that this basis is not the only possible choice but intrinsic to GRAPE (as well as to other gradient based algorithms like Krotov~\citep{Krotov1993}). We can now reformulate the cost function as
    \begin{equation}
        J =1-|\bra{\phi_t}U^{(N)}U^{(N-1)}...\,U^{(1)}U^{(0)}\ket{\phi_0}|^2.
    \end{equation}
    We start by calculating the derivatives of $\bra{\phi_t} \hat{T}\prod_k U^{(k)}\ket{\phi_0}$ w.r.t. the control parameters.
    \begin{equation}
       \frac{\partial}{\partial u_i^{(k)}} \bra{\phi_t}U^{(N)}U^{(N-1)}\quad.\;.\;.\quad \;U^{(k)}\quad.\;.\;.\quad U^{(1)}U^{(0)}\ket{\phi_0}\\
    \end{equation}
    \begin{equation}
       = \quad \underbrace{\bra{\phi_t}U^{(N)}U^{(N-1)}...\,U^{(k+1)}}_{\bra{\xi^{(k)}}}\frac{\partial \, U^{(k)}}{\partial u_i^{(k)}}\underbrace{U^{(k-1)}...\,U^{(1)}U^{(0)}\ket{\phi_0}}_{\ket{\phi^{(k)}}}
    \end{equation}
    \begin{equation}
    \begin{split}
       &= \qquad \; \bra{\xi^{(k)}}\frac{\partial \, U^{(k)}}{\partial u_i^{(k)}}\ket{\phi^{(k)}},\\
       \frac{\partial J}{\partial u_i^{(k)}} &= 2 \Re{\bra{\xi^{(k)}}\frac{\partial \, U^{(k)}}{\partial u_i^{(k)}}\ket{\phi^{(k)}}}.
    \end{split}
    \end{equation}
    We have defined the forward propagated state $\ket{\phi^{(k)}}$ and the backward propagated state $\bra{\xi^{(k)}}$, which is usually referred to as the adjoint state. They can both be easily calculated by solving Schr{\"o}dinger's equation. A graphical representation is given in Fig.~\ref{Fig. 15}. We applied the chain rule to find the gradient of $J$. In the case that the control pulses are mapped to a restricted subspace (see section \ref{sec:ControlSpaceRestrictions}), the chain rule can be used again.\\
    The last thing left to evaluate are the following directional derivatives:
    \begin{equation}
        \frac{\partial \, U^{(k)}}{\partial u_i^{(k)}} = \frac{\partial }{\partial u_i^{(k)}}\left[\exp({-\frac{i\Delta t}{\hbar}\left(\hat{H}^d+\sum_i u_i^{(k)} \hat{H}^c_i\right)})\right].
    \end{equation}
\end{tcolorbox}

\begin{figure}[h!]
    \centering
    \includegraphics[width=\textwidth]{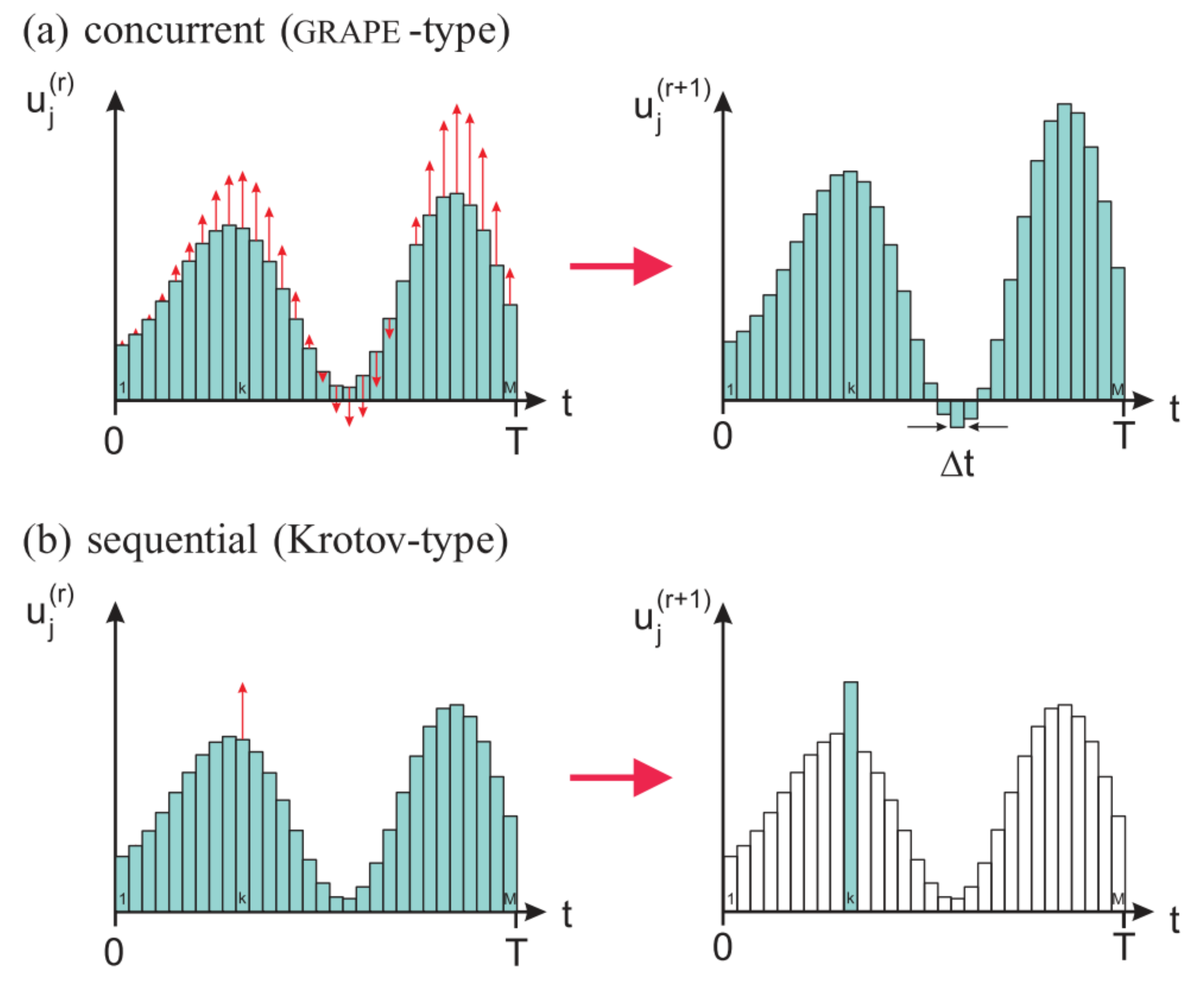}
    \caption{The difference between concurrent (a) and sequential (b) QOC algorithms is illustrated. For concurrent algorithms, the update is calculated at once for the entire time grid. For sequential algorithms, the pulse's basis components i.e.\ time slices are updated sequentially, meaning that in each iteration the forward propagated state is calculated with the latest version of the pulse.Adapted and modified with permission from S. Machnes, U. Sander, S. J. Glaser, P. de Fouquières, A. Gruslys, S. Schirmer, and T. Schulte-Herbrüggen, Physical Review A84, 22305 (2011). Copyright 2011 by the American Physical Society.
}
    \label{Fig. 16}
\end{figure}

All in all, any gradient-based optimisation algorithm relies on calculating the first derivative of the cost function with respect to the control parameters. On top of this specific example of GRAPE, we list below the most commonly used gradient-based QOC algorithms and their natural features (for an illustration of the terms ``sequential'' and ``concurrent'' see Fig.~\ref{Fig. 16}):

\begin{itemize}
    \item GRAPE \citep{Khaneja2005, DeFouquieres2011}\\
        GRadient Ascent Pulse Engineering concurrently optimises in the piece-wise constant basis.
    \item Krotov \citep{Krotov1993, Konnov1999_global_methods, Reich2012,  Eitan2011, Palao2003, Sklarz2002}\\
        Krotov's method sequentially optimises one control parameter after the other. It also relies on the piece-wise constant basis.
    \item GROUP \citep{Sorensen2018}\\
        GRadient Optimization Using Parametrisation is based on GRAPE combined with the chain rule and optimises concurrently. Its chopped basis is flexible (also see ``CRAB'' in section \ref{sec:CRAB}) but relies on an initial piece-wise constant basis.
    \item GOAT \citep{Machnes_2018}\\
        Gradient Optimisation of Analytic conTrols is based on a system of equations of motion obtained by differentiating the full propagator with respect to the control parameters. The parameters are optimised concurrently and its chopped basis is flexible (also see ``CRAB'' in section \ref{sec:CRAB}).
\end{itemize}

\subsubsection{Gradient-free Optimisation}
\label{sec:GradientFreeOptimisation}

In an experiment, the gradients described above cannot be calculated analytically. Certain finite-difference methods help to find them regardless.~\citep{Feng2018,Ferrie2015,Sun2014} However, if the control landscape is not smooth this method might prove inefficient or very costly in terms of measurements. This is where gradient-free optimisation algorithms shine. Even for certain open-loop optimisations they can offer an alternative, when their gradient-based counterparts fail: If e.g.\ the dynamics of a system are significantly more complicated than described in section \ref{sec:GradientbasedOptimisation}, the gradient of the cost function might be hard or impossible to find analytically. One example for such a case is the CRAB algorithm, described below, which was initially introduced to optimise many-body problems using tensor networks to simulate the time dynamics.\citep{Doria2011}\\
The first step then is to choose an optimisation basis. In the following, we will focus on the CRAB algorithm (see section \ref{sec:CRAB}) and consider the basis of trigonometric functions but it should be noted that we could also use Slepians, Chebyshev polynomials or indeed piece-wise constant elements. Broadly following Caneva et al.\citep{Caneva2011} the expanded control pulses each take the form
\begin{equation}
\label{eqn:GradientFree}
    u^{n}= \sum_{\ell=1}^{N_\text{be}}[A^{n}_\ell \sin(\omega_{\ell}t)+B^{n}_\ell \cos(\omega_\ell t)].
\end{equation}
Each pulse is composed of a sum of $N_\text{be}$ basis elements. Each basis element is defined by a frequency $\omega_\ell$ and the control parameters $[A_\ell^{n},B_\ell^{n}]$. The index $n$ stands for the iteration number.\\
If the number of available basis elements is restricted (i.e. only a certain region of frequency space is accessible), this is called a chopped basis (CB). Especially in the case of bandwidth limitations, only optimising in the accessible restricted control space can be a powerful tool to avoid introducing distorting penalty terms (see section \ref{sec:ControlObjectives}). Decreasing the number of parameters, also shrinks the size of the search space, potentially making the optimisation a lot more efficient.

\subsubsection{CRAB}
\label{sec:CRAB}

\begin{figure}[h!]
    \centering
    \includegraphics[width=\textwidth]{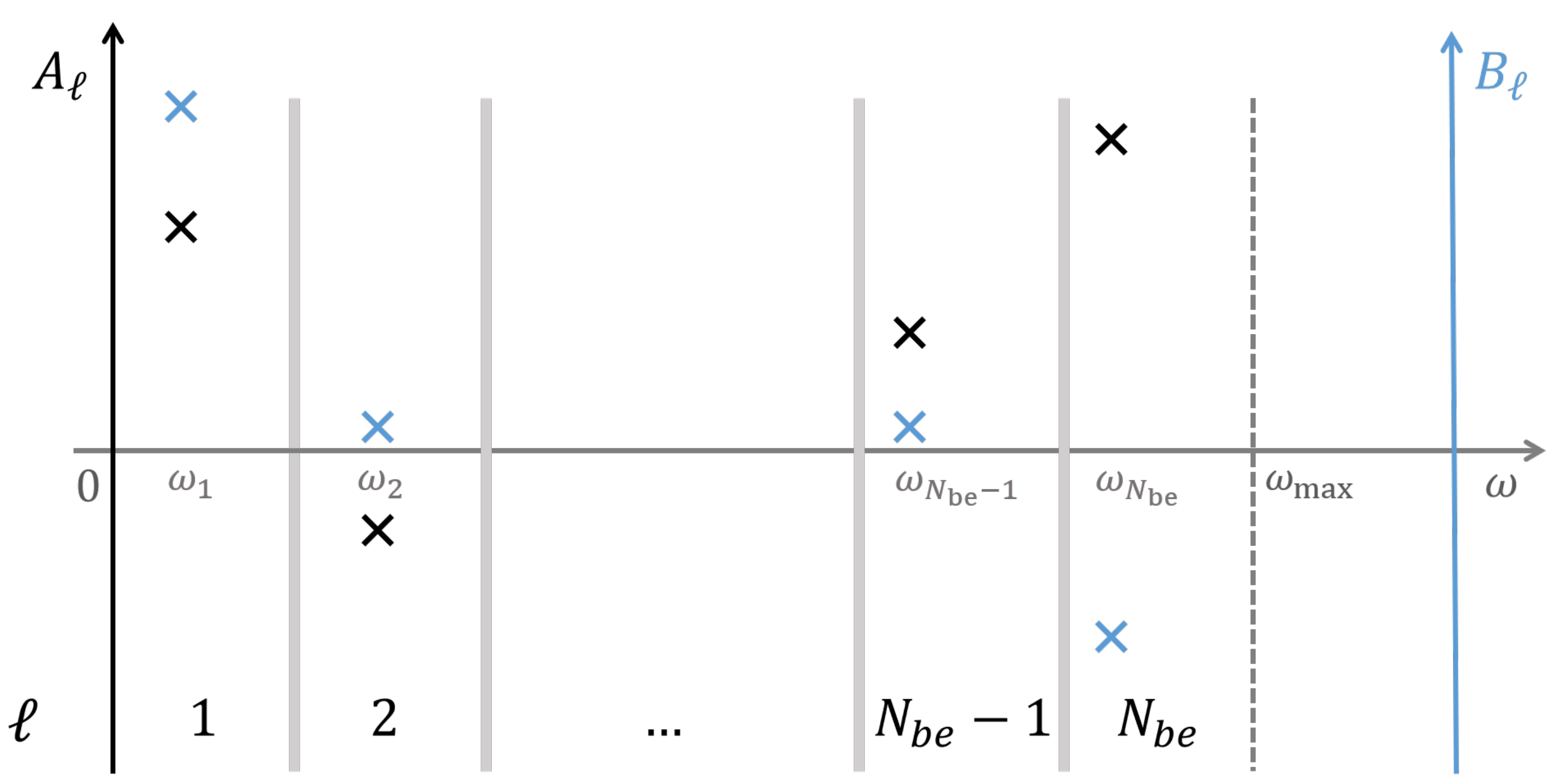}
    \caption{Illustration of the chopped random basis. A frequency space is segmented into $N_\text{be}$ parts. In each part $\ell$, a frequency $\omega_\ell$ is randomly selected according to Eq.~\eqref{eqn:CRAB_w}. Two corresponding parameters, $A_\ell$ and $B_\ell$, are optimised. They are defined as in Eq.~\eqref{eqn:GradientFree} and represented in this plot by black and blue crosses.}
    \label{Fig. 17}
\end{figure}

The Chopped RAndom Basis algorithm (CRAB) \citep{Caneva2011, Doria2011} is defined by the optimisation of a random choice of basis elements taken from a truncated function space. Intuitively, one might instead chose the basis elements to coincide with the principal harmonics of the pulse. However, Caneva et al.~\citep{Caneva2011} showed that randomness can be surprisingly effective, especially if the energy scales of the system are not fully known. Indeed, a larger function space is covered, if multiple optimizations are done with different randomized bases. The elements that make up the basis of our example in Eq.~\eqref{eqn:GradientFree} are defined by the frequencies $\omega_\ell$ and are illustrated in Fig.~\ref{Fig. 17}. These frequencies are chosen according to
\begin{equation}
    \omega_\ell = \frac{\omega_\text{max}}{N_\text{be}} \left(\ell+r_\ell-\frac{1}{2}\right),
\label{eqn:CRAB_w}
\end{equation} 
where $\omega_\text{max}$ is the maximum admissible frequency and $\ell=\{1,2,...,N_\text{be}\}$ is an index which selects the chunk of the frequency space that $\omega_\ell$ is chosen from. Let us further choose the random numbers $r_\ell$ from an interval $[-0.5,0.5]$. Then the bandwidth of the control pulses is automatically limited to $[0, \omega_\text{max}]$, where a typical choice is $\omega_\text{max} = 2\pi N_\text{be}/T$ ($T$ refers to the length of the pulse). By changing the $\omega_\text{max}$, we can change the bandwidth. Moreover, we can see that the available frequency space has been split into $N_\text{be}$ regions permitting the optimisation to make use of the entire space. It also conditions the optimisation problem to have clearly distinct control parameters. It should be noted that the number of basis elements should be dependent on the number of degrees of freedom inherent to the system.~\citep{Rach2015, Lloyd2014, Ohtsuki2004, Moore2012}\\
During the optimisation the $2N_\text{be}$-dimensional landscape will be followed using any updating algorithm (it could even be gradient-based as in GOAT and GROUP). The most common choice is the gradient-free Nelder-Mead algorithm\citep{Nelder1965} (hence the description of this algorithm under gradient-free algorithms) but others such as CMA-ES \citep{Beyer2002}, genetic algorithms or reinforcement learning are possible.

\begin{figure}[h!]
    \centering
    \includegraphics[width=\textwidth]{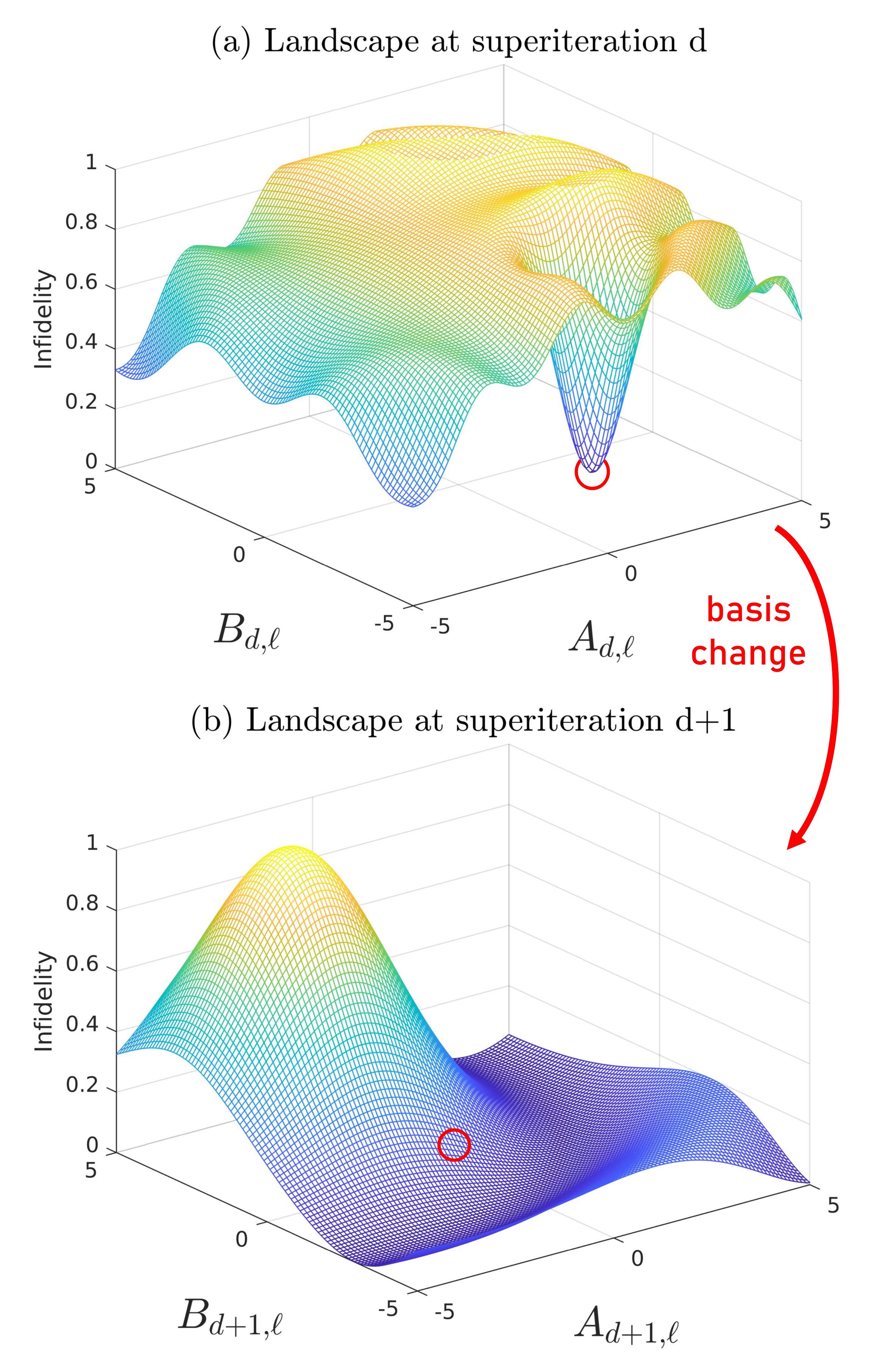}
    \caption{By changing the basis of the optimisation from (a) to (b) the landscape is transformed. The prior minimum (red circle), is relocated, making it possible to escape local minima and reduce the convergence time.}
    \label{Fig. 18}
\end{figure}

\subsubsection{dCRAB}
\label{sec:dCRAB}

In the basic version of CRAB, the basis elements are fixed and the local control landscape is explored for all $N_\text{be}$ frequencies simultaneously. This leads to a restriction in the number of frequencies that can efficiently be optimised. Using the dressed Chopped RAndom Basis algorithm (dCRAB), much fewer basis elements with $\omega_{d,\ell}$ need to be optimised at a time ($N_\text{be}($dCRAB$)<N_\text{be}($CRAB$)$). Instead, when one CRAB routine converges, we move on to $\omega_{d+1,\ell}$. This enables the method to include an arbitrarily large number of bases and to derive the solutions without -- whenever no other constraints are present -- being trapped by local optima.
The extra iterations changing up the basis after each CRAB-run are called superiterations and the index $d$ refers to the $d^\text{th}$ superiteration. Their effect is illustrated in Fig.~\ref{Fig. 18}. If their number is fixed to $N_\text{SI}$, the full description of the pulse can be summed up at the end of the optimisation $u^\text{opt}$, with all optimised parameters $A^\text{opt}_{d,\ell},B^\text{opt}_{d,\ell}$ as
\begin{equation}
    u^\text{opt}= \sum_{d=1}^{N_\text{SI}} \sum_{\ell=1}^{N_\text{be}}[A^\text{opt}_{d,\ell} \sin(\omega_{d,\ell}t)+B^\text{opt}_{d,\ell} \cos(\omega_{d,\ell} t)].
\end{equation}
In each superiteration only the parameters with corresponding index $d$ are optimized.
By repeatedly changing the basis, dCRAB does not get caught in local minima for most control problems and thus allows to retain this advantageous property of unconstrained control algorithms in a parametrized (e.g. bandwidth limited) setting. Rach et al.~\citep{Rach2015} explored the improvement from CRAB to dCRAB in detail considering the random Ising model. They found that convergence may be achieved by taking enough parameters to fix the degrees of freedom present in the optimisation problem. The underlying algorithm used for both CRAB and dCRAB performed best for a basis with $10-20$ parameters. This allowed dCRAB to outperform CRAB as it requires less optimisation parameters per optimisation (i.e. superiteration).\\
All in all, dCRAB, promises faster convergence with respect to CRAB as fewer parameters are optimised in parallel and instead, new basis elements are chosen sequentially. An example for its experimental application to NV centers, among others, can be found in the work of Frank et al.~\citep{Frank2017} where a Hadamard gate was optimised.\\

\subsubsection{Optimal Control Packages}
\label{sec:OCPackages}

\begin{table*}
\begin{minipage}{\textwidth}
\caption{\label{Tab. 2}Quantum Optimal Control Packages. In this table four widely-used Optimal Control software packages are presented which implement some of the previously described algorithms. Note that the list is not exhaustive.}
\resizebox{\textwidth}{!}{%
    \begin{ruledtabular}
        \begin{tabular}{lcccc}
        Name\blfootnote{Environment:}  & \makecell{QOC\\ Algorithm} &  \makecell{Gradient\\ required} & Access & Specialty\\
        \hline
        &&&\\
        RedCRAB\footnote{MATLAB\label{matlab}}\footnote{python\label{python}}\footnote{command line} \citep{Heck2018,Zoller2018}          &  dCRAB & no & on request & \makecell{allows connection \\directly to experiment}\\
        &&&\\
        DYNAMO\textsuperscript{\ref{matlab}} \citep{Machnes_2011}           & \makecell{GRAPE,\\Krotov}   & yes & \href{https://github.com/shaimach/Dynamo}{github}  & \makecell{many pre-programmed \\optimisation options}\\
        &&&\\
        QuTiP\textsuperscript{\ref{python}} \citep{Johansson2013_qutip,Johansson2012_qutip}            & \makecell{GRAPE,\\ CRAB}  & \makecell{yes,\\ no} & pip, conda, etc. & \makecell{all-round quantum\\ simulation}\\
        &&&\\
        Krotov Package\textsuperscript{\ref{python}} \citep{Goerz2019_krotov}    & Krotov & yes & pip, conda, etc. &\makecell{Connects to QuTip, \\many pre-programmed \\optimisation options} \\
        \end{tabular}
    \end{ruledtabular}}
    \end{minipage}
\end{table*}
In the past years, a number of QOC algorithms were implemented in ready-to-use software packages. In this section, we present four of these packages that we deem to be closest to applications with NV centers. An overview over some of their distinguishing features is given in Table \ref{Tab. 2}. Nevertheless, more solutions exist.\\
RedCRAB~\citep{Heck2018,Zoller2018} is a python based programme, aiming to remotely optimise any experiment or simulation with gradient-free methods. It can be linked to the experiment setup via MATLAB, python, terminal or simple file transfer and is hence very versatile. RedCRAB makes use of the dCRAB alogithm and provides pulse updates. As it does not require any knowledge about the quantum system itself, it is compatible even with more complicated many-body systems and tensor-network simulations. RedCRAB is available from the authors on request.\\
DYNAMO~\citep{Machnes_2011} was originally developed as a GRAPE (and Krotov) implementation in MATLAB. It allows the user to choose their own Hamiltonian and dissipator terms as well as one of the available figures of merit. Hence, it combines simulation and optimisation for certain problems dealing with small quantum systems. It allows for the optimisation of robust pulses and includes a large number of examples. The full version is available on github.\\
QuTiP\citep{Johansson2013_qutip,Johansson2012_qutip} is an open source python library for simulating quantum systems. One of its features is a quantum optimal control implementation. As such it offers limited optimisation techniques with GRAPE and CRAB. Conveniently, the optimisation settings are defined with the usual QuTiP structure. The library is available for example via pip or conda.\footnote{Popular python package managers}\\
The Krotov package\citep{Goerz2019_krotov} is an open source python library built on top of QuTiP. As such it offers optimisation via Krotov's method. It includes an extended range of settings in comparison to QuTiP's own QOC implementation. The library is available for example via pip and conda.\\
Other QOC packages include Spinach \citep{Hogben2011} and SIMPSON, \citep{Tosner2009} which focus on NMR applications, as well as QEngine \citep{Sherson2018_GROUP,Sorensen2019_qengine} which includes a GROUP implementation designed especially for ultra-cold atom physics. GRAPE was also recently implemented in the GRAPE-Tensorflow python package, \citep{Leung2017_grape_tensorflow} using methods known from machine learning to calculate the gradients.

\subsection{Limits of Control: Controllability and the QSL}
\label{sec:LimitsOfControl}

Whether or not a QOC problem is (approximately) solvable, is not always simple to answer. However, by examining a number of characteristics of the Hamiltonian, some general predictions can be made.\\
First of all, one may ask whether the control objective is in principle reachable. This can be addressed by examining the controllability of the system.\cite{DomenicoDAlessandro2007} The drift and control Hamiltonians define a certain state (and also gate) space that is reachable. A system is called controllable when all states (gates) in the Hilbert space are accessible in finite time. It has been shown that, if the rank of the dynamical Lie algebra generated by the different terms of the Hamiltonian corresponds to the rank of the control space (and fulfills certain symmetry criteria), the system is fully controllable. Alternatively, the question of state-controllability can be examined via a geometric approach based on graph theory, which can be more convenient to check, especially for larger systems. \cite{Turinici2003}
For more information on controllability, please refer to the following books.\citep{DomenicoDAlessandro2007,VelimirJurdjevic1996} For open quantum systems, the deleterious effect of the environment usually can not be completely canceled and only a subset of the whole set of states (gates) can be reached.\cite{Koch2016} \\
If the controllability criteria are fulfilled, the question remains whether the controls are complex and energetic enough to navigate the Hilbert space to the specified target.
In general, the quantum speed limit (QSL), i.e.\ the smallest possible control time needed for a system to reach its target, is influenced by two factors. First, the dynamical equation determines how fast the system may change. This is usually quantified by the so-called Schatten p-norm of the dynamical operator (see reference \citep{Deffner2017} for details). Second, the exact distance between the initial system to the objective needs to be taken into account.
\begin{tcolorbox}[breakable, enhanced,colback=palecornflowerblue,boxrule=0pt,title=Quantum Speed Limit]
The minimum time it takes to evolve a system into a target state is mostly dependent on two things: The Hamiltonian $\hat{H}$ and the distance between the initial and the target state $\braket{\phi_0}{\phi_t}$. For a time-independent Hamiltonian, we obtain the Bhattacharyya-bound:\citep{Bhattacharyya1983,Caneva2009}
    \begin{equation}
        T_\text{QSL}\ge \Delta E^{-1} \arccos|\braket{\phi_0}{\phi_t}|
    \end{equation}
    This time $T_\text{QSL}$ is called the quantum speed limit (QSL). It can be interpreted as follows:
    If the Hamiltonian has a high energy variance calculated on the initial state $\Delta E = \sqrt{\bra{\phi_0}\hat{H}^2\ket{\phi_0}-\bra{\phi_0}\hat{H}\ket{\phi_0}^2}$, then any other state is reached more quickly. It might be more intuitive to consider the case that $\phi_0$ is an eigenstate of $\hat{H}$, hence it will never change and as $\Delta E=0$, the speed limit will go towards infinity. The distance to the target state finally determines the exact time scale.
\end{tcolorbox}
For a more general and complete picture of the QSL the reader is advised to refer to the following references.\citep{Caneva2009,Deffner2017}\\
Similarly to the QSL, the information speed limit (ISL) can also restrict the minimum length of the control pulse. Behind this is the idea that the information encoded in the control pulse has to be sufficient to steer the system to the target. For example, in the noiseless case the degrees of freedom in the control (the number of independent frequencies in a bandwidth-limited control field or the number of kicks in a bang-bang control~\citep{Viola1998}) should at least reflect the dimension of the system.\citep{Lloyd2014} Note that in the presence of noise in the system or in the controls, more degrees of freedom are required to transmit the same amount of information.

\section{QOC for NV Centers}\label{sec:QOC_forNVs}

We now take a look at various applications of QOC to NV center-based systems including quantum sensing (section \ref{sec:Optimalcontrolforquantumsensing}), quantum information and quantum computation (section \ref{sec:QOC_QI}).

\subsection{\label{sec:Optimalcontrolforquantumsensing}Quantum Sensing}

The sensitivity of NV center-based sensors depends on the coherence times ($T_2$) of the spin states, which in turn are theoretically only strictly limited by their lifetimes. Other limiting factors include efficient and coherent spin manipulation, which is inherently subjected to experimental imperfections and limitations. 
Several material synthesis techniques\citep{Barry2019} have been developed to fabricate NV centers in ultrapure host crystals with minimal noise due to paramagnetic impurities and crystal field induced phenomena (strain) that diminish the coherence time (T$_2$) of the spin qubit. Additionally, dynamical decoupling protocols discussed in section \ref{sec:Magneticfieldsensing} have shown promising results enhancing the coherence time. Yet, the currently achieved sensitivities fall short of the theoretical limits set by the spin state lifetimes.
 
\begin{figure}[h!]
    \centering
    \includegraphics[width = \textwidth]{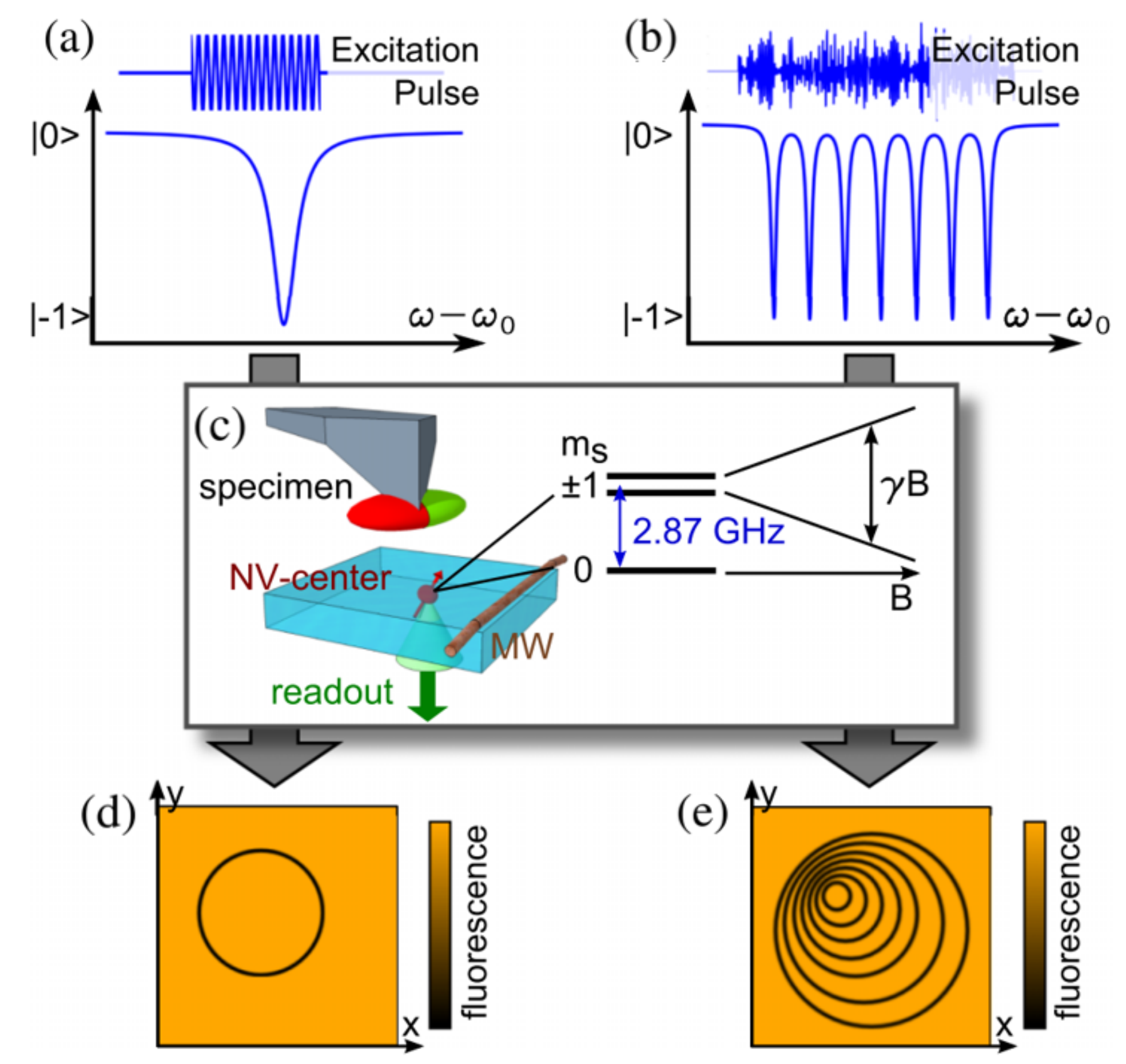}
    \caption{Detuning dependent excitation of NV spins: (a) and (b) show the excitation of NV centers dependent on their detuning via pulsed MW spectroscopy using rectangular and optimally designed pulses respectively. (c) shows the diamond scanning probe based experimental setup. (d) and (e) show the simulated fluorescence images for spectroscopy using the pulses from (a) and (b) respectively. Refer to H{\"a}berle et al.\citep{Haberle2013} for details. Adapted and modified with permission from T. Häberle, D. Schmid-Lorch, K. Karrai, F. Reinhard, and J. Wrachtrup, Phys. Rev. Lett. 111, 170801 (2013). Copyright 2013 by the American Physical Society.}
    \label{Fig. 19}
\end{figure}
 
A material independent approach, which is the focus of this review, is to design specialized spin manipulation protocols that are optimised for efficiency in consideration of noise and experimental limitations/imperfections. The critical processes in the sensing techniques discussed so far are spin state initialization, state-to-state transfer and spin state readout. All these steps require the system to evolve in a specified way given a certain set of constraints. Broadly speaking, this catches the essence of quantum optimal control (QOC) theory, as described in section \ref{sec:Theory}. QOC techniques have shown promising results in fields like NMR\citep{Peirce1988,Brif2010} and atomic physics related experiments.\citep{VanFrank2016,Omran2019}
In the past decade, a number of interesting results have been obtained by using optimal control for NV spin system.
 
   \begin{figure}[h!]
    \centering
    \includegraphics[width = \textwidth]{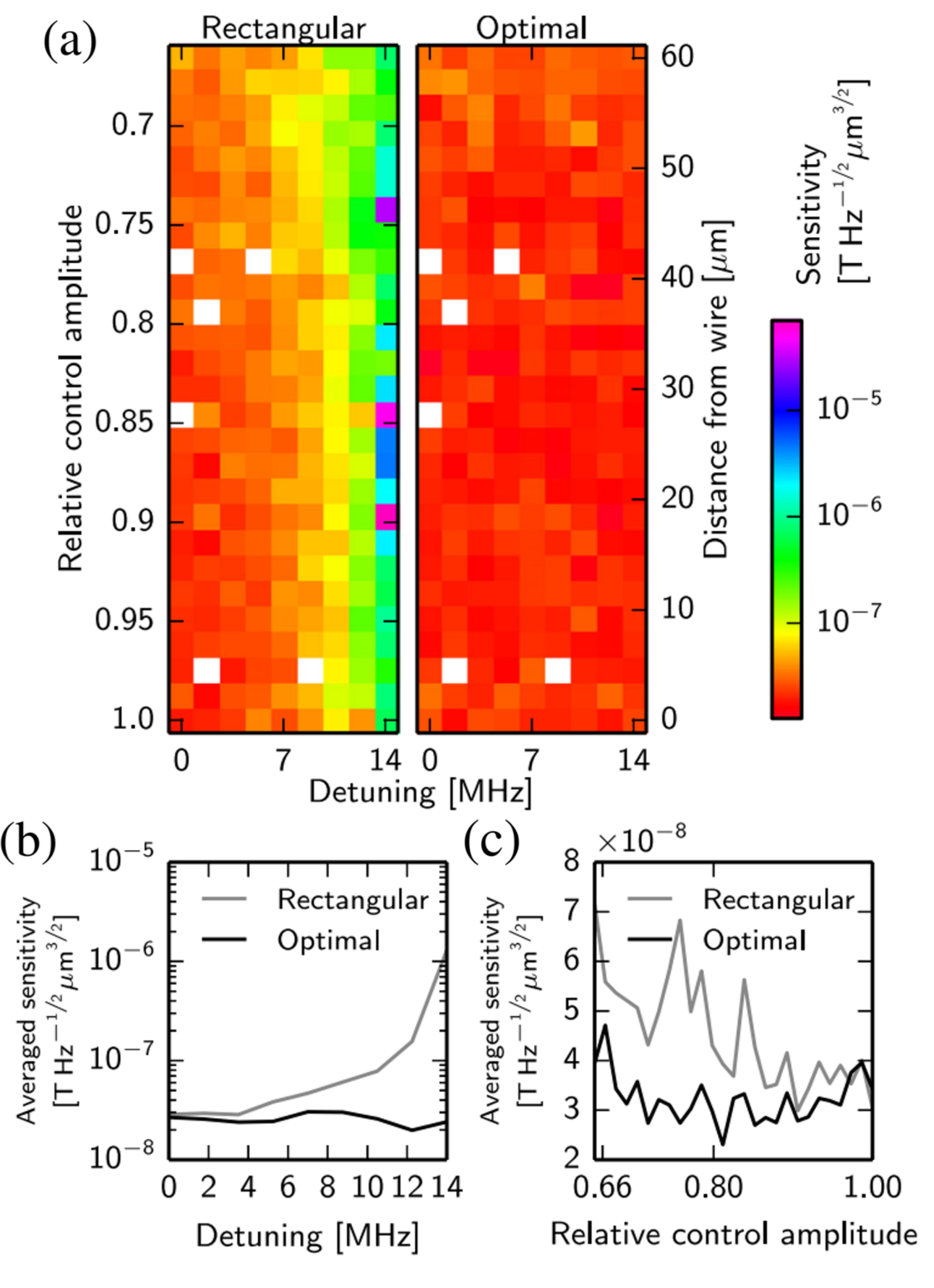}
    \caption{Smooth QOC for robust solid-state spin magnetometry: (a) shows a matrix plot of sensitivity against detuning (horizontal) and relative amplitude variation (vertical) for rectangular (left) and optimised (right) MW control pulses. In the lower panels the sensitivity of two sensing sequences (using rectangular and optimised pulses) is compared for a range of detunings in (b), and relative control amplitudes in (c). For strong detuning, the optimised pulses show almost two order of magnitude enhancement in sensitivity. Refer to N{\"o}bauer et al. \citep{Nobauer2015} for details. Adapted and modified from Tobias Nöbauer, Andreas Angerer, Björn Bartels, Michael Trupke, Stefan Rotter, Jörg Schmiedmayer, Florian Mintert, and Johannes Majer,
Phys. Rev. Lett. 115, 190801, (2015) under the terms of the Creative Commons Attribution 4.0 International License. }
    \label{Fig. 20}
\end{figure}
 
\begin{figure}[h!]
    \centering
    \includegraphics[width = 0.8\textwidth]{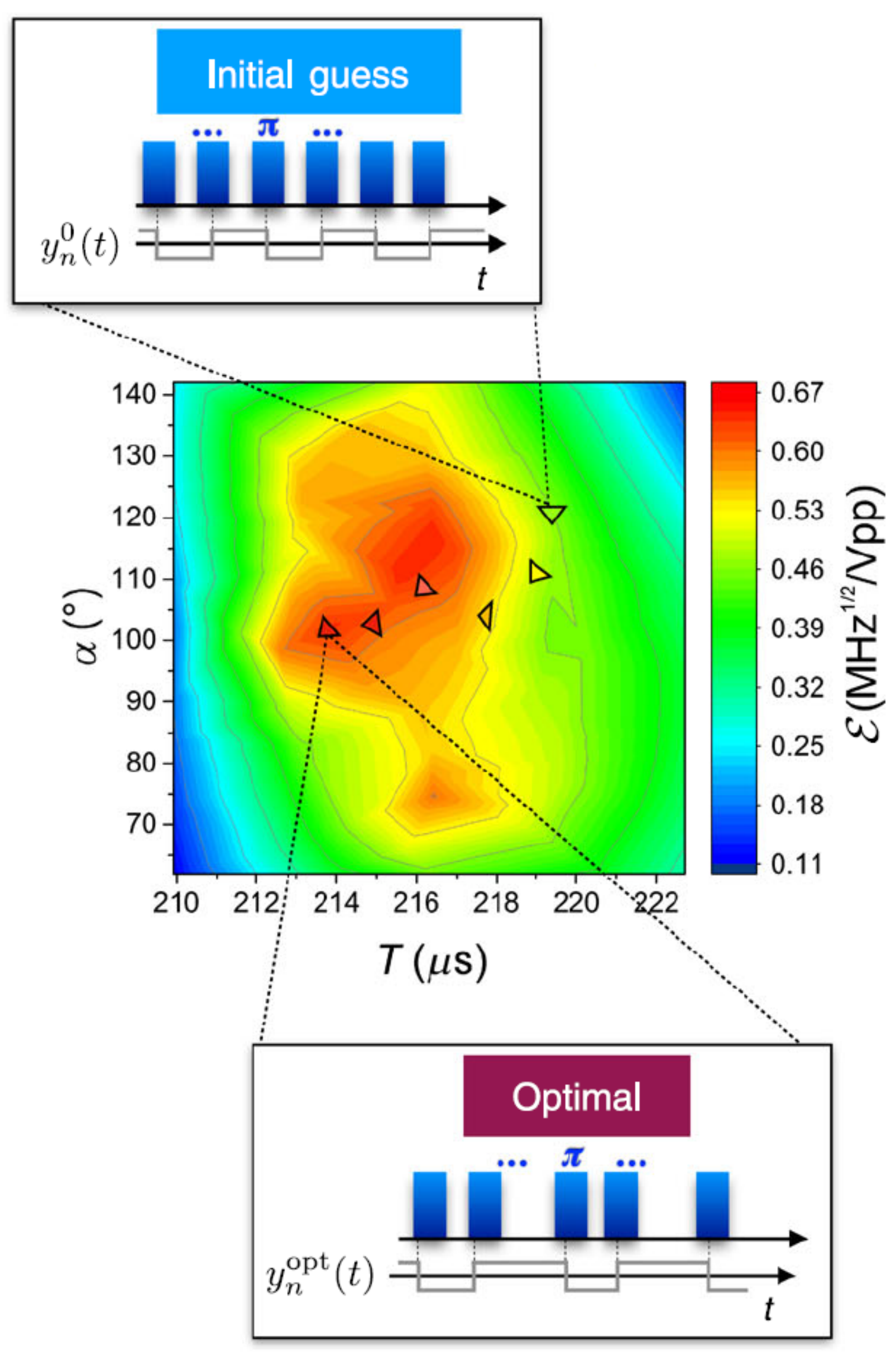}
    \caption{Optimal control for one-qubit quantum sensing. Illustration of the optimisation process in a two parameter (sensing time $T$ and phase shift $\alpha$) landscape (see Fig.~\ref{Fig .14}). The initial guess (see section \ref{sec:InitAndStop}) was chosen to be the CPMG\citep{Carr1954} pulse sequence. It should be noted that in this figure $\mathcal{E}$ refers to the sensitivity. Refer to Poggiali et al. \citep{Poggiali2018} for details. Adapted and modified from F. Poggiali, P. Cappellaro, and N. Fabbri, Phys. Rev. X 8, 021059, (2018) under the terms of the Creative Commons Attribution 4.0 International License. }
    \label{Fig. 21}
\end{figure}

H{\"a}berle et al.\citep{Haberle2013} showed that quantum limited sensitivities for magnetic field sensing can be achieved using QOC based spectroscopy techniques (Fig.~\ref{Fig. 19}). In this work, the team used a single NV qubit system to image nanoscale magnetic fields with scanning probes. The aim of the work was to obtain a wider dynamic range for the spectroscopic microwave (MW) pulses using QOC . An open-loop numerical optimisation technique (GRAPE, see section \ref{sec:GradientbasedOptimisation} for more details) was used to obtain frequency selective, high bandwidth MW pulses for state transfer. Their results showed photo-shot-noise limited sensitivities of around 4.5~$\mu\text{T}\sqrt{\text{Hz}}$ and a dynamic range of more than 2.2~mT, as well as improved robustness against fluctuations in MW power.\\
N{\"o}bauer et al. \citep{Nobauer2015} exploited a Floquet theory\citep{ASENS_1883_2_12__47_0} based approach for open-loop optimisation limiting MW pulses to a certain frequency domain (Fig.~\ref{Fig. 20}). They obtained pulses that were robust against MW amplitude variation and frequency detuning. They first demonstrated the working principle employing a single NV using different MW detunings from the resonant frequency. They further concluded that the optimised pulse is ideal for NV ensembles that require the same spin manipulation pulse to be effective for a number of systems with different resonant frequencies and hence detunings as seen in the matrix plot in Fig.~\ref{Fig. 20} shows the comparison of regular rectangular control pulses and the optimised pulses, demonstrating two orders of magnitude enhancement in sensitivity. This is demonstrated in a proof-of-principle experiment with a spin echo sequence for magnetic field sensing.\\
Poggiali et al. \cite{Poggiali2018} demonstrated a different approach (Fig.~\ref{Fig. 21}) making use of a modified cost function based on the Fisher information to enhance the sensitivity of NV centers to a wide range of AC magnetic fields (Fig.~\ref{Fig. 21}). The Fisher information of the measurement (see section \ref{sec:ControlObjectives} specifically Eq.~\eqref{eqn:FisherInfo}) takes into account the signal of interest as well as the noise. The sensitivity is related to the Fisher information via Eq.~\eqref{eqn:Sensitivity}. 
A gradient-free minimisation technique (see section \ref{sec:GradientFreeOptimisation}) was used to find the optimal temporal distance between pulses and adjust the initial phase of the signal. Finally, they experimentally demonstrated enhanced sensitivity by up to a factor of two for a single qubit system compared to a CPMG~\citep{Carr1954} pulse sequences (a type of dynamical decoupling sequence, see section~\ref{sec:Magneticfieldsensing}). Along this line, M{\"u}ller et al.~\cite{Muller2018} showed how optimal control-designed frequency filter functions allow a speed-up of the measurement and fast detection of fluctuating signals.\\
In another work, Scheuer et al.\citep{Scheuer2014} demonstrated a novel technique for spin qubit control in the ultra fast driving regime beyond the rotating wave approximation (see Appendix~\ref{app:RWA}). When the Rabi frequency of the drive becomes comparable to the transition frequency (in the reference about 30~MHz), the RWA breaks down and the system is no longer described by Eq.~\eqref{eqn:Hamiltonian}. To overcome this artificial constraint on the pulse duration, an optimal control pulse considering the counter-rotating terms was designed. They used the CRAB algorithm (see section \ref{sec:CRAB}) to optimise standard $\pi$- and $\frac{\pi}{2}$-pulses in this regime and experimentally demonstrated Ramsey and spin-echo sensing protocols.

\begin{figure}[h!]
    \centering
    \includegraphics[width = \textwidth]{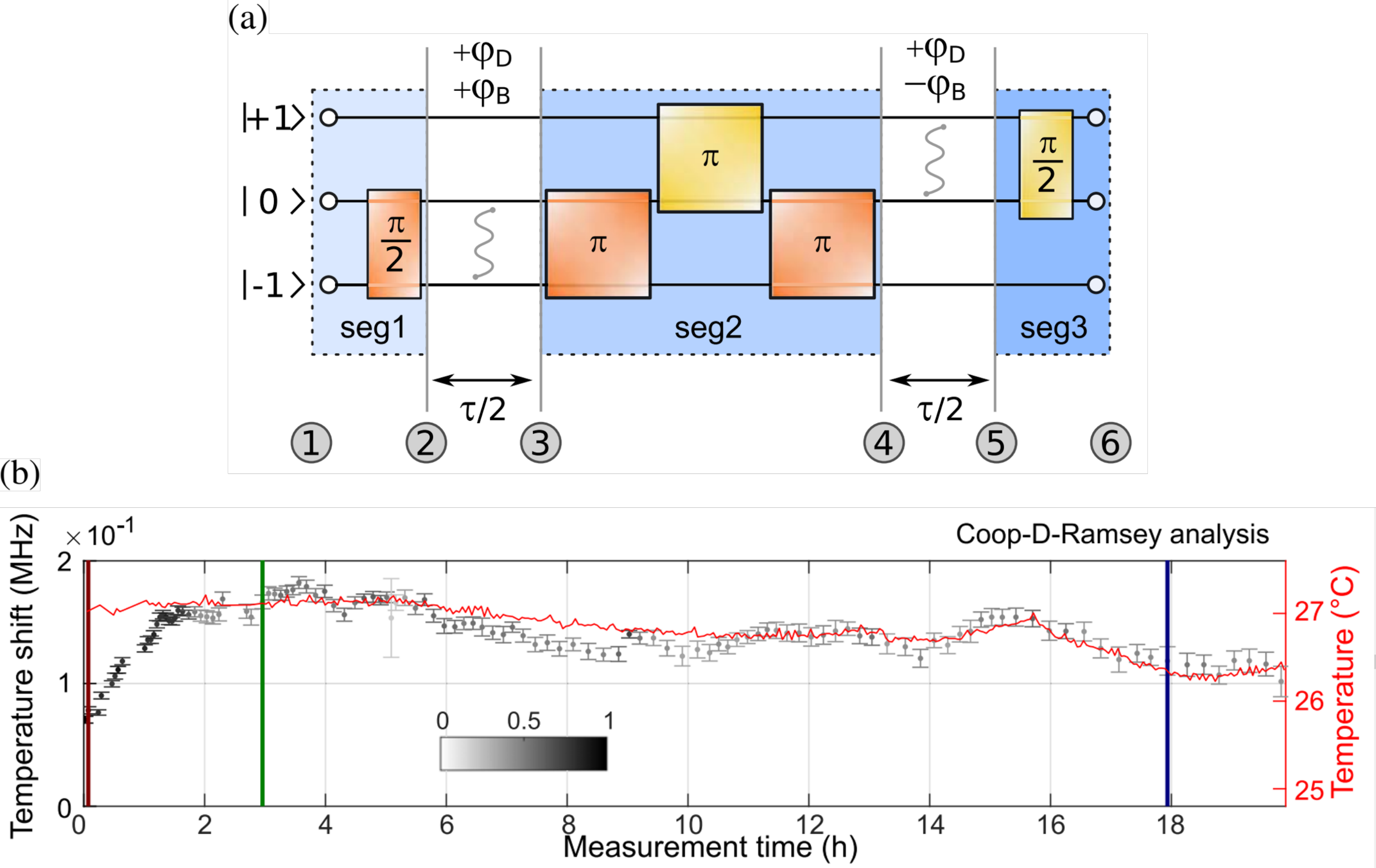}
    \caption{Robust and efficient quantum optimal control of spin probes in a complex (biological) environment: (a) Working principle for the Cooperative D-Ramsey pulse sequence for temperature sensing based on the NV center ground state three level system. The sequence is sliced into 3 segments (seg1, seg2 and seg3), each separated in time by $\tau/2$. $\varphi_D$ and $\varphi_B$ denote phases accumulated due to change in the zero-field parameter $D$ and due to the external magnetic field, respectively. Optimal control pulses are used for spin projection to include cooperative design in the protocol to cancel out the undesired effects in phase accumulation. (b) The plot shows the measurements of temperature fluctuations over time using this protocol. Refer to Konzelmann et al.\citep{Konzelmann2018} for more details. Adapted and modified from New J. Phys. 20 123013, under the terms of the Creative Commons Attribution 3.0 International License}
    \label{Fig. 22}
\end{figure}

Ziem et al.\citep{Ziem2019} optimised resonance imaging of $^{19}$F nuclei using GRAPE. The nuclei were part of patterned calcium fluoride on the diamond surface. The optimised pulses significantly improved the robustness of their DD protocol against variations in the driving field in comparison to a standard XY16-N sequence.\\
Recently, Konzelmann et al. \citep{Konzelmann2018} showed that QOC pulses can enhance the robustness of temperature sensing for biological applications. They used nanodiamonds in an agarose matrix to demonstrate enhanced signal quality for fast temperature fluctuation measurements in dynamic biological media. For this purpose, a cooperative D-Ramsey pulse sequence, specifically designed for temperature measurements, \cite{Braun2010} was optimised using the MATLAB based DYNAMO package that uses a GRAPE optimisation algorithm (see section \ref{sec:OCPackages} for more details). Here, the typical state-to-state transfer fidelity serves as the cost function for the optimisation (Fig.~\ref{Fig. 22}).

\subsection{Quantum Information and Computation}
\label{sec:QOC_QI}

\begin{figure}[h!]
    \centering
    \includegraphics[width = 0.8\textwidth]{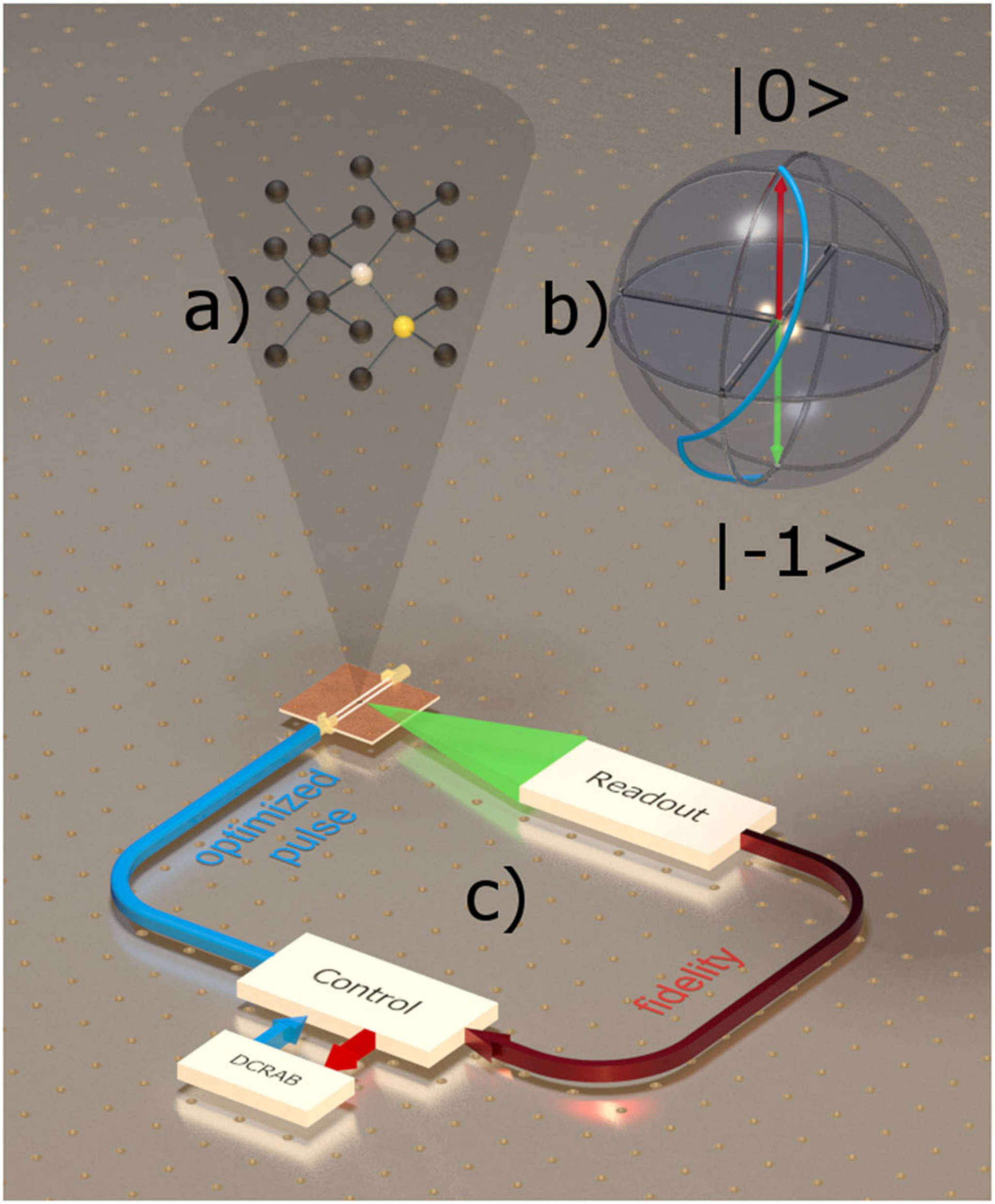}
    \caption{Closed-loop optimisation scheme: (a) Structure of the NV center. (b) Optimised spin inversion pulse represented on a Bloch sphere. (c) The controls (blue) were iteratively updated using the dCRAB algorithm. After each control attempt the fidelity (red) was read out and fed back into the algorithm. See section \ref{sec:dCRAB} for more information on dCRAB and refer to Frank et al.\citep{Frank2017} for more details on the experiment. Adapted and modified from F.  Frank,  T.  Unden,  J.  Zoller,  R.  S.  Said,  T.  Calarco,  S.  Montangero,B. Naydenov,  and F. Jelezko, npj Quantum Information 3, 48 (2017) under the terms of the Creative Commons Attribution 4.0 International License.}
    \label{Fig. 23}
\end{figure}

The usefulness of QOC techniques has been demonstrated repeatedly in the field of quantum information and computation. As the basic theoretical protocols and experimental setup need to fulfill similar fundamental demands as for the described sensing techniques, similar optimisation strategies come into play.\\
Dolde et al.\citep{Dolde2014} realised entanglement of two nuclear spins over a distance of 25~nm by entangling them with one NV center each which in turn were coupled through dipole-dipole interaction. They used GRAPE (see section \ref{sec:GradientbasedOptimisation}) to realise optimised PSWAP and NOT gates which worked despite the hyperfine interactions which interfere with standard controls. The optimised pulses allowed for 20 NOT gate repetitions without significant loss of fidelity, while standard pulses already showed poor performance after a single gate.\\
Waldherr et al. \citep{Waldherr2014} demonstrated three-qubit phase-flip error correction using three nuclear spins and the NV center as an ancilla. While the system was entirely manipulated via the NV center spin, the hyperfine couplings complicated its control. This was solved by applying microwave pulses of two different frequencies simultaneously, finding their shape with a GRAPE-implementation.\\
In order to construct optimal $\pi/2$- as well as spin-inversion pulses, Frank et al.~\citep{Frank2017} designed and experimentally implemented closed-loop optimisation for single NV centers (Fig.~\ref{Fig. 23}). Their dCRAB implementation autonomously found an optimal solution for the desired goals within the experimental error. Their techniques are translatable to a number of quantum computing applications as $\pi/2$-pulses form the building blocks of most common quantum gates. In case of moderate detuning, the closed-loop pulses outperformed any open-loop-generated sequences. More details on the algorithm can be found in section \ref{sec:dCRAB}.\\
Said et al.\citep{Said2009} compared different strategies to entangle an NV center with a $^{13}$C nuclear spin, namely sequential pulses, composite pulses, and numerically-optimised pulses (GRAPE). They concluded that the optimised pulse, did not only outperform the others in robustness but was also faster than the composite pulse.\\
Tsurumoto et al.\cite{Tsurumoto2019} utilised a GRAPE algorithm to optimise a state-to-state transfer pulse applied to a NV center, which was utilised to transfer photon polarization to a nuclear spin qubit.\\
Recently, Chen et al.\citep{Chen2020} proposed a QOC technique based on a combination of open-loop and closed-loop optimisation to demonstrate the working principle of a NV-based quantum processor.\\

\section{Conclusion}

NV centers are a valuable platform for a whole family of quantum applications. As the boundaries of existing implementations of different quantum sensing and quantum computation schemes are being explored, QOC offers a route to transverse these limits. We have reviewed a number of existing applications of QOC to NV center-based systems and provided a recipe for the application of QOC with a special focus on NV centers. It has been shown that QOC methods can increase the precision in qubit control and manipulation, especially in the presence of environmental factors such as surrounding nuclei. These advantages have also been exploited to improve NV-based sensors. Their versatility in combination with QOC shows to be especially effective when the limiting factor is defined by the experimental constraints and/or limited knowledge of the system. Enhanced control also provides a mean to efficiently implement quantum algorithms in NV center-based quantum registers. All combined, QOC enhances the capabilities which are crucial to building a NV center-based quantum computer. QOC methods provide the boost that may lead diamond-based quantum systems into the realm of commercial quantum technologies.

\begin{acknowledgments}
This work has received funding from the European Union's Horizon 2020 research and innovation programme under the Marie Sk\l{}odowska-Curie grant agreement N$^\circ$ ``765267'' (QuSCo), N$^\circ$ ``820394'' (AsteriQs), and N$^\circ$ ``817482'' (PASQuanS). This work was partially supported by the Italian PRIN 2017, the QuantERA project QTFLAG and QuantHEP, the DFG via the TWITTER project.
\end{acknowledgments}
\section*{Data Availability}
The data that support the findings of this study are available from the corresponding author upon reasonable request.
\appendix
\section{Spin Hamiltonian}
\label{app:SpinHam}
In a solid state system, spin interactions are mostly mediated through magnetic fields. In general the spin interaction energies are very small in comparison to electronic interaction energies (Fig.~\ref{Fig .3}) for systems like diamond crystal. The energy of the system $\mathbb{E}$ can be expanded as 

\begin{equation}
\mathbb{E} = \mathbb{E}_0 + \mathlarger{\sum}_{n} \pdv{\mathbb{E}}{\hat{s}_n}\hat{s}_n + \frac{1}{2}\mathlarger{\sum}_{n,m} \pdv{\mathbb{E}}{\hat{s}_n}{\hat{s}_m}\hat{s}_n \hat{s}_m + ...
\label{eqn:taylor_expansion}
\end{equation}

$\mathbb{E}_0$ represents all the non spin-interaction energies and $\hat{s}_n$ are the relevant spin bases (i.e. of different atoms, the NV center itself etc.). The Hamiltonian for such spin interaction energies is called the \textbf{spin Hamiltonian}. For a NV center like system, the spin Hamiltonian can be obtained in terms of spin operators for the corresponding singlet spin state basis:

\begin{equation}
    \hat{S}_X = \frac{1}{\sqrt{2}}\begin{pmatrix}
    0 & 1 & 0\\
    1 & 0 & 1\\
    0 & 1 & 0
    \end{pmatrix}, \: \hat{S}_Y = \frac{1}{\sqrt{2}}\begin{pmatrix}
    0 & -i & 0\\
    i & 0 & -i\\
    0 & i & 0
    \end{pmatrix}, \: \hat{S}_Z = \frac{1}{\sqrt{2}}\begin{pmatrix}
    1 & 0 & 0\\
    0 & 0 & 0\\
    0 & 0 & -1
    \end{pmatrix}.  
\end{equation}{} 
Terms of different order in the expansion \ref{eqn:taylor_expansion} represent different kinds of spin interactions. For our purposes, it suffices to consider the terms up to second-order. Consequently, all the different types of interactions can be formalised in the same way through interaction tensors and spin operators. We now look into the details of these different types, for simplicity $\hbar$ is assumed to be equal to 1 in the following equations.\\

\noindent\textbf{Linear terms:} These terms mainly represent the interaction of the spin with external fields, whose origin may also lie within the crystal.

\begin{equation}
    \hat{H}_{li} = \hat{\Vec{S}}^\intercal \cdot \mathcal{Z} \cdot\Vec{B} = (\hat{S}_X \; \hat{S}_Y \; \hat{S}_Z)
    \begin{bmatrix}
    \mathcal{Z}_{XX} & \mathcal{Z}_{XY} & \mathcal{Z}_{XZ}\\
    \mathcal{Z}_{YX} & \mathcal{Z}_{YY} & \mathcal{Z}_{YZ}\\
    \mathcal{Z}_{ZX} & \mathcal{Z}_{ZY} & \mathcal{Z}_{ZZ}
    \end{bmatrix}
     \begin{pmatrix}
    B_{X}\\
    B_{Y}\\
    B_{Z}
    \end{pmatrix},
\end{equation}
where, $\hat{\Vec{S}}$ is the spin operator, $\Vec{B}$ is the magnetic field, and $\mathcal{Z}$ is the Zeeman interaction tensor. In the Hamiltonian discussed in section \ref{sec:SpinHamiltonian} the magnetic interaction term is of this nature. The coupling tensor in that case is a diagonal matrix multiplied with the gyromagnetic ratio of the electron.\\

\noindent\textbf{Bi-linear terms:} These terms represent spin-spin interactions such as dipolar coupling, exchange interactions, and hyperfine couplings. 

\begin{equation}
    \hat{H}_{bl} = \hat{\Vec{S}}^\intercal \cdot \mathcal{N} \cdot \hat{\Vec{I}} = (\hat{S}_X \; \hat{S}_Y \; \hat{S}_Z)
    \begin{bmatrix}
    \mathcal{N}_{XX} & \mathcal{N}_{XY} & \mathcal{N}_{XZ}\\
    \mathcal{N}_{YX} & \mathcal{N}_{YY} & \mathcal{N}_{YZ}\\
    \mathcal{N}_{ZX} & \mathcal{N}_{ZY} & \mathcal{N}_{ZZ}
    \end{bmatrix}
     \begin{pmatrix}
    \hat{I}_X\\
    \hat{I}_Y\\
    \hat{I}_Z
    \end{pmatrix},
\end{equation}
where, $\hat{\Vec{I}}$ is the nuclear spin operator vector, and $\mathcal{N}$ is the hyperfine coupling tensor. In the Hamiltonian presented in Eq.~\eqref{eqn:fullHamiltonian} in section \ref{sec:SpinHamiltonian} the interaction of the NV spin with other spins is bi-linear in nature. In this case, the hyperfine coupling tensor may be simplified and described by the two main contributing interactions; the axial coupling constant $\mathcal{N}_\text{axial}$ and the transverse coupling constant $\mathcal{N}_\text{tran}$:
\begin{equation}
    \mathcal{N} =     
    \begin{bmatrix}
    \mathcal{N}_\text{tran} & 0 & 0\\
    0 & \mathcal{N}_\text{tran} & 0\\
    0 & 0 & \mathcal{N}_\text{axial}
    \end{bmatrix}.
\end{equation}{}
Note that the Fermi contact interaction term is given by\citep{Felton2009}
\begin{equation}
    \mathit{f}_\mathcal{N} =  \frac{\mathcal{N}_\text{axial} + 2\mathcal{N}_\text{tran}}{3},
\end{equation}{}
and the dipole interaction term is given by
\begin{equation}
    \mathit{d}_\mathcal{N} = \frac{\mathcal{N}_\text{axial} - \mathcal{N}_\text{tran}}{3},
\end{equation}{}
$\mathit{f}_\mathcal{N}$ is an order of magnitude larger than $\mathit{d}_\mathcal{N}$ for the $^{14}$N and $^{15}$N nuclei.\citep{Felton2009}\\
We may specifically consider the interaction between two different NV center spins. The Hamiltonian for these dipole-dipole interactions can be written as
\begin{equation}
    \hat{H}_{NV-NV} = \frac{\mu_0}{4\pi}\frac{\gamma_{nv}}{\Vec{r}\;^3}\left(\hat{\Vec{S}}_1\cdot\hat{\Vec{S}}_2 - 3(\hat{\Vec{S}}_1\cdot\Vec{r})(\hat{\Vec{S}}_2\cdot\Vec{r})\right),
\end{equation}
where $\hat{S}_1$ and $\hat{S}_2$ are the spin operators for the respective NV centers and $\Vec{r}$ is the spatial vector which joins them. This interaction is relatively weak in comparison to other spin interactions. Nevertheless, NV-NV interactions have been a matter of interest in several works in literature (for example see Dolde et al.\citep{Dolde2014}).\\

\noindent\textbf{Quadratic terms:} These terms usually represented the interaction of spin with itself, although the source of the interaction can indirectly be due to an external field. Nuclear quadrupole interaction in NMR and electron zero-field splitting terms are of this nature. They are represented as:

\begin{equation}
    H = \hat{\Vec{S}}^\intercal \cdot \mathcal{D} \cdot \hat{\Vec{S}} = (\hat{S}_X \; \hat{S}_Y \; \hat{S}_Z)
    \begin{bmatrix}
    \mathcal{D}_{XX} & \mathcal{D}_{XY} & \mathcal{D}_{XZ}\\
    \mathcal{D}_{YX} & \mathcal{D}_{YY} & \mathcal{D}_{YZ}\\
    \mathcal{D}_{ZX} & \mathcal{D}_{ZY} & \mathcal{D}_{ZZ}
    \end{bmatrix}
     \begin{pmatrix}
    \hat{S}_X\\
    \hat{S}_Y\\
    \hat{S}_Z
    \end{pmatrix},
\end{equation}
where, $\mathcal{D}$ is the quadrupole coupling tensor. $\mathcal{D}$ is in good approximation a symmetric matrix. The zero-field splitting terms for the Hamiltonian in Eq.~\eqref{eqn:fullHamiltonian} in section \ref{sec:SpinHamiltonian} are of this nature. For a NV center this tensor can, to a good approximation,\citep{Doherty2011} be written in the basis of spin matrices as:
\begin{equation}
    \mathcal{D} = \begin{bmatrix}+
    \frac{D}{3} & 0 & E\\
    0 & -\frac{2D}{3} & 0\\
    E & 0 & \frac{D}{3}
    \end{bmatrix},
\end{equation}{}
where $D$ is the zero-field splitting and $E$ is the non-axial zero-field parameter. For most practical purposes one can consider $E\approx0$.\\
The effect of an electric field on the spin is rather more complicated and for a solid state system it depends on symmetry conditions. For a $C_{3v}$ symmetry system like the nitrogen-vacancy center, the effect of a linear electric field can be described by the following approximate Hamiltonian:\citep{Kiel1972,VanOort1990}

\begin{align*}
    \hat{H}_\text{elec} =\; & \mathcal{E}_X\left[-\mathcal{P}_{11}(\hat{S}_X\hat{S}_Y+\hat{S}_Y\hat{S}_X) + \mathcal{P}_{15}(\hat{S}_Y\hat{S}_Z+\hat{S}_Z\hat{S}_Y\right]\\
    & + \mathcal{E}_Y\left[\mathcal{P}_{11}(\hat{S}_X^2 - \hat{S}_Y^2) + \mathcal{P}_{15}(\hat{S}_X\hat{S}_Z+\hat{S}_Z\hat{S}_X)\right]\\
    & + \mathcal{E}_Z\mathcal{P}_{31}(\hat{S}_Z^2 - \frac{1}{3}S(S+1))
\end{align*}

where, $\mathcal{E}_i$ are the component of the effective electric field. External strain/pressure is manifested as a crystal strain electric field and is incorporated in the effective electric field. $\mathcal{P}_{ij}$ are the components of a third rank coupling tensor\citep{Kiel1972}. These coupling constants can be determined by electron paramagnetic resonance experiments. In practice $\mathcal{P}_{15}$ arises from the mixing of the $m_s=\pm1$ and $m_s=0$ spin states\citep{Kiel1972}, and is negligible with respect to the other coupling terms.

\section{The Rotating Wave Approximation}
\label{app:RWA}

In the following, the approximate Hamiltonian for a NV qubit is derived using the rotating wave approximation (RWA). It is a simplified version of the second term of Eq.~\eqref{eqn:fullHamiltonian}.\\
First, let us consider a single NV center in a magnetic field. The magnetic field is composed of two parts: The static component $B_{\parallel}$ points along $z$ and a microwave field with amplitude $2 B_\perp$ oscillating along $x$ with a frequency $\omega_{mw}$ and phase $\varphi$. The magnetic field $\Vec{B}$ can thus be written in a slightly odd form, the usefulness of which will become apparent in the following steps.
\begin{equation}
\begin{split}
    \Vec{B} &= 
        \begin{pmatrix}
        2 B_\perp  \cos(\omega_{mw}t+\varphi)\\
        0\\
        B_{\parallel}
        \end{pmatrix}\\
    &= \begin{pmatrix}
        B_\perp (\cos(\omega_{mw}t+\varphi)+\cos(\omega_{mw}t+\varphi))\\
        B_\perp (\sin(\omega_{mw}t+\varphi)-\sin(\omega_{mw}t+\varphi))\\
        B_{\parallel}
        \end{pmatrix}
\end{split}
\end{equation}
The splitting of the energy levels of the NV center ground state due to the static magnetic field allows the description as a two level system only considering $\ket{0}$ and $\ket{1}$. $\hat{\Vec{\sigma}}$ are the Pauli matrices used to describe the corresponding spin operators which are denoted with $\hat{\Vec{\boldsymbol{s}}}=\frac{\hbar}{2}\hat{\Vec{\sigma}}$ to distinguish them from the three level system $\hat{\Vec{S}}$ and generic $\hat{\Vec{s}}$ spin operators. The resulting Hamiltonian reflects the interaction between the field and the NV spin. 
\begin{equation}
    \begin{split}
        \hat{H}_{NV-B} =& \gamma_{nv}\Vec{B}\cdot\hat{\Vec{\boldsymbol{s}}}\\
        =& \gamma_{nv} (B_\perp(\cos(\omega_{mw}t+\varphi)\hat{\boldsymbol{s}}_X+\sin(\omega_{mw}t+\varphi)\hat{\boldsymbol{s}}_Y)\\
        & + B_\perp(\cos(\omega_{mw}t+\varphi)\hat{\boldsymbol{s}}_X-\sin(\omega_{mw}t+\varphi)\hat{\boldsymbol{s}}_Y)\\
        & + B_{\parallel} \hat{\boldsymbol{s}}_Z)\\
        =&\gamma_{nv}(\hat{A}_{+} + \hat{A}_{-} + B_{\parallel} \hat{\boldsymbol{s}}_Z),
    \end{split}
\end{equation}
with $\hat{A}_{+}=B_\perp\left(\cos(\omega_{mw}t+\varphi)\hat{\boldsymbol{s}}_X+\sin(\omega_{mw}t+\varphi)\hat{\boldsymbol{s}}_Y\right)$ and $\hat{A}_{-}=B_\perp\left(\cos(\omega_{mw}t+\varphi)\hat{\boldsymbol{s}}_X-\sin(\omega_{mw}t+\varphi)\hat{\boldsymbol{s}}_Y\right)$.\\
In the rotating frame of the microwave field with the transformation $U = \text{exp}(i \omega_{mw} \hat{\boldsymbol{s}}_Z t/\hbar)$ this Hamiltonian becomes:
\begin{equation}
    \begin{split}
        \Tilde{H}_{NV-B} &= U \hat{H}_{NV-B} U^\dagger -i \hbar U\dot{U}^{\dagger}\\
        &= \gamma_{nv} \left( U \hat{A}_{+} U^\dagger + U \hat{A}_{-} U^\dagger + U B_{\parallel} \hat{\boldsymbol{s}}_Z U^\dagger \right) -i \hbar U\dot{U}^{\dagger}.
    \end{split}
\end{equation}
Using the following relations the corresponding spin operators may be determined.
\begin{equation}
    \begin{split}
        U \hat{\boldsymbol{s}}_X U^\dagger &= \cos(\omega_{mw}t)\hat{\boldsymbol{s}}_X - \sin(\omega_{mw}t )\hat{\boldsymbol{s}}_Y\\
        U \hat{\boldsymbol{s}}_Y U^\dagger &= \sin(\omega_{mw}t )\hat{\boldsymbol{s}}_X + \cos(\omega_{mw}t )\hat{\boldsymbol{s}}_Y\\
        U \hat{\boldsymbol{s}}_Z U^\dagger &= \hat{\boldsymbol{s}}_Z\\
        U \dot{U}^\dagger &= -\frac{i}{\hbar} \omega_{mw}\hat{\boldsymbol{s}}_Z
    \end{split}
\end{equation}
The different terms in the rotating frame, after some algebra and trigonometric identities, can be written as
\begin{equation}
    \begin{split}
        U \hat{A}_{+} U^\dagger &= B_\perp (\cos((\omega_{mw}-\omega_{mw})t+\varphi)\hat{\boldsymbol{s}}_X\\
        &\quad +\sin((\omega_{mw}-\omega_{mw})t+\varphi)\hat{\boldsymbol{s}}_Y),\\
        U \hat{A}_{-} U^\dagger &= B_\perp (\cos((\omega_{mw}+\omega_{mw})t+\varphi)\hat{\boldsymbol{s}}_X\\
        &\quad -\sin((\omega_{mw}+\omega_{mw})t+\varphi)\hat{\boldsymbol{s}}_Y),\\
        U B_{\parallel} \hat{\boldsymbol{s}}_Z U^\dagger &= B_{\parallel} \hat{\boldsymbol{s}}_Z,\\
        -i \hbar U\dot{U}^{\dagger} &= - \omega_{mw}\hat{\boldsymbol{s}}_Z.
    \end{split}
\end{equation}
One may notice that the terms in $U \hat{A}_{-} U^\dagger$ are of much higher frequency than the rest. Hence, in the rotating wave approximation, it is assumed that they average out and their contribution is negligible. Under this approximation, the Hamiltonian in the rotating frame depends only on the Rabi frequency $\Omega = \gamma_{nv} B_\perp$, the detuning $\Delta = \omega_{nv}-\omega_{mw}$, with the NV center's resonant frequency $\omega_{nv}=\gamma_{nv} B_{\parallel}$, and the phase~$\varphi$:
\begin{equation}
    \begin{split}
         \hat{H}_\text{RWA}&\approx \Tilde{H}_{NV-B}\\
        &= \Delta \hat{\boldsymbol{s}}_Z + \Omega\left(\cos(\varphi)\hat{\boldsymbol{s}}_X + \sin(\varphi) \hat{\boldsymbol{s}}_Y\right).
    \end{split}
\end{equation}
\section*{References}


%

\end{document}